\documentclass[longauth]{aa}

\usepackage{graphicx}
\usepackage{natbib}
\usepackage{scalerel}

\usepackage[table]{xcolor}

\usepackage{txfonts}

\usepackage{hyperref}
\hypersetup{
    colorlinks=true,
    linkcolor=blue,
    filecolor=magenta,      
    urlcolor=blue,
    citecolor=blue
}
\makeatletter
\renewcommand*\aa@pageof{, page \thepage{} of \pageref*{LastPage}}
\makeatother

\usepackage[utf8]{inputenc}
\usepackage{multirow}

\usepackage[switch, modulo]{lineno}

\usepackage{euclid}
\usepackage{orcidlink}
\let\orcid\orcidlink

\begin{document} 

\title{Euclid Quick Data Release (Q1)}
\subtitle{From images to multi-wavelength catalogues: The \Euclid MERge Processing Function\thanks{We wish to remember our colleagues Mario Nonino and Stefano Pilo, who
played an important role in OU-MER, and, more importantly, in our lives.}}

\author{Euclid Collaboration: E.~Romelli\orcid{0000-0003-3069-9222}\thanks{\email{erik.romelli@inaf.it}}\inst{\ref{aff1}}
\and M.~K\"ummel\orcid{0000-0003-2791-2117}\inst{\ref{aff2}}
\and H.~Dole\orcid{0000-0002-9767-3839}\inst{\ref{aff3}}
\and J.~Gracia-Carpio\inst{\ref{aff4}}
\and E.~Merlin\orcid{0000-0001-6870-8900}\inst{\ref{aff5}}
\and S.~Galeotta\orcid{0000-0002-3748-5115}\inst{\ref{aff1}}
\and Y.~Fang\inst{\ref{aff2}}
\and M.~Castellano\orcid{0000-0001-9875-8263}\inst{\ref{aff5}}
\and F.~Caro\inst{\ref{aff5}}
\and E.~Soubrie\orcid{0000-0001-9295-1863}\inst{\ref{aff3}}
\and L.~Maurin\orcid{0000-0002-8406-0857}\inst{\ref{aff3}}
\and R.~Cabanac\orcid{0000-0001-6679-2600}\inst{\ref{aff6}}
\and P.~Dimauro\orcid{0000-0001-7399-2854}\inst{\ref{aff5},\ref{aff7}}
\and M.~Huertas-Company\orcid{0000-0002-1416-8483}\inst{\ref{aff8},\ref{aff9},\ref{aff10},\ref{aff11}}
\and M.~D.~Lepinzan\orcid{0000-0003-1287-9801}\inst{\ref{aff12},\ref{aff1}}
\and T.~Vassallo\orcid{0000-0001-6512-6358}\inst{\ref{aff2},\ref{aff1}}
\and M.~Walmsley\orcid{0000-0002-6408-4181}\inst{\ref{aff13},\ref{aff14}}
\and I.~A.~Zinchenko\orcid{0000-0002-2944-2449}\inst{\ref{aff2}}
\and A.~Boucaud\orcid{0000-0001-7387-2633}\inst{\ref{aff15}}
\and A.~Calabro\orcid{0000-0003-2536-1614}\inst{\ref{aff5}}
\and V.~Roscani\inst{\ref{aff5}}
\and A.~Tramacere\orcid{0000-0002-8186-3793}\inst{\ref{aff16}}
\and M.~Douspis\orcid{0000-0003-4203-3954}\inst{\ref{aff3}}
\and A.~Fontana\orcid{0000-0003-3820-2823}\inst{\ref{aff5}}
\and N.~Aghanim\orcid{0000-0002-6688-8992}\inst{\ref{aff3}}
\and B.~Altieri\orcid{0000-0003-3936-0284}\inst{\ref{aff17}}
\and A.~Amara\inst{\ref{aff18}}
\and S.~Andreon\orcid{0000-0002-2041-8784}\inst{\ref{aff19}}
\and N.~Auricchio\orcid{0000-0003-4444-8651}\inst{\ref{aff20}}
\and H.~Aussel\orcid{0000-0002-1371-5705}\inst{\ref{aff21}}
\and C.~Baccigalupi\orcid{0000-0002-8211-1630}\inst{\ref{aff22},\ref{aff1},\ref{aff23},\ref{aff24}}
\and M.~Baldi\orcid{0000-0003-4145-1943}\inst{\ref{aff25},\ref{aff20},\ref{aff26}}
\and A.~Balestra\orcid{0000-0002-6967-261X}\inst{\ref{aff27}}
\and S.~Bardelli\orcid{0000-0002-8900-0298}\inst{\ref{aff20}}
\and A.~Basset\inst{\ref{aff28}}
\and P.~Battaglia\orcid{0000-0002-7337-5909}\inst{\ref{aff20}}
\and A.~N.~Belikov\inst{\ref{aff29},\ref{aff30}}
\and A.~Biviano\orcid{0000-0002-0857-0732}\inst{\ref{aff1},\ref{aff22}}
\and A.~Bonchi\orcid{0000-0002-2667-5482}\inst{\ref{aff31}}
\and D.~Bonino\orcid{0000-0002-3336-9977}\inst{\ref{aff32}}
\and E.~Branchini\orcid{0000-0002-0808-6908}\inst{\ref{aff33},\ref{aff34},\ref{aff19}}
\and M.~Brescia\orcid{0000-0001-9506-5680}\inst{\ref{aff35},\ref{aff36}}
\and J.~Brinchmann\orcid{0000-0003-4359-8797}\inst{\ref{aff37},\ref{aff38}}
\and S.~Camera\orcid{0000-0003-3399-3574}\inst{\ref{aff39},\ref{aff40},\ref{aff32}}
\and G.~Ca\~nas-Herrera\orcid{0000-0003-2796-2149}\inst{\ref{aff41},\ref{aff42},\ref{aff43}}
\and V.~Capobianco\orcid{0000-0002-3309-7692}\inst{\ref{aff32}}
\and C.~Carbone\orcid{0000-0003-0125-3563}\inst{\ref{aff44}}
\and J.~Carretero\orcid{0000-0002-3130-0204}\inst{\ref{aff45},\ref{aff46}}
\and S.~Casas\orcid{0000-0002-4751-5138}\inst{\ref{aff47}}
\and F.~J.~Castander\orcid{0000-0001-7316-4573}\inst{\ref{aff48},\ref{aff49}}
\and G.~Castignani\orcid{0000-0001-6831-0687}\inst{\ref{aff20}}
\and S.~Cavuoti\orcid{0000-0002-3787-4196}\inst{\ref{aff36},\ref{aff50}}
\and K.~C.~Chambers\orcid{0000-0001-6965-7789}\inst{\ref{aff51}}
\and A.~Cimatti\inst{\ref{aff52}}
\and C.~Colodro-Conde\inst{\ref{aff8}}
\and G.~Congedo\orcid{0000-0003-2508-0046}\inst{\ref{aff53}}
\and C.~J.~Conselice\orcid{0000-0003-1949-7638}\inst{\ref{aff14}}
\and L.~Conversi\orcid{0000-0002-6710-8476}\inst{\ref{aff54},\ref{aff17}}
\and Y.~Copin\orcid{0000-0002-5317-7518}\inst{\ref{aff55}}
\and F.~Courbin\orcid{0000-0003-0758-6510}\inst{\ref{aff56},\ref{aff57}}
\and H.~M.~Courtois\orcid{0000-0003-0509-1776}\inst{\ref{aff58}}
\and M.~Cropper\orcid{0000-0003-4571-9468}\inst{\ref{aff59}}
\and J.-G.~Cuby\orcid{0000-0002-8767-1442}\inst{\ref{aff60},\ref{aff61}}
\and A.~Da~Silva\orcid{0000-0002-6385-1609}\inst{\ref{aff62},\ref{aff63}}
\and R.~da~Silva\orcid{0000-0003-4788-677X}\inst{\ref{aff5},\ref{aff31}}
\and H.~Degaudenzi\orcid{0000-0002-5887-6799}\inst{\ref{aff16}}
\and S.~de~la~Torre\inst{\ref{aff61}}
\and G.~De~Lucia\orcid{0000-0002-6220-9104}\inst{\ref{aff1}}
\and A.~M.~Di~Giorgio\orcid{0000-0002-4767-2360}\inst{\ref{aff64}}
\and F.~Dubath\orcid{0000-0002-6533-2810}\inst{\ref{aff16}}
\and C.~A.~J.~Duncan\orcid{0009-0003-3573-0791}\inst{\ref{aff53},\ref{aff14}}
\and X.~Dupac\inst{\ref{aff17}}
\and S.~Dusini\orcid{0000-0002-1128-0664}\inst{\ref{aff65}}
\and S.~Escoffier\orcid{0000-0002-2847-7498}\inst{\ref{aff66}}
\and M.~Fabricius\orcid{0000-0002-7025-6058}\inst{\ref{aff4},\ref{aff2}}
\and M.~Farina\orcid{0000-0002-3089-7846}\inst{\ref{aff64}}
\and R.~Farinelli\inst{\ref{aff20}}
\and F.~Faustini\orcid{0000-0001-6274-5145}\inst{\ref{aff5},\ref{aff31}}
\and S.~Ferriol\inst{\ref{aff55}}
\and F.~Finelli\orcid{0000-0002-6694-3269}\inst{\ref{aff20},\ref{aff67}}
\and S.~Fotopoulou\orcid{0000-0002-9686-254X}\inst{\ref{aff68}}
\and N.~Fourmanoit\orcid{0009-0005-6816-6925}\inst{\ref{aff66}}
\and M.~Frailis\orcid{0000-0002-7400-2135}\inst{\ref{aff1}}
\and E.~Franceschi\orcid{0000-0002-0585-6591}\inst{\ref{aff20}}
\and K.~George\orcid{0000-0002-1734-8455}\inst{\ref{aff2}}
\and W.~Gillard\orcid{0000-0003-4744-9748}\inst{\ref{aff66}}
\and B.~Gillis\orcid{0000-0002-4478-1270}\inst{\ref{aff53}}
\and C.~Giocoli\orcid{0000-0002-9590-7961}\inst{\ref{aff20},\ref{aff26}}
\and B.~R.~Granett\orcid{0000-0003-2694-9284}\inst{\ref{aff19}}
\and A.~Grazian\orcid{0000-0002-5688-0663}\inst{\ref{aff27}}
\and F.~Grupp\inst{\ref{aff4},\ref{aff2}}
\and S.~Gwyn\orcid{0000-0001-8221-8406}\inst{\ref{aff69}}
\and S.~V.~H.~Haugan\orcid{0000-0001-9648-7260}\inst{\ref{aff70}}
\and J.~Hoar\inst{\ref{aff17}}
\and H.~Hoekstra\orcid{0000-0002-0641-3231}\inst{\ref{aff43}}
\and W.~Holmes\inst{\ref{aff71}}
\and I.~M.~Hook\orcid{0000-0002-2960-978X}\inst{\ref{aff72}}
\and F.~Hormuth\inst{\ref{aff73}}
\and A.~Hornstrup\orcid{0000-0002-3363-0936}\inst{\ref{aff74},\ref{aff75}}
\and K.~Jahnke\orcid{0000-0003-3804-2137}\inst{\ref{aff76}}
\and M.~Jhabvala\inst{\ref{aff77}}
\and B.~Joachimi\orcid{0000-0001-7494-1303}\inst{\ref{aff78}}
\and E.~Keih\"anen\orcid{0000-0003-1804-7715}\inst{\ref{aff79}}
\and S.~Kermiche\orcid{0000-0002-0302-5735}\inst{\ref{aff66}}
\and A.~Kiessling\orcid{0000-0002-2590-1273}\inst{\ref{aff71}}
\and M.~Kilbinger\orcid{0000-0001-9513-7138}\inst{\ref{aff21}}
\and B.~Kubik\orcid{0009-0006-5823-4880}\inst{\ref{aff55}}
\and K.~Kuijken\orcid{0000-0002-3827-0175}\inst{\ref{aff43}}
\and M.~Kunz\orcid{0000-0002-3052-7394}\inst{\ref{aff80}}
\and H.~Kurki-Suonio\orcid{0000-0002-4618-3063}\inst{\ref{aff81},\ref{aff82}}
\and R.~Laureijs\inst{\ref{aff29},\ref{aff41}}
\and Q.~Le~Boulc'h\inst{\ref{aff83}}
\and A.~M.~C.~Le~Brun\orcid{0000-0002-0936-4594}\inst{\ref{aff84}}
\and D.~Le~Mignant\orcid{0000-0002-5339-5515}\inst{\ref{aff61}}
\and P.~Liebing\inst{\ref{aff59}}
\and S.~Ligori\orcid{0000-0003-4172-4606}\inst{\ref{aff32}}
\and P.~B.~Lilje\orcid{0000-0003-4324-7794}\inst{\ref{aff70}}
\and V.~Lindholm\orcid{0000-0003-2317-5471}\inst{\ref{aff81},\ref{aff82}}
\and I.~Lloro\orcid{0000-0001-5966-1434}\inst{\ref{aff85}}
\and G.~Mainetti\orcid{0000-0003-2384-2377}\inst{\ref{aff83}}
\and D.~Maino\inst{\ref{aff86},\ref{aff44},\ref{aff87}}
\and E.~Maiorano\orcid{0000-0003-2593-4355}\inst{\ref{aff20}}
\and O.~Mansutti\orcid{0000-0001-5758-4658}\inst{\ref{aff1}}
\and S.~Marcin\inst{\ref{aff88}}
\and O.~Marggraf\orcid{0000-0001-7242-3852}\inst{\ref{aff89}}
\and M.~Martinelli\orcid{0000-0002-6943-7732}\inst{\ref{aff5},\ref{aff90}}
\and N.~Martinet\orcid{0000-0003-2786-7790}\inst{\ref{aff61}}
\and F.~Marulli\orcid{0000-0002-8850-0303}\inst{\ref{aff91},\ref{aff20},\ref{aff26}}
\and R.~Massey\orcid{0000-0002-6085-3780}\inst{\ref{aff92}}
\and S.~Maurogordato\inst{\ref{aff93}}
\and H.~J.~McCracken\orcid{0000-0002-9489-7765}\inst{\ref{aff94}}
\and E.~Medinaceli\orcid{0000-0002-4040-7783}\inst{\ref{aff20}}
\and S.~Mei\orcid{0000-0002-2849-559X}\inst{\ref{aff15},\ref{aff95}}
\and M.~Melchior\inst{\ref{aff88}}
\and Y.~Mellier\inst{\ref{aff96},\ref{aff94}}
\and M.~Meneghetti\orcid{0000-0003-1225-7084}\inst{\ref{aff20},\ref{aff26}}
\and G.~Meylan\inst{\ref{aff97}}
\and A.~Mora\orcid{0000-0002-1922-8529}\inst{\ref{aff98}}
\and M.~Moresco\orcid{0000-0002-7616-7136}\inst{\ref{aff91},\ref{aff20}}
\and L.~Moscardini\orcid{0000-0002-3473-6716}\inst{\ref{aff91},\ref{aff20},\ref{aff26}}
\and R.~Nakajima\orcid{0009-0009-1213-7040}\inst{\ref{aff89}}
\and C.~Neissner\orcid{0000-0001-8524-4968}\inst{\ref{aff99},\ref{aff46}}
\and R.~C.~Nichol\orcid{0000-0003-0939-6518}\inst{\ref{aff18}}
\and S.-M.~Niemi\inst{\ref{aff41}}
\and J.~W.~Nightingale\orcid{0000-0002-8987-7401}\inst{\ref{aff100}}
\and C.~Padilla\orcid{0000-0001-7951-0166}\inst{\ref{aff99}}
\and S.~Paltani\orcid{0000-0002-8108-9179}\inst{\ref{aff16}}
\and F.~Pasian\orcid{0000-0002-4869-3227}\inst{\ref{aff1}}
\and K.~Pedersen\inst{\ref{aff101}}
\and W.~J.~Percival\orcid{0000-0002-0644-5727}\inst{\ref{aff102},\ref{aff103},\ref{aff104}}
\and V.~Pettorino\inst{\ref{aff41}}
\and S.~Pires\orcid{0000-0002-0249-2104}\inst{\ref{aff21}}
\and G.~Polenta\orcid{0000-0003-4067-9196}\inst{\ref{aff31}}
\and M.~Poncet\inst{\ref{aff28}}
\and L.~A.~Popa\inst{\ref{aff105}}
\and L.~Pozzetti\orcid{0000-0001-7085-0412}\inst{\ref{aff20}}
\and G.~D.~Racca\inst{\ref{aff41},\ref{aff43}}
\and F.~Raison\orcid{0000-0002-7819-6918}\inst{\ref{aff4}}
\and A.~Renzi\orcid{0000-0001-9856-1970}\inst{\ref{aff106},\ref{aff65}}
\and J.~Rhodes\orcid{0000-0002-4485-8549}\inst{\ref{aff71}}
\and G.~Riccio\inst{\ref{aff36}}
\and M.~Roncarelli\orcid{0000-0001-9587-7822}\inst{\ref{aff20}}
\and E.~Rossetti\orcid{0000-0003-0238-4047}\inst{\ref{aff25}}
\and R.~Saglia\orcid{0000-0003-0378-7032}\inst{\ref{aff2},\ref{aff4}}
\and Z.~Sakr\orcid{0000-0002-4823-3757}\inst{\ref{aff107},\ref{aff6},\ref{aff108}}
\and A.~G.~S\'anchez\orcid{0000-0003-1198-831X}\inst{\ref{aff4}}
\and D.~Sapone\orcid{0000-0001-7089-4503}\inst{\ref{aff109}}
\and B.~Sartoris\orcid{0000-0003-1337-5269}\inst{\ref{aff2},\ref{aff1}}
\and J.~A.~Schewtschenko\orcid{0000-0002-4913-6393}\inst{\ref{aff53}}
\and M.~Schirmer\orcid{0000-0003-2568-9994}\inst{\ref{aff76}}
\and P.~Schneider\orcid{0000-0001-8561-2679}\inst{\ref{aff89}}
\and M.~Scodeggio\inst{\ref{aff44}}
\and A.~Secroun\orcid{0000-0003-0505-3710}\inst{\ref{aff66}}
\and E.~Sefusatti\orcid{0000-0003-0473-1567}\inst{\ref{aff1},\ref{aff22},\ref{aff23}}
\and G.~Seidel\orcid{0000-0003-2907-353X}\inst{\ref{aff76}}
\and M.~Seiffert\orcid{0000-0002-7536-9393}\inst{\ref{aff71}}
\and S.~Serrano\orcid{0000-0002-0211-2861}\inst{\ref{aff49},\ref{aff110},\ref{aff48}}
\and P.~Simon\inst{\ref{aff89}}
\and C.~Sirignano\orcid{0000-0002-0995-7146}\inst{\ref{aff106},\ref{aff65}}
\and G.~Sirri\orcid{0000-0003-2626-2853}\inst{\ref{aff26}}
\and A.~Spurio~Mancini\orcid{0000-0001-5698-0990}\inst{\ref{aff111}}
\and L.~Stanco\orcid{0000-0002-9706-5104}\inst{\ref{aff65}}
\and J.~Steinwagner\orcid{0000-0001-7443-1047}\inst{\ref{aff4}}
\and P.~Tallada-Cresp\'{i}\orcid{0000-0002-1336-8328}\inst{\ref{aff45},\ref{aff46}}
\and D.~Tavagnacco\orcid{0000-0001-7475-9894}\inst{\ref{aff1}}
\and A.~N.~Taylor\inst{\ref{aff53}}
\and H.~I.~Teplitz\orcid{0000-0002-7064-5424}\inst{\ref{aff112}}
\and I.~Tereno\inst{\ref{aff62},\ref{aff113}}
\and N.~Tessore\orcid{0000-0002-9696-7931}\inst{\ref{aff78}}
\and S.~Toft\orcid{0000-0003-3631-7176}\inst{\ref{aff114},\ref{aff115}}
\and R.~Toledo-Moreo\orcid{0000-0002-2997-4859}\inst{\ref{aff116}}
\and F.~Torradeflot\orcid{0000-0003-1160-1517}\inst{\ref{aff46},\ref{aff45}}
\and A.~Tsyganov\inst{\ref{aff117}}
\and I.~Tutusaus\orcid{0000-0002-3199-0399}\inst{\ref{aff6}}
\and E.~A.~Valentijn\inst{\ref{aff29}}
\and L.~Valenziano\orcid{0000-0002-1170-0104}\inst{\ref{aff20},\ref{aff67}}
\and J.~Valiviita\orcid{0000-0001-6225-3693}\inst{\ref{aff81},\ref{aff82}}
\and G.~Verdoes~Kleijn\orcid{0000-0001-5803-2580}\inst{\ref{aff29}}
\and A.~Veropalumbo\orcid{0000-0003-2387-1194}\inst{\ref{aff19},\ref{aff34},\ref{aff33}}
\and Y.~Wang\orcid{0000-0002-4749-2984}\inst{\ref{aff112}}
\and J.~Weller\orcid{0000-0002-8282-2010}\inst{\ref{aff2},\ref{aff4}}
\and G.~Zamorani\orcid{0000-0002-2318-301X}\inst{\ref{aff20}}
\and F.~M.~Zerbi\inst{\ref{aff19}}
\and E.~Zucca\orcid{0000-0002-5845-8132}\inst{\ref{aff20}}
\and V.~Allevato\orcid{0000-0001-7232-5152}\inst{\ref{aff36}}
\and M.~Ballardini\orcid{0000-0003-4481-3559}\inst{\ref{aff118},\ref{aff119},\ref{aff20}}
\and M.~Bolzonella\orcid{0000-0003-3278-4607}\inst{\ref{aff20}}
\and E.~Bozzo\orcid{0000-0002-8201-1525}\inst{\ref{aff16}}
\and C.~Burigana\orcid{0000-0002-3005-5796}\inst{\ref{aff120},\ref{aff67}}
\and A.~Cappi\inst{\ref{aff20},\ref{aff93}}
\and P.~Casenove\orcid{0009-0006-6736-1670}\inst{\ref{aff28}}
\and D.~Di~Ferdinando\inst{\ref{aff26}}
\and J.~A.~Escartin~Vigo\inst{\ref{aff4}}
\and L.~Gabarra\orcid{0000-0002-8486-8856}\inst{\ref{aff121}}
\and W.~G.~Hartley\inst{\ref{aff16}}
\and H.~Israel\orcid{0000-0002-3045-4412}\inst{\ref{aff122}}
\and J.~Mart\'{i}n-Fleitas\orcid{0000-0002-8594-569X}\inst{\ref{aff98}}
\and S.~Matthew\orcid{0000-0001-8448-1697}\inst{\ref{aff53}}
\and N.~Mauri\orcid{0000-0001-8196-1548}\inst{\ref{aff52},\ref{aff26}}
\and R.~B.~Metcalf\orcid{0000-0003-3167-2574}\inst{\ref{aff91},\ref{aff20}}
\and A.~Pezzotta\orcid{0000-0003-0726-2268}\inst{\ref{aff123},\ref{aff4}}
\and M.~P\"ontinen\orcid{0000-0001-5442-2530}\inst{\ref{aff81}}
\and C.~Porciani\orcid{0000-0002-7797-2508}\inst{\ref{aff89}}
\and I.~Risso\orcid{0000-0003-2525-7761}\inst{\ref{aff124}}
\and V.~Scottez\inst{\ref{aff96},\ref{aff125}}
\and M.~Sereno\orcid{0000-0003-0302-0325}\inst{\ref{aff20},\ref{aff26}}
\and M.~Tenti\orcid{0000-0002-4254-5901}\inst{\ref{aff26}}
\and M.~Viel\orcid{0000-0002-2642-5707}\inst{\ref{aff22},\ref{aff1},\ref{aff24},\ref{aff23},\ref{aff126}}
\and M.~Wiesmann\orcid{0009-0000-8199-5860}\inst{\ref{aff70}}
\and Y.~Akrami\orcid{0000-0002-2407-7956}\inst{\ref{aff127},\ref{aff128}}
\and S.~Alvi\orcid{0000-0001-5779-8568}\inst{\ref{aff118}}
\and I.~T.~Andika\orcid{0000-0001-6102-9526}\inst{\ref{aff129},\ref{aff130}}
\and S.~Anselmi\orcid{0000-0002-3579-9583}\inst{\ref{aff65},\ref{aff106},\ref{aff131}}
\and M.~Archidiacono\orcid{0000-0003-4952-9012}\inst{\ref{aff86},\ref{aff87}}
\and F.~Atrio-Barandela\orcid{0000-0002-2130-2513}\inst{\ref{aff132}}
\and P.~Bergamini\orcid{0000-0003-1383-9414}\inst{\ref{aff86},\ref{aff20}}
\and D.~Bertacca\orcid{0000-0002-2490-7139}\inst{\ref{aff106},\ref{aff27},\ref{aff65}}
\and M.~Bethermin\orcid{0000-0002-3915-2015}\inst{\ref{aff133}}
\and A.~Blanchard\orcid{0000-0001-8555-9003}\inst{\ref{aff6}}
\and L.~Blot\orcid{0000-0002-9622-7167}\inst{\ref{aff134},\ref{aff84}}
\and H.~B\"ohringer\orcid{0000-0001-8241-4204}\inst{\ref{aff4},\ref{aff135},\ref{aff136}}
\and S.~Borgani\orcid{0000-0001-6151-6439}\inst{\ref{aff12},\ref{aff22},\ref{aff1},\ref{aff23},\ref{aff126}}
\and M.~L.~Brown\orcid{0000-0002-0370-8077}\inst{\ref{aff14}}
\and S.~Bruton\orcid{0000-0002-6503-5218}\inst{\ref{aff137}}
\and B.~Camacho~Quevedo\orcid{0000-0002-8789-4232}\inst{\ref{aff49},\ref{aff48}}
\and C.~S.~Carvalho\inst{\ref{aff113}}
\and T.~Castro\orcid{0000-0002-6292-3228}\inst{\ref{aff1},\ref{aff23},\ref{aff22},\ref{aff126}}
\and R.~Chary\orcid{0000-0001-7583-0621}\inst{\ref{aff112},\ref{aff138}}
\and F.~Cogato\orcid{0000-0003-4632-6113}\inst{\ref{aff91},\ref{aff20}}
\and S.~Conseil\orcid{0000-0002-3657-4191}\inst{\ref{aff55}}
\and A.~R.~Cooray\orcid{0000-0002-3892-0190}\inst{\ref{aff139}}
\and O.~Cucciati\orcid{0000-0002-9336-7551}\inst{\ref{aff20}}
\and S.~Davini\orcid{0000-0003-3269-1718}\inst{\ref{aff34}}
\and F.~De~Paolis\orcid{0000-0001-6460-7563}\inst{\ref{aff140},\ref{aff141},\ref{aff142}}
\and G.~Desprez\orcid{0000-0001-8325-1742}\inst{\ref{aff29}}
\and A.~D\'iaz-S\'anchez\orcid{0000-0003-0748-4768}\inst{\ref{aff143}}
\and J.~J.~Diaz\inst{\ref{aff8}}
\and S.~Di~Domizio\orcid{0000-0003-2863-5895}\inst{\ref{aff33},\ref{aff34}}
\and J.~M.~Diego\orcid{0000-0001-9065-3926}\inst{\ref{aff144}}
\and P.-A.~Duc\orcid{0000-0003-3343-6284}\inst{\ref{aff133}}
\and A.~Enia\orcid{0000-0002-0200-2857}\inst{\ref{aff25},\ref{aff20}}
\and A.~M.~N.~Ferguson\inst{\ref{aff53}}
\and A.~Finoguenov\orcid{0000-0002-4606-5403}\inst{\ref{aff81}}
\and A.~Franco\orcid{0000-0002-4761-366X}\inst{\ref{aff141},\ref{aff140},\ref{aff142}}
\and K.~Ganga\orcid{0000-0001-8159-8208}\inst{\ref{aff15}}
\and J.~Garc\'ia-Bellido\orcid{0000-0002-9370-8360}\inst{\ref{aff127}}
\and T.~Gasparetto\orcid{0000-0002-7913-4866}\inst{\ref{aff1}}
\and R.~Gavazzi\orcid{0000-0002-5540-6935}\inst{\ref{aff61},\ref{aff94}}
\and E.~Gaztanaga\orcid{0000-0001-9632-0815}\inst{\ref{aff48},\ref{aff49},\ref{aff145}}
\and F.~Giacomini\orcid{0000-0002-3129-2814}\inst{\ref{aff26}}
\and F.~Gianotti\orcid{0000-0003-4666-119X}\inst{\ref{aff20}}
\and G.~Gozaliasl\orcid{0000-0002-0236-919X}\inst{\ref{aff146},\ref{aff81}}
\and A.~Gregorio\orcid{0000-0003-4028-8785}\inst{\ref{aff12},\ref{aff1},\ref{aff23}}
\and M.~Guidi\orcid{0000-0001-9408-1101}\inst{\ref{aff25},\ref{aff20}}
\and C.~M.~Gutierrez\orcid{0000-0001-7854-783X}\inst{\ref{aff147}}
\and A.~Hall\orcid{0000-0002-3139-8651}\inst{\ref{aff53}}
\and C.~Hern\'andez-Monteagudo\orcid{0000-0001-5471-9166}\inst{\ref{aff148},\ref{aff8}}
\and H.~Hildebrandt\orcid{0000-0002-9814-3338}\inst{\ref{aff149}}
\and J.~Hjorth\orcid{0000-0002-4571-2306}\inst{\ref{aff101}}
\and J.~J.~E.~Kajava\orcid{0000-0002-3010-8333}\inst{\ref{aff150},\ref{aff151}}
\and Y.~Kang\orcid{0009-0000-8588-7250}\inst{\ref{aff16}}
\and V.~Kansal\orcid{0000-0002-4008-6078}\inst{\ref{aff152},\ref{aff153}}
\and D.~Karagiannis\orcid{0000-0002-4927-0816}\inst{\ref{aff118},\ref{aff154}}
\and K.~Kiiveri\inst{\ref{aff79}}
\and C.~C.~Kirkpatrick\inst{\ref{aff79}}
\and S.~Kruk\orcid{0000-0001-8010-8879}\inst{\ref{aff17}}
\and V.~Le~Brun\orcid{0000-0002-5027-1939}\inst{\ref{aff61}}
\and J.~Le~Graet\orcid{0000-0001-6523-7971}\inst{\ref{aff66}}
\and L.~Legrand\orcid{0000-0003-0610-5252}\inst{\ref{aff155},\ref{aff156}}
\and M.~Lembo\orcid{0000-0002-5271-5070}\inst{\ref{aff118},\ref{aff119}}
\and F.~Lepori\orcid{0009-0000-5061-7138}\inst{\ref{aff157}}
\and G.~F.~Lesci\orcid{0000-0002-4607-2830}\inst{\ref{aff91},\ref{aff20}}
\and J.~Lesgourgues\orcid{0000-0001-7627-353X}\inst{\ref{aff47}}
\and L.~Leuzzi\orcid{0009-0006-4479-7017}\inst{\ref{aff91},\ref{aff20}}
\and T.~I.~Liaudat\orcid{0000-0002-9104-314X}\inst{\ref{aff158}}
\and A.~Loureiro\orcid{0000-0002-4371-0876}\inst{\ref{aff159},\ref{aff160}}
\and J.~Macias-Perez\orcid{0000-0002-5385-2763}\inst{\ref{aff161}}
\and G.~Maggio\orcid{0000-0003-4020-4836}\inst{\ref{aff1}}
\and M.~Magliocchetti\orcid{0000-0001-9158-4838}\inst{\ref{aff64}}
\and E.~A.~Magnier\orcid{0000-0002-7965-2815}\inst{\ref{aff51}}
\and C.~Mancini\orcid{0000-0002-4297-0561}\inst{\ref{aff44}}
\and F.~Mannucci\orcid{0000-0002-4803-2381}\inst{\ref{aff162}}
\and R.~Maoli\orcid{0000-0002-6065-3025}\inst{\ref{aff163},\ref{aff5}}
\and C.~J.~A.~P.~Martins\orcid{0000-0002-4886-9261}\inst{\ref{aff164},\ref{aff37}}
\and M.~Miluzio\inst{\ref{aff17},\ref{aff165}}
\and P.~Monaco\orcid{0000-0003-2083-7564}\inst{\ref{aff12},\ref{aff1},\ref{aff23},\ref{aff22}}
\and C.~Moretti\orcid{0000-0003-3314-8936}\inst{\ref{aff24},\ref{aff126},\ref{aff1},\ref{aff22},\ref{aff23}}
\and G.~Morgante\inst{\ref{aff20}}
\and S.~Nadathur\orcid{0000-0001-9070-3102}\inst{\ref{aff145}}
\and K.~Naidoo\orcid{0000-0002-9182-1802}\inst{\ref{aff145}}
\and A.~Navarro-Alsina\orcid{0000-0002-3173-2592}\inst{\ref{aff89}}
\and S.~Nesseris\orcid{0000-0002-0567-0324}\inst{\ref{aff127}}
\and F.~Passalacqua\orcid{0000-0002-8606-4093}\inst{\ref{aff106},\ref{aff65}}
\and K.~Paterson\orcid{0000-0001-8340-3486}\inst{\ref{aff76}}
\and L.~Patrizii\inst{\ref{aff26}}
\and A.~Pisani\orcid{0000-0002-6146-4437}\inst{\ref{aff66},\ref{aff166}}
\and D.~Potter\orcid{0000-0002-0757-5195}\inst{\ref{aff157}}
\and S.~Quai\orcid{0000-0002-0449-8163}\inst{\ref{aff91},\ref{aff20}}
\and M.~Radovich\orcid{0000-0002-3585-866X}\inst{\ref{aff27}}
\and P.~Reimberg\orcid{0000-0003-3410-0280}\inst{\ref{aff96}}
\and P.-F.~Rocci\inst{\ref{aff3}}
\and G.~Rodighiero\orcid{0000-0002-9415-2296}\inst{\ref{aff106},\ref{aff27}}
\and S.~Sacquegna\orcid{0000-0002-8433-6630}\inst{\ref{aff140},\ref{aff141},\ref{aff142}}
\and M.~Sahl\'en\orcid{0000-0003-0973-4804}\inst{\ref{aff167}}
\and D.~B.~Sanders\orcid{0000-0002-1233-9998}\inst{\ref{aff51}}
\and E.~Sarpa\orcid{0000-0002-1256-655X}\inst{\ref{aff24},\ref{aff126},\ref{aff23}}
\and C.~Scarlata\orcid{0000-0002-9136-8876}\inst{\ref{aff168}}
\and A.~Schneider\orcid{0000-0001-7055-8104}\inst{\ref{aff157}}
\and M.~Schultheis\inst{\ref{aff93}}
\and D.~Sciotti\orcid{0009-0008-4519-2620}\inst{\ref{aff5},\ref{aff90}}
\and E.~Sellentin\inst{\ref{aff169},\ref{aff43}}
\and L.~C.~Smith\orcid{0000-0002-3259-2771}\inst{\ref{aff170}}
\and K.~Tanidis\orcid{0000-0001-9843-5130}\inst{\ref{aff121}}
\and G.~Testera\inst{\ref{aff34}}
\and R.~Teyssier\orcid{0000-0001-7689-0933}\inst{\ref{aff166}}
\and S.~Tosi\orcid{0000-0002-7275-9193}\inst{\ref{aff33},\ref{aff34},\ref{aff19}}
\and A.~Troja\orcid{0000-0003-0239-4595}\inst{\ref{aff106},\ref{aff65}}
\and M.~Tucci\inst{\ref{aff16}}
\and C.~Valieri\inst{\ref{aff26}}
\and A.~Venhola\orcid{0000-0001-6071-4564}\inst{\ref{aff171}}
\and D.~Vergani\orcid{0000-0003-0898-2216}\inst{\ref{aff20}}
\and G.~Verza\orcid{0000-0002-1886-8348}\inst{\ref{aff172}}
\and P.~Vielzeuf\orcid{0000-0003-2035-9339}\inst{\ref{aff66}}
\and N.~A.~Walton\orcid{0000-0003-3983-8778}\inst{\ref{aff170}}
\and J.~R.~Weaver\orcid{0000-0003-1614-196X}\inst{\ref{aff173}}
\and J.~G.~Sorce\orcid{0000-0002-2307-2432}\inst{\ref{aff174},\ref{aff3}}
\and D.~Scott\orcid{0000-0002-6878-9840}\inst{\ref{aff175}}}
										   
\institute{INAF-Osservatorio Astronomico di Trieste, Via G. B. Tiepolo 11, 34143 Trieste, Italy\label{aff1}
\and
Universit\"ats-Sternwarte M\"unchen, Fakult\"at f\"ur Physik, Ludwig-Maximilians-Universit\"at M\"unchen, Scheinerstrasse 1, 81679 M\"unchen, Germany\label{aff2}
\and
Universit\'e Paris-Saclay, CNRS, Institut d'astrophysique spatiale, 91405, Orsay, France\label{aff3}
\and
Max Planck Institute for Extraterrestrial Physics, Giessenbachstr. 1, 85748 Garching, Germany\label{aff4}
\and
INAF-Osservatorio Astronomico di Roma, Via Frascati 33, 00078 Monteporzio Catone, Italy\label{aff5}
\and
Institut de Recherche en Astrophysique et Plan\'etologie (IRAP), Universit\'e de Toulouse, CNRS, UPS, CNES, 14 Av. Edouard Belin, 31400 Toulouse, France\label{aff6}
\and
Observatorio Nacional, Rua General Jose Cristino, 77-Bairro Imperial de Sao Cristovao, Rio de Janeiro, 20921-400, Brazil\label{aff7}
\and
Instituto de Astrof\'{\i}sica de Canarias, V\'{\i}a L\'actea, 38205 La Laguna, Tenerife, Spain\label{aff8}
\and
Instituto de Astrof\'isica de Canarias (IAC); Departamento de Astrof\'isica, Universidad de La Laguna (ULL), 38200, La Laguna, Tenerife, Spain\label{aff9}
\and
Universit\'e PSL, Observatoire de Paris, Sorbonne Universit\'e, CNRS, LERMA, 75014, Paris, France\label{aff10}
\and
Universit\'e Paris-Cit\'e, 5 Rue Thomas Mann, 75013, Paris, France\label{aff11}
\and
Dipartimento di Fisica - Sezione di Astronomia, Universit\`a di Trieste, Via Tiepolo 11, 34131 Trieste, Italy\label{aff12}
\and
David A. Dunlap Department of Astronomy \& Astrophysics, University of Toronto, 50 St George Street, Toronto, Ontario M5S 3H4, Canada\label{aff13}
\and
Jodrell Bank Centre for Astrophysics, Department of Physics and Astronomy, University of Manchester, Oxford Road, Manchester M13 9PL, UK\label{aff14}
\and
Universit\'e Paris Cit\'e, CNRS, Astroparticule et Cosmologie, 75013 Paris, France\label{aff15}
\and
Department of Astronomy, University of Geneva, ch. d'Ecogia 16, 1290 Versoix, Switzerland\label{aff16}
\and
ESAC/ESA, Camino Bajo del Castillo, s/n., Urb. Villafranca del Castillo, 28692 Villanueva de la Ca\~nada, Madrid, Spain\label{aff17}
\and
School of Mathematics and Physics, University of Surrey, Guildford, Surrey, GU2 7XH, UK\label{aff18}
\and
INAF-Osservatorio Astronomico di Brera, Via Brera 28, 20122 Milano, Italy\label{aff19}
\and
INAF-Osservatorio di Astrofisica e Scienza dello Spazio di Bologna, Via Piero Gobetti 93/3, 40129 Bologna, Italy\label{aff20}
\and
Universit\'e Paris-Saclay, Universit\'e Paris Cit\'e, CEA, CNRS, AIM, 91191, Gif-sur-Yvette, France\label{aff21}
\and
IFPU, Institute for Fundamental Physics of the Universe, via Beirut 2, 34151 Trieste, Italy\label{aff22}
\and
INFN, Sezione di Trieste, Via Valerio 2, 34127 Trieste TS, Italy\label{aff23}
\and
SISSA, International School for Advanced Studies, Via Bonomea 265, 34136 Trieste TS, Italy\label{aff24}
\and
Dipartimento di Fisica e Astronomia, Universit\`a di Bologna, Via Gobetti 93/2, 40129 Bologna, Italy\label{aff25}
\and
INFN-Sezione di Bologna, Viale Berti Pichat 6/2, 40127 Bologna, Italy\label{aff26}
\and
INAF-Osservatorio Astronomico di Padova, Via dell'Osservatorio 5, 35122 Padova, Italy\label{aff27}
\and
Centre National d'Etudes Spatiales -- Centre spatial de Toulouse, 18 avenue Edouard Belin, 31401 Toulouse Cedex 9, France\label{aff28}
\and
Kapteyn Astronomical Institute, University of Groningen, PO Box 800, 9700 AV Groningen, The Netherlands\label{aff29}
\and
ATG Europe BV, Huygensstraat 34, 2201 DK Noordwijk, The Netherlands\label{aff30}
\and
Space Science Data Center, Italian Space Agency, via del Politecnico snc, 00133 Roma, Italy\label{aff31}
\and
INAF-Osservatorio Astrofisico di Torino, Via Osservatorio 20, 10025 Pino Torinese (TO), Italy\label{aff32}
\and
Dipartimento di Fisica, Universit\`a di Genova, Via Dodecaneso 33, 16146, Genova, Italy\label{aff33}
\and
INFN-Sezione di Genova, Via Dodecaneso 33, 16146, Genova, Italy\label{aff34}
\and
Department of Physics "E. Pancini", University Federico II, Via Cinthia 6, 80126, Napoli, Italy\label{aff35}
\and
INAF-Osservatorio Astronomico di Capodimonte, Via Moiariello 16, 80131 Napoli, Italy\label{aff36}
\and
Instituto de Astrof\'isica e Ci\^encias do Espa\c{c}o, Universidade do Porto, CAUP, Rua das Estrelas, PT4150-762 Porto, Portugal\label{aff37}
\and
Faculdade de Ci\^encias da Universidade do Porto, Rua do Campo de Alegre, 4150-007 Porto, Portugal\label{aff38}
\and
Dipartimento di Fisica, Universit\`a degli Studi di Torino, Via P. Giuria 1, 10125 Torino, Italy\label{aff39}
\and
INFN-Sezione di Torino, Via P. Giuria 1, 10125 Torino, Italy\label{aff40}
\and
European Space Agency/ESTEC, Keplerlaan 1, 2201 AZ Noordwijk, The Netherlands\label{aff41}
\and
Institute Lorentz, Leiden University, Niels Bohrweg 2, 2333 CA Leiden, The Netherlands\label{aff42}
\and
Leiden Observatory, Leiden University, Einsteinweg 55, 2333 CC Leiden, The Netherlands\label{aff43}
\and
INAF-IASF Milano, Via Alfonso Corti 12, 20133 Milano, Italy\label{aff44}
\and
Centro de Investigaciones Energ\'eticas, Medioambientales y Tecnol\'ogicas (CIEMAT), Avenida Complutense 40, 28040 Madrid, Spain\label{aff45}
\and
Port d'Informaci\'{o} Cient\'{i}fica, Campus UAB, C. Albareda s/n, 08193 Bellaterra (Barcelona), Spain\label{aff46}
\and
Institute for Theoretical Particle Physics and Cosmology (TTK), RWTH Aachen University, 52056 Aachen, Germany\label{aff47}
\and
Institute of Space Sciences (ICE, CSIC), Campus UAB, Carrer de Can Magrans, s/n, 08193 Barcelona, Spain\label{aff48}
\and
Institut d'Estudis Espacials de Catalunya (IEEC),  Edifici RDIT, Campus UPC, 08860 Castelldefels, Barcelona, Spain\label{aff49}
\and
INFN section of Naples, Via Cinthia 6, 80126, Napoli, Italy\label{aff50}
\and
Institute for Astronomy, University of Hawaii, 2680 Woodlawn Drive, Honolulu, HI 96822, USA\label{aff51}
\and
Dipartimento di Fisica e Astronomia "Augusto Righi" - Alma Mater Studiorum Universit\`a di Bologna, Viale Berti Pichat 6/2, 40127 Bologna, Italy\label{aff52}
\and
Institute for Astronomy, University of Edinburgh, Royal Observatory, Blackford Hill, Edinburgh EH9 3HJ, UK\label{aff53}
\and
European Space Agency/ESRIN, Largo Galileo Galilei 1, 00044 Frascati, Roma, Italy\label{aff54}
\and
Universit\'e Claude Bernard Lyon 1, CNRS/IN2P3, IP2I Lyon, UMR 5822, Villeurbanne, F-69100, France\label{aff55}
\and
Institut de Ci\`{e}ncies del Cosmos (ICCUB), Universitat de Barcelona (IEEC-UB), Mart\'{i} i Franqu\`{e}s 1, 08028 Barcelona, Spain\label{aff56}
\and
Instituci\'o Catalana de Recerca i Estudis Avan\c{c}ats (ICREA), Passeig de Llu\'{\i}s Companys 23, 08010 Barcelona, Spain\label{aff57}
\and
UCB Lyon 1, CNRS/IN2P3, IUF, IP2I Lyon, 4 rue Enrico Fermi, 69622 Villeurbanne, France\label{aff58}
\and
Mullard Space Science Laboratory, University College London, Holmbury St Mary, Dorking, Surrey RH5 6NT, UK\label{aff59}
\and
Canada-France-Hawaii Telescope, 65-1238 Mamalahoa Hwy, Kamuela, HI 96743, USA\label{aff60}
\and
Aix-Marseille Universit\'e, CNRS, CNES, LAM, Marseille, France\label{aff61}
\and
Departamento de F\'isica, Faculdade de Ci\^encias, Universidade de Lisboa, Edif\'icio C8, Campo Grande, PT1749-016 Lisboa, Portugal\label{aff62}
\and
Instituto de Astrof\'isica e Ci\^encias do Espa\c{c}o, Faculdade de Ci\^encias, Universidade de Lisboa, Campo Grande, 1749-016 Lisboa, Portugal\label{aff63}
\and
INAF-Istituto di Astrofisica e Planetologia Spaziali, via del Fosso del Cavaliere, 100, 00100 Roma, Italy\label{aff64}
\and
INFN-Padova, Via Marzolo 8, 35131 Padova, Italy\label{aff65}
\and
Aix-Marseille Universit\'e, CNRS/IN2P3, CPPM, Marseille, France\label{aff66}
\and
INFN-Bologna, Via Irnerio 46, 40126 Bologna, Italy\label{aff67}
\and
School of Physics, HH Wills Physics Laboratory, University of Bristol, Tyndall Avenue, Bristol, BS8 1TL, UK\label{aff68}
\and
NRC Herzberg, 5071 West Saanich Rd, Victoria, BC V9E 2E7, Canada\label{aff69}
\and
Institute of Theoretical Astrophysics, University of Oslo, P.O. Box 1029 Blindern, 0315 Oslo, Norway\label{aff70}
\and
Jet Propulsion Laboratory, California Institute of Technology, 4800 Oak Grove Drive, Pasadena, CA, 91109, USA\label{aff71}
\and
Department of Physics, Lancaster University, Lancaster, LA1 4YB, UK\label{aff72}
\and
Felix Hormuth Engineering, Goethestr. 17, 69181 Leimen, Germany\label{aff73}
\and
Technical University of Denmark, Elektrovej 327, 2800 Kgs. Lyngby, Denmark\label{aff74}
\and
Cosmic Dawn Center (DAWN), Denmark\label{aff75}
\and
Max-Planck-Institut f\"ur Astronomie, K\"onigstuhl 17, 69117 Heidelberg, Germany\label{aff76}
\and
NASA Goddard Space Flight Center, Greenbelt, MD 20771, USA\label{aff77}
\and
Department of Physics and Astronomy, University College London, Gower Street, London WC1E 6BT, UK\label{aff78}
\and
Department of Physics and Helsinki Institute of Physics, Gustaf H\"allstr\"omin katu 2, 00014 University of Helsinki, Finland\label{aff79}
\and
Universit\'e de Gen\`eve, D\'epartement de Physique Th\'eorique and Centre for Astroparticle Physics, 24 quai Ernest-Ansermet, CH-1211 Gen\`eve 4, Switzerland\label{aff80}
\and
Department of Physics, P.O. Box 64, 00014 University of Helsinki, Finland\label{aff81}
\and
Helsinki Institute of Physics, Gustaf H{\"a}llstr{\"o}min katu 2, University of Helsinki, Helsinki, Finland\label{aff82}
\and
Centre de Calcul de l'IN2P3/CNRS, 21 avenue Pierre de Coubertin 69627 Villeurbanne Cedex, France\label{aff83}
\and
Laboratoire d'etude de l'Univers et des phenomenes eXtremes, Observatoire de Paris, Universit\'e PSL, Sorbonne Universit\'e, CNRS, 92190 Meudon, France\label{aff84}
\and
SKA Observatory, Jodrell Bank, Lower Withington, Macclesfield, Cheshire SK11 9FT, UK\label{aff85}
\and
Dipartimento di Fisica "Aldo Pontremoli", Universit\`a degli Studi di Milano, Via Celoria 16, 20133 Milano, Italy\label{aff86}
\and
INFN-Sezione di Milano, Via Celoria 16, 20133 Milano, Italy\label{aff87}
\and
University of Applied Sciences and Arts of Northwestern Switzerland, School of Engineering, 5210 Windisch, Switzerland\label{aff88}
\and
Universit\"at Bonn, Argelander-Institut f\"ur Astronomie, Auf dem H\"ugel 71, 53121 Bonn, Germany\label{aff89}
\and
INFN-Sezione di Roma, Piazzale Aldo Moro, 2 - c/o Dipartimento di Fisica, Edificio G. Marconi, 00185 Roma, Italy\label{aff90}
\and
Dipartimento di Fisica e Astronomia "Augusto Righi" - Alma Mater Studiorum Universit\`a di Bologna, via Piero Gobetti 93/2, 40129 Bologna, Italy\label{aff91}
\and
Department of Physics, Institute for Computational Cosmology, Durham University, South Road, Durham, DH1 3LE, UK\label{aff92}
\and
Universit\'e C\^{o}te d'Azur, Observatoire de la C\^{o}te d'Azur, CNRS, Laboratoire Lagrange, Bd de l'Observatoire, CS 34229, 06304 Nice cedex 4, France\label{aff93}
\and
Institut d'Astrophysique de Paris, UMR 7095, CNRS, and Sorbonne Universit\'e, 98 bis boulevard Arago, 75014 Paris, France\label{aff94}
\and
CNRS-UCB International Research Laboratory, Centre Pierre Binetruy, IRL2007, CPB-IN2P3, Berkeley, USA\label{aff95}
\and
Institut d'Astrophysique de Paris, 98bis Boulevard Arago, 75014, Paris, France\label{aff96}
\and
Institute of Physics, Laboratory of Astrophysics, Ecole Polytechnique F\'ed\'erale de Lausanne (EPFL), Observatoire de Sauverny, 1290 Versoix, Switzerland\label{aff97}
\and
Aurora Technology for European Space Agency (ESA), Camino bajo del Castillo, s/n, Urbanizacion Villafranca del Castillo, Villanueva de la Ca\~nada, 28692 Madrid, Spain\label{aff98}
\and
Institut de F\'{i}sica d'Altes Energies (IFAE), The Barcelona Institute of Science and Technology, Campus UAB, 08193 Bellaterra (Barcelona), Spain\label{aff99}
\and
School of Mathematics, Statistics and Physics, Newcastle University, Herschel Building, Newcastle-upon-Tyne, NE1 7RU, UK\label{aff100}
\and
DARK, Niels Bohr Institute, University of Copenhagen, Jagtvej 155, 2200 Copenhagen, Denmark\label{aff101}
\and
Waterloo Centre for Astrophysics, University of Waterloo, Waterloo, Ontario N2L 3G1, Canada\label{aff102}
\and
Department of Physics and Astronomy, University of Waterloo, Waterloo, Ontario N2L 3G1, Canada\label{aff103}
\and
Perimeter Institute for Theoretical Physics, Waterloo, Ontario N2L 2Y5, Canada\label{aff104}
\and
Institute of Space Science, Str. Atomistilor, nr. 409 M\u{a}gurele, Ilfov, 077125, Romania\label{aff105}
\and
Dipartimento di Fisica e Astronomia "G. Galilei", Universit\`a di Padova, Via Marzolo 8, 35131 Padova, Italy\label{aff106}
\and
Institut f\"ur Theoretische Physik, University of Heidelberg, Philosophenweg 16, 69120 Heidelberg, Germany\label{aff107}
\and
Universit\'e St Joseph; Faculty of Sciences, Beirut, Lebanon\label{aff108}
\and
Departamento de F\'isica, FCFM, Universidad de Chile, Blanco Encalada 2008, Santiago, Chile\label{aff109}
\and
Satlantis, University Science Park, Sede Bld 48940, Leioa-Bilbao, Spain\label{aff110}
\and
Department of Physics, Royal Holloway, University of London, TW20 0EX, UK\label{aff111}
\and
Infrared Processing and Analysis Center, California Institute of Technology, Pasadena, CA 91125, USA\label{aff112}
\and
Instituto de Astrof\'isica e Ci\^encias do Espa\c{c}o, Faculdade de Ci\^encias, Universidade de Lisboa, Tapada da Ajuda, 1349-018 Lisboa, Portugal\label{aff113}
\and
Cosmic Dawn Center (DAWN)\label{aff114}
\and
Niels Bohr Institute, University of Copenhagen, Jagtvej 128, 2200 Copenhagen, Denmark\label{aff115}
\and
Universidad Polit\'ecnica de Cartagena, Departamento de Electr\'onica y Tecnolog\'ia de Computadoras,  Plaza del Hospital 1, 30202 Cartagena, Spain\label{aff116}
\and
Centre for Information Technology, University of Groningen, P.O. Box 11044, 9700 CA Groningen, The Netherlands\label{aff117}
\and
Dipartimento di Fisica e Scienze della Terra, Universit\`a degli Studi di Ferrara, Via Giuseppe Saragat 1, 44122 Ferrara, Italy\label{aff118}
\and
Istituto Nazionale di Fisica Nucleare, Sezione di Ferrara, Via Giuseppe Saragat 1, 44122 Ferrara, Italy\label{aff119}
\and
INAF, Istituto di Radioastronomia, Via Piero Gobetti 101, 40129 Bologna, Italy\label{aff120}
\and
Department of Physics, Oxford University, Keble Road, Oxford OX1 3RH, UK\label{aff121}
\and
Ernst-Reuter-Str. 4e, 31224 Peine, Germany\label{aff122}
\and
INAF - Osservatorio Astronomico di Brera, via Emilio Bianchi 46, 23807 Merate, Italy\label{aff123}
\and
INAF-Osservatorio Astronomico di Brera, Via Brera 28, 20122 Milano, Italy, and INFN-Sezione di Genova, Via Dodecaneso 33, 16146, Genova, Italy\label{aff124}
\and
ICL, Junia, Universit\'e Catholique de Lille, LITL, 59000 Lille, France\label{aff125}
\and
ICSC - Centro Nazionale di Ricerca in High Performance Computing, Big Data e Quantum Computing, Via Magnanelli 2, Bologna, Italy\label{aff126}
\and
Instituto de F\'isica Te\'orica UAM-CSIC, Campus de Cantoblanco, 28049 Madrid, Spain\label{aff127}
\and
CERCA/ISO, Department of Physics, Case Western Reserve University, 10900 Euclid Avenue, Cleveland, OH 44106, USA\label{aff128}
\and
Technical University of Munich, TUM School of Natural Sciences, Physics Department, James-Franck-Str.~1, 85748 Garching, Germany\label{aff129}
\and
Max-Planck-Institut f\"ur Astrophysik, Karl-Schwarzschild-Str.~1, 85748 Garching, Germany\label{aff130}
\and
Laboratoire Univers et Th\'eorie, Observatoire de Paris, Universit\'e PSL, Universit\'e Paris Cit\'e, CNRS, 92190 Meudon, France\label{aff131}
\and
Departamento de F{\'\i}sica Fundamental. Universidad de Salamanca. Plaza de la Merced s/n. 37008 Salamanca, Spain\label{aff132}
\and
Universit\'e de Strasbourg, CNRS, Observatoire astronomique de Strasbourg, UMR 7550, 67000 Strasbourg, France\label{aff133}
\and
Center for Data-Driven Discovery, Kavli IPMU (WPI), UTIAS, The University of Tokyo, Kashiwa, Chiba 277-8583, Japan\label{aff134}
\and
Ludwig-Maximilians-University, Schellingstrasse 4, 80799 Munich, Germany\label{aff135}
\and
Max-Planck-Institut f\"ur Physik, Boltzmannstr. 8, 85748 Garching, Germany\label{aff136}
\and
California Institute of Technology, 1200 E California Blvd, Pasadena, CA 91125, USA\label{aff137}
\and
University of California, Los Angeles, CA 90095-1562, USA\label{aff138}
\and
Department of Physics \& Astronomy, University of California Irvine, Irvine CA 92697, USA\label{aff139}
\and
Department of Mathematics and Physics E. De Giorgi, University of Salento, Via per Arnesano, CP-I93, 73100, Lecce, Italy\label{aff140}
\and
INFN, Sezione di Lecce, Via per Arnesano, CP-193, 73100, Lecce, Italy\label{aff141}
\and
INAF-Sezione di Lecce, c/o Dipartimento Matematica e Fisica, Via per Arnesano, 73100, Lecce, Italy\label{aff142}
\and
Departamento F\'isica Aplicada, Universidad Polit\'ecnica de Cartagena, Campus Muralla del Mar, 30202 Cartagena, Murcia, Spain\label{aff143}
\and
Instituto de F\'isica de Cantabria, Edificio Juan Jord\'a, Avenida de los Castros, 39005 Santander, Spain\label{aff144}
\and
Institute of Cosmology and Gravitation, University of Portsmouth, Portsmouth PO1 3FX, UK\label{aff145}
\and
Department of Computer Science, Aalto University, PO Box 15400, Espoo, FI-00 076, Finland\label{aff146}
\and
Instituto de Astrof\'\i sica de Canarias, c/ Via Lactea s/n, La Laguna 38200, Spain. Departamento de Astrof\'\i sica de la Universidad de La Laguna, Avda. Francisco Sanchez, La Laguna, 38200, Spain\label{aff147}
\and
Universidad de La Laguna, Departamento de Astrof\'{\i}sica, 38206 La Laguna, Tenerife, Spain\label{aff148}
\and
Ruhr University Bochum, Faculty of Physics and Astronomy, Astronomical Institute (AIRUB), German Centre for Cosmological Lensing (GCCL), 44780 Bochum, Germany\label{aff149}
\and
Department of Physics and Astronomy, Vesilinnantie 5, 20014 University of Turku, Finland\label{aff150}
\and
Serco for European Space Agency (ESA), Camino bajo del Castillo, s/n, Urbanizacion Villafranca del Castillo, Villanueva de la Ca\~nada, 28692 Madrid, Spain\label{aff151}
\and
ARC Centre of Excellence for Dark Matter Particle Physics, Melbourne, Australia\label{aff152}
\and
Centre for Astrophysics \& Supercomputing, Swinburne University of Technology,  Hawthorn, Victoria 3122, Australia\label{aff153}
\and
Department of Physics and Astronomy, University of the Western Cape, Bellville, Cape Town, 7535, South Africa\label{aff154}
\and
DAMTP, Centre for Mathematical Sciences, Wilberforce Road, Cambridge CB3 0WA, UK\label{aff155}
\and
Kavli Institute for Cosmology Cambridge, Madingley Road, Cambridge, CB3 0HA, UK\label{aff156}
\and
Department of Astrophysics, University of Zurich, Winterthurerstrasse 190, 8057 Zurich, Switzerland\label{aff157}
\and
IRFU, CEA, Universit\'e Paris-Saclay 91191 Gif-sur-Yvette Cedex, France\label{aff158}
\and
Oskar Klein Centre for Cosmoparticle Physics, Department of Physics, Stockholm University, Stockholm, SE-106 91, Sweden\label{aff159}
\and
Astrophysics Group, Blackett Laboratory, Imperial College London, London SW7 2AZ, UK\label{aff160}
\and
Univ. Grenoble Alpes, CNRS, Grenoble INP, LPSC-IN2P3, 53, Avenue des Martyrs, 38000, Grenoble, France\label{aff161}
\and
INAF-Osservatorio Astrofisico di Arcetri, Largo E. Fermi 5, 50125, Firenze, Italy\label{aff162}
\and
Dipartimento di Fisica, Sapienza Universit\`a di Roma, Piazzale Aldo Moro 2, 00185 Roma, Italy\label{aff163}
\and
Centro de Astrof\'{\i}sica da Universidade do Porto, Rua das Estrelas, 4150-762 Porto, Portugal\label{aff164}
\and
HE Space for European Space Agency (ESA), Camino bajo del Castillo, s/n, Urbanizacion Villafranca del Castillo, Villanueva de la Ca\~nada, 28692 Madrid, Spain\label{aff165}
\and
Department of Astrophysical Sciences, Peyton Hall, Princeton University, Princeton, NJ 08544, USA\label{aff166}
\and
Theoretical astrophysics, Department of Physics and Astronomy, Uppsala University, Box 515, 751 20 Uppsala, Sweden\label{aff167}
\and
Minnesota Institute for Astrophysics, University of Minnesota, 116 Church St SE, Minneapolis, MN 55455, USA\label{aff168}
\and
Mathematical Institute, University of Leiden, Einsteinweg 55, 2333 CA Leiden, The Netherlands\label{aff169}
\and
Institute of Astronomy, University of Cambridge, Madingley Road, Cambridge CB3 0HA, UK\label{aff170}
\and
Space physics and astronomy research unit, University of Oulu, Pentti Kaiteran katu 1, FI-90014 Oulu, Finland\label{aff171}
\and
Center for Computational Astrophysics, Flatiron Institute, 162 5th Avenue, 10010, New York, NY, USA\label{aff172}
\and
Department of Astronomy, University of Massachusetts, Amherst, MA 01003, USA\label{aff173}
\and
Univ. Lille, CNRS, Centrale Lille, UMR 9189 CRIStAL, 59000 Lille, France\label{aff174}
\and
Department of Physics and Astronomy, University of British Columbia, Vancouver, BC V6T 1Z1, Canada\label{aff175}}

\date{...}

  \abstract 
   {The \Euclid satellite is an ESA mission that was launched in July 2023. \Euclid is working in its regular observing mode with the target of observing an area of $14\,000~\text{deg}^2$ with two instruments, the Visible Camera (VIS) and the Near IR Spectrometer and Photometer (NISP) down to $\text{\IE} = 24.5~\text{mag}$ ($10\, \sigma$) in the Euclid Wide Survey. Ground-based imaging data in the \textit{ugriz} bands complement the \Euclid data to enable photo-$z$ determination and VIS PSF modelling for weak lensing analysis. \Euclid investigates the distance-redshift relation and the evolution of cosmic structures by measuring the shapes and redshifts of galaxies and clusters of galaxies out to $z\sim 2$.
   Generating the multi-wavelength catalogues from \Euclid and ground-based data is an essential part of the \Euclid data processing system. In the framework of the \Euclid Science Ground Segment (SGS), the aim of the MERge Processing Function (MER PF) pipeline is to detect objects in the \Euclid imaging data, measure their properties, and merge them into a single multi-wavelength catalogue.
   The MER PF pipeline performs source detection on both visible (VIS) and near-infrared (NIR) images and offers four different photometric measurements: Kron total flux, aperture photometry on PSF-matched images, template fitting photometry, and S\'ersic fitting photometry. Furthermore, the MER PF pipeline measures a set of ancillary quantities, spanning from morphology to quality flags, to better characterise all detected sources.
   In this paper, we show how the MER PF pipeline is designed, detailing its main steps, and we show that the pipeline products meet the tight requirements that \Euclid aims to achieve on photometric accuracy. We also present the other measurements (e.g., morphology) that are included in the OU-MER output catalogues and we list all output products coming out of the MER PF pipeline.}

\keywords{Galaxies:photometry, Galaxies:morphology, Catalogues, Methods:data analysis}
\authorrunning{}   
\titlerunning{Euclid Quick Data Release (Q1): The MERge processing function}   

\maketitle

\section{Introduction}\label{sect:intro}

The European Space Agency’s \Euclid mission \citep{EuclidSkyOverview} is dedicated to studying the dark Universe through a photometric and spectroscopic survey of the extragalactic sky at visible and near-infrared wavelengths. Successfully launched from Cape Canaveral on July 1 2023, the \Euclid satellite aims to gather data for the Euclid Wide Survey (EWS) and the Euclid Deep Survey (EDS), both described in detail in \citet{Scaramella-EP1}, by means of two instruments, the Visible Camera \citep[VIS,][]{EuclidSkyVIS} and the Near-Infrared Spectrometer and Photometer \citep[NISP,][]{EuclidSkyNISP}.

To efficiently manage and process the immense volume of data generated by \Euclid, the Euclid Consortium (EC) has developed a suite of software processing functions (PFs) forming the \Euclid Science Ground Segment (SGS). Organisational units (OUs) are responsible for developing, maintaining, and validating the PFs. A detailed description of how the SGS processing is organised and how the OUs interact with each other is found in \citet{Q1-TP001}.

The MER PF realises the merging of photometric information from VIS and NISP, together with information from external photometric surveys: DES \citep{Abbott2021}, CFIS \citep{Ibata2017}, Pan-STARRS \citep{Magnier2020}, WHIGS and WHISHES \citep[see][]{Aihara2017}, and, in the future, Rubin-LSST \citep{Ivezic2019}. The main objective of the MER PF is to provide source catalogues containing object information relevant for \Euclid's science goals. 

The primary requirement for which the MER PF has been designed is the production of image mosaics, and a photometric catalogue of sources. To provide this, it starts from VIS and NISP calibrated frames, processed by OU-VIS \citep{Q1-TP002} and OU-NIR \citep{Q1-TP003}, respectively, and external surveys' stacked frames, processed by OU-EXT \citep{Q1-TP001}, together with related ancillary data and PSF models. Separate object detections are performed on both the photometric VIS and NIR (stacking the three NISP bands \JE, \HE, and \YE) mosaics. A multi-wavelength flux determination is performed on detected objects with suitable photometric techniques. The final product of the MER PF is a catalogue of sources with unique identification numbers, multi-wavelength photometric measurements, and ancillary information (morphological estimates, star-galaxy separation, etc.).

The MER PF also provides the required input, as additional information integrated in the final \Euclid catalogue, for photometric redshift computation and for spectra extraction for which, respectively, OU-PHZ \citep{Q1-TP005} and OU-SIR \citep{Q1-TP006} are in charge as customer OUs of OU-MER. The MER PF catalogues also represent a primary input for shear estimation, which is performed by OU-SHE. The MER products are among the core products delivered in the first \Euclid data release, Q1 \citep{Q1cite}, and are used in many science papers. The Q1 data release covers $63.1\,\rm deg^2$ of the Euclid Deep Fields (EDFs) to the nominal wide-survey depth and contains nearly 30 million objects. For a detailed description of the Q1 data release, we refer the reader to \citet{Q1-TP001}.\\

\section{The \Euclid MER software pipeline overview}\label{sect:pipeline}

The MER PF is designed around the concept of a pipeline as a chain of processing elements (PEs). Each PE represents an atomic block that cannot be split further from a functional point of view. The tasks related to each PE are implemented by one or more software components. Source codes are in the official \Euclid \texttt{GitLab} repository and are available throughout the EC via the COmmon DEvelopment ENvironment (CODEEN).

\begin{figure*}[h!]
    \centering
    \includegraphics[width=0.75\textwidth]{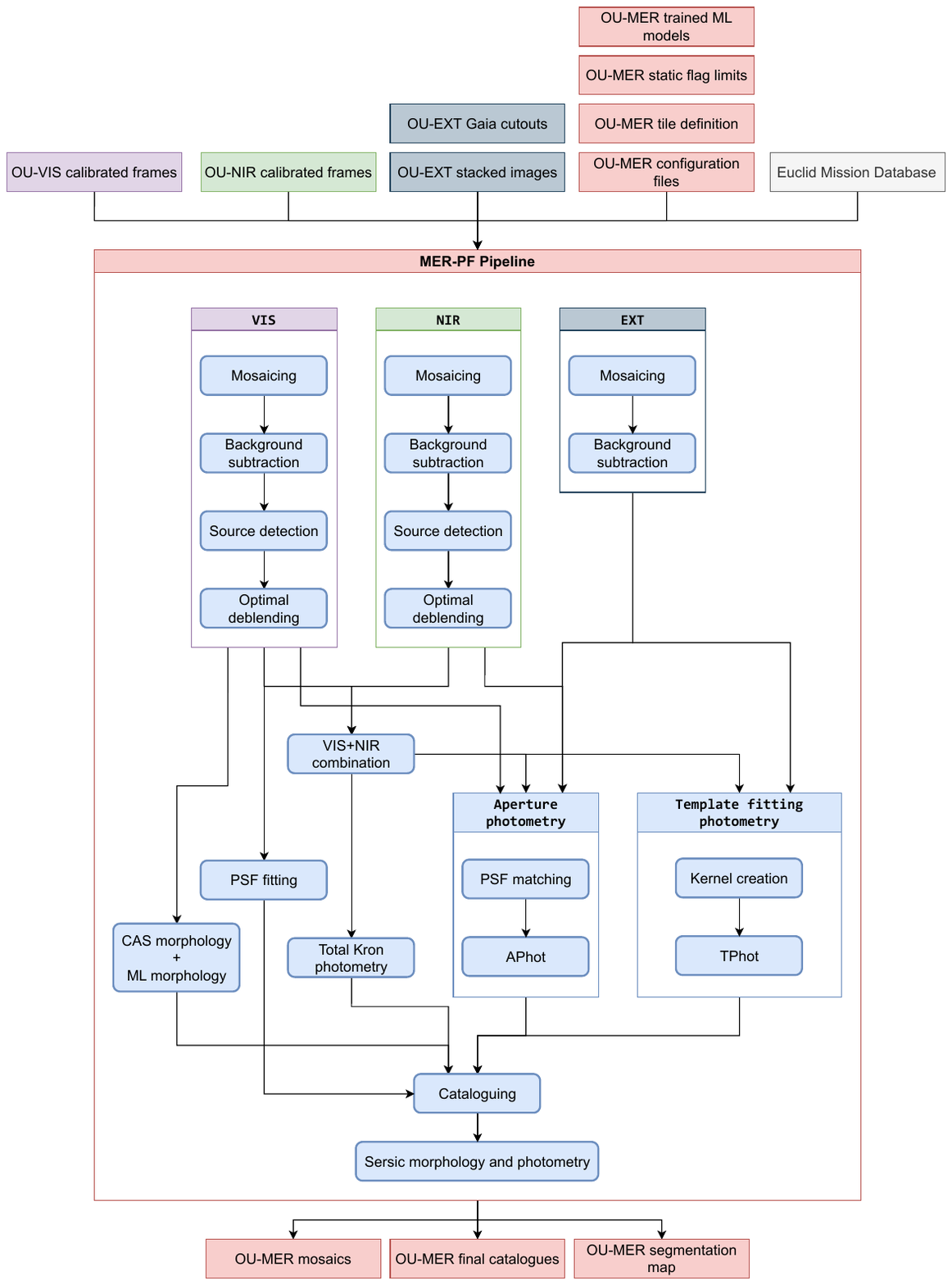}
    \caption{Simplified workflow of the OU-MER pipeline. The pipeline is fed with a list of input products coming from OU-VIS, OU-NIR, and OU-EXT, together with a set of configuration files and auxiliary products. The pipeline performs a mosaicing of input images before generating the source list after a combination of detection and optimal deblending procedures. The pipeline computes photometry and morphology measurements and collects all this information in a set of output catalogues. Those catalogues, together with the reference images, represent the output of the pipeline.}
    \label{pipeline_workflow}
\end{figure*}

The pipeline is in charge of running PEs on the \Euclid SGS infrastructure and controlling the parallelisation scheme. The orchestration is performed via the \Euclid Infrastructure Abstraction Layer (IAL), which is also in charge of managing the input and output data stream from and to the \Euclid Archive System (EAS). A simplified MER PF pipeline workflow is shown in Fig.~\ref{pipeline_workflow}.

A standard run of the MER PF pipeline is performed on sky regions called MER tiles, described in Sect.~\ref{subsect:tile}, fed with the following input:
\begin{itemize}
    \item the definition of the MER tile to be processed (see Sect. \ref{subsect:tile});
    \item the list of VIS calibrated frames \citep[both nominal and short exposures, see][]{Q1-TP002} intersecting the tile;
    \item the list of NIR calibrated frames intersecting the tile;
    \item the list of EXT stacked frames covering the tile;
    \item a cut-out of the Gaia Data Release 3 catalogue \citep{GAIADR3} with the sources that fall inside the tile;
    \item a set of configuration files, specifying the parametrisation of the pipeline PEs;
    \item a set of ancillary files, stored in the Mission Data Base (MDB) data container;
    \item the trained machine-learning models to be used for source classification;
    \item a list of quality parameter ranges used to flag invalid output data products.
\end{itemize}
The MER PF outputs the following products:
\begin{itemize}
    \item the list of background-subtracted mosaiced images, one for each band, used to measure the source photometry;
    \item the VIS and NIR background-subtracted mosaiced images used to obtain the list of detected sources;
    \item the segmentation map containing the spatial pixels associated with each individual source;
    \item a set of catalogues, storing all the source information measured by the pipeline.
\end{itemize}

After a first initialisation step, aimed at decompressing and sorting the input images, the MER PF pipeline produces the mosaiced images. The mosaicing step (Sect.~\ref{subsect:mosaicing}) co-adds the VIS and NIR calibrated frames. VIS and NIR stacked images, together with the EXT stacks, are then re-binned so that all the mosaiced images share the same dimensions, image centre, and pixel size ($0\overset{\prime\prime}{.}1$). A background subtraction algorithm (Sect.~\ref{subsect:background}) evaluates the background of each input mosaic, subtracts it from the original mosaic, and stores the background-subtracted image in a FITS file. The goal of this step is to remove any remaining residual background, because a flat background is fundamental to obtain precise photometry measurements. 

The list of source candidates is extracted from the VIS mosaic and from a stack of the three NIR bands independently (Sect.~\ref{subsect:detection}). This preliminary segmentation of the image is then refined by an optimal deblending algorithm (Sect.~\ref{subsect:deblending}) in order to obtain the list of sources detected in the VIS mosaic and the one detected on the NIR stack. The two lists are then merged into one single list, taking care of source duplicates (Sect.~\ref{subsect:det_combination}); this represents the formal list of sources detected by the OU-MER pipeline.

At this point, the input point spread functions (PSFs) are propagated for each mosaiced image and evaluated at the detected source positions, together with the PSF convolution kernels required by the subsequent photometry steps (Sect.~\ref{sect:PSF_calculation}).

The list of VIS- and NIR-detected sources is used as a baseline to force the photometry measurements in the detected loci (Sect.~\ref{sect:photometry}). For each source, the MER PF pipeline measures:
\begin{itemize}
    \item the Kron total flux, measured on the detection mosaics;
    \item PSF-matched aperture photometry on all the input bands;
    \item template fitting photometry on all the input bands.
\end{itemize}

For all the sources detected in the \IE band, the MER PF pipeline provides a measure of the concentration, asymmetry, and smoothness (CAS) parameters (Sect.~\ref{subsect:morpho_cas}), a list of morphological features measured via a machine-learning approach (Sect.~\ref{subsect:morpho_zoobot}), and a point-like probability (that can be used for star-galaxy separation, see Sect. ~\ref{subsect:sg_separation}). 

A cataloguing unit then carries out the merging of all the information provided by the previous steps (Sect.~\ref{sect:catalog}). The main output of this step is the OU-MER final catalogue of sources with unique identification number, containing object information relevant in the \Euclid project, in particular to OU-MER's customer OUs, namely OU-PHZ, OU-SIR, and OU-SHE. Before releasing the output products to the Euclid Archive System (EAS), a final step runs on all the mosaiced images and the final source catalogue, performing a S\'ersic fit and adding further photometry and morphology information (Sect.~\ref{subsect:sersic}). Initially designed as a separate pipeline to be run on the output products of the nominal pipeline, the S\'ersic fit module has been integrated rather late into the MER PF processing workflow, after extensive testing and optimisation.

With the current Q1 release configuration, taking into account that many PEs run in parallel and some of them make use of more than one processing cores, the total execution time is equivalent to approximately 140 core hours per tile. The mosaicing, background-subtraction, and PSF calculation PEs account for 13\% of the processing time. On average, 10\% of the time is spent on the detection and deblending PEs. The large majority (75\%) of the processing time is spent running the different photometry methods and on the calculation of the PSF transformation kernels.

\section{Imaging input and mosaic creation}\label{sect:imaging}

\subsection{Euclid tiling}\label{subsect:tile}
\begin{figure}
    \centering
    \includegraphics[width=0.5\textwidth]{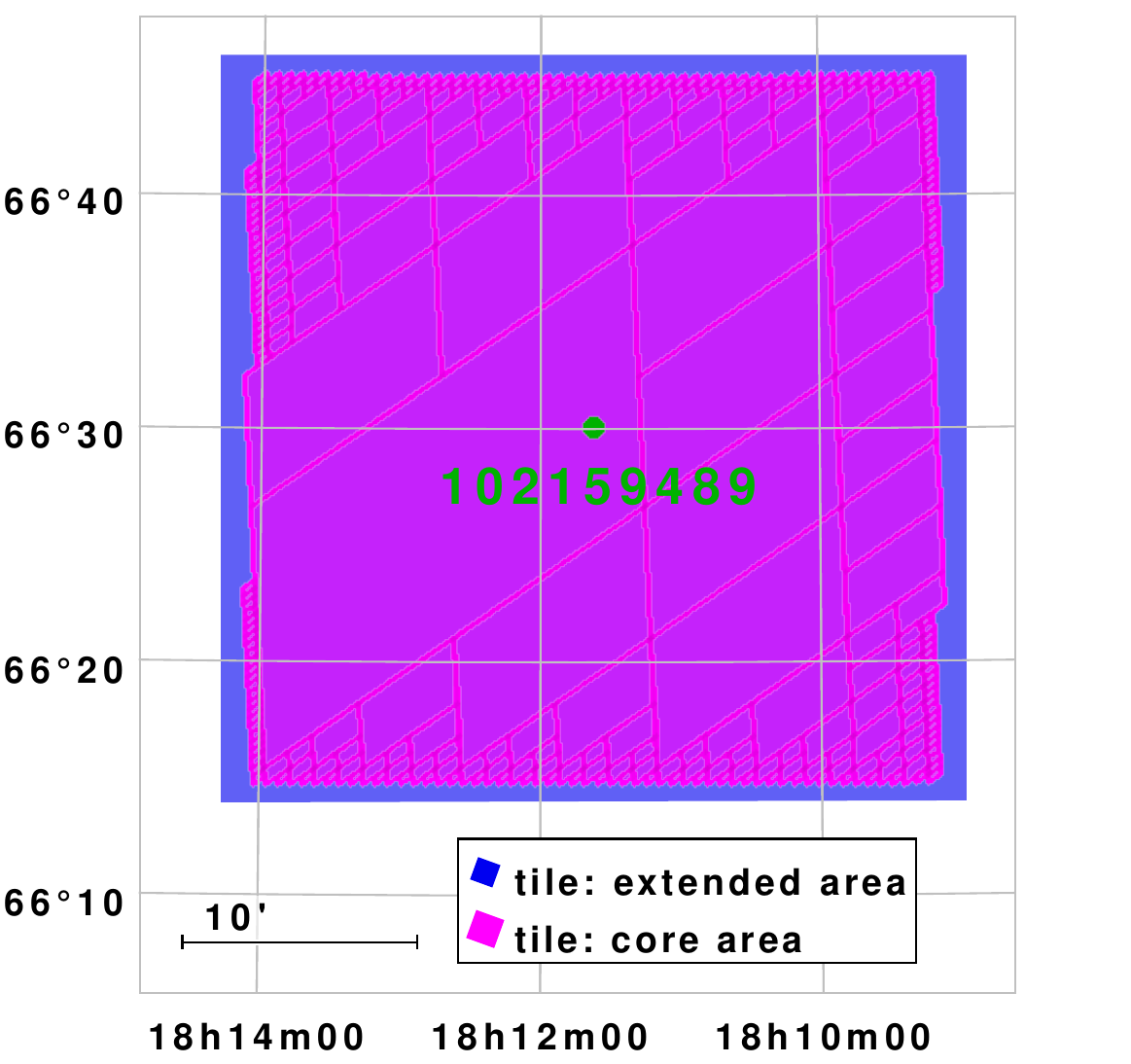}
    \caption{Example of tile geometry, specifically for EDF-N tile 10215949: centre (green), core area (magenta), and extended area (blue).}
    \label{figure:tiling}
\end{figure}

The MER PF pipeline runs on chunks of data called `tiles', with each tile covering a predefined area on the sky. In every OU-MER PF run, all the available \Euclid data covering the tile region are included in the processing. The \Euclid tiling solution \citep{2022ASPC..532..329K} has a tile placement that is independent of the pattern of the \Euclid observations and covers the entire sky, $-88\overset{\circ}{.}0 < {\rm Dec} < +88\overset{\circ}{.}0$, to become independent of the \Euclid survey plan \citep{Scaramella-EP1}. The tiling solution associates each sky position with exactly one tile. The catalogues resulting from the processing of a set of adjacent tiles do not contain the same object in more than one tile. The \Euclid tiling assures a certain overlap between two adjacent tiles to guarantee that the tile boundaries do not interfere with computation of the objects' properties.

In the \Euclid tiling solution, each tile is defined by its central position in $({\rm RA, Dec})$ and size (width, height), and identified by a unique tile identifier. Every tile has an extended area and a core area. While the extended areas of adjacent tiles have an overlap to avoid border problems in the measurements, the core areas of any two tiles do not overlap. The extended areas and the core areas of all tiles cover the entire sky, with the limits mentioned above. The imaging output of OU-MER covers the extended tile area and object detection and property measurements are done on the extended tile area; however, the catalogue output is delivered only for the core area, thus avoiding multiple outputs of the same object. The core areas of the tiles are defined as \texttt{HEALPix} indices \citep{2005ApJ...622..759G}. Figure \ref{figure:tiling} shows tile 10215949 in Euclid Deep Field North (EDF-N), with $\rm (RA,Dec) = (272\overset{\circ}{.}9038602,66\overset{\circ}{.}5)$ and its centre (green), extended area (blue), and the \texttt{HEALPix} pixels (magenta) defining the core area.

\begin{figure}
    \centering
    \includegraphics[width=0.5\textwidth]{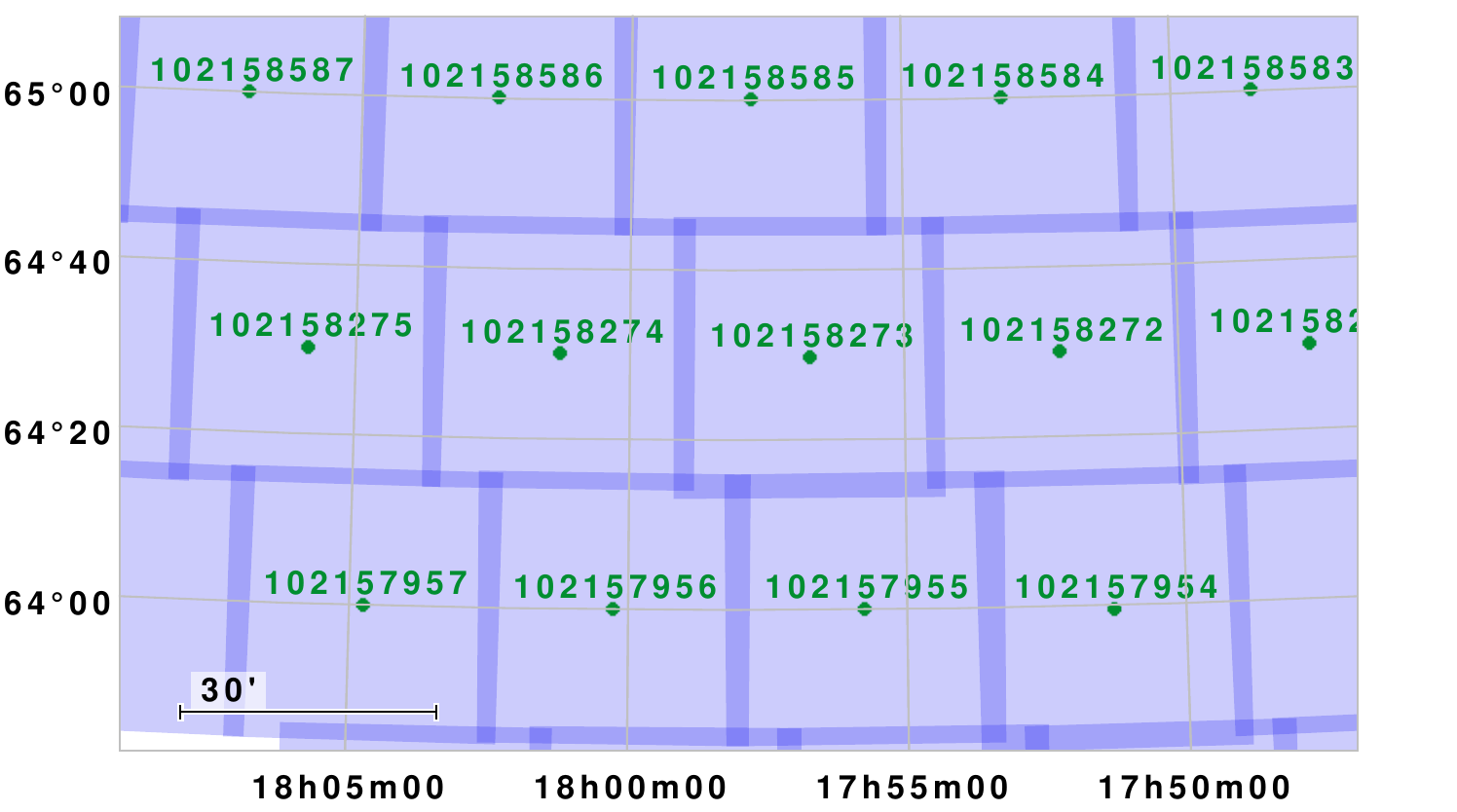}
    \caption{Example tiling pattern. Notice that the height of tile 102158273 in the EDF-N is enlarged to fully cover a larger galaxy.}
    \label{figure:edfn_tiling}
\end{figure}
The \Euclid tiling solution:
\begin{itemize}
\item has an extended tile area of $32'\times32'$ in tangential projection;
\item increments the tile centres by $30'$, which leads to an overlap  of $2'$ for adjacent tiles;
\item starts the central tile positions at $\rm (RA,Dec) = (0\overset{\circ}{.}0,0\overset{\circ}{.}0)$ and places them in strips of constant Dec;
\item adjusts the tile width when closing a strip at the Meridian;
\item defines the core area using \texttt{HEALPix} indices with $N_{\text{side}} = 13$.
\end{itemize}
To define the core areas, we first started, for each tile, with the \texttt{HEALPix} indices that are inside of the extended area. Then we identified the indices located in the overlap of extended areas, which appear in more than one tile. For those, we measured the minimum distance to the border of each tile in which they appear; then each index was attached to the tile that maximises this distance and removed from all others. This procedure avoids duplicated \texttt{HEALPix} indices in several tiles and assures a contiguous coverage of the core areas.

With a strict tiling as outlined so far, some of the galaxies with sizes larger than the overlap size will inevitably extend beyond the tile that its centre is located on. To solve these cases, we identified, from publicly available catalogues, all large galaxies ($\rm size>2'$) located in the Q1 area and store their position and size. We identified the critical cases in which a large galaxy extends between more than one tile. We allowed the tile with the galaxy centre to `breathe', which means to enlarge its size until it contains the entire galaxy. This procedure covered all large galaxies in Q1. Figure \ref{figure:edfn_tiling} shows the tiling in the lower part of the EDF-N. In this particular case, the height of tile 102158273 was enlarged to $33'$ in order to fully cover the galaxy PGC61126.

\subsection{Mosaicing}\label{subsect:mosaicing}
For the later measurement steps, specifically the photometry with \texttt{A-PHOT} and \texttt{T-PHOT} (see Sect.\,\ref{sect:photometry}), we need a multi-band set of co-aligned images, where each pixel represent the same position on the sky.
The layout of these co-aligned images is given by setting a pixel scale of $0\overset{\prime\prime}{.}1\,{\rm pixel}^{-1}$ (to facilitate critical Nyquist sampling for the VIS data, which have a resolution of $\sim 0\overset{\prime\prime}{.}2$) and the dimensions of the \Euclid tiles presented in Sect.~\ref{subsect:tile}. 

EXT imaging data were already provided for each tile as stacked frames in the native scale of the respective camera; for example, $0\overset{\prime\prime}{.}264\,{\rm pixel}^{-1}$ for the DES data. These data were re-binned to the common pixel scale using a \texttt{BILINEAR} interpolation kernel.

VIS and NIR input data are the set of fully calibrated frames covering a tile in their native pixel scale of $0\overset{\prime\prime}{.}1\,{\rm pixel}^{-1}$ and $0\overset{\prime\prime}{.}3\,{\rm pixel}^{-1}$, respectively. After subtracting the background estimate provided with the input data \citep[see][]{Q1-TP002, Q1-TP003}, these calibrated frames were re-binned and stacked to deep co-added images with a \texttt{BILINEAR} interpolation kernel and the \texttt{MEDIAN} combination type. For the stacking we applied the RMS data provided with the VIS and NIR data as weights. The different zero points for the VIS and NIR data were translated to global weighting values that are used in the co-add process. In addition, we scaled these global weights such that the zero point of the resulting co-added images corresponds to the median value of the input zero points, rounded to one decimal place. For all re-binning and co-adding operations we used the software \texttt{CT\_SWarp}, which is a version of the \texttt{SWarp} \citep{2002ASPC..281..228B} software that is fully integrated into the \Euclid processing environment.

In \texttt{CT\_SWarp} we also implemented the interpolation kernel \texttt{BILINEAR\_FLAGS}, which allows one to combine the flagging information for the VIS and NIR data with the bitwise \texttt{XOR} operation. The resulting flag image shows, for every pixel, the flag values that were associated with the corresponding input pixels, which usually resulted in the exclusion of the flagged pixels. 

The result of this processing step is, for every filter band, the set of co-added image plus the corresponding RMS and flag images. The astrometric validation of the co-added images shows that the non-linear astrometric solutions provided with the VIS and NIR data fulfil the requirements (see Sect.\ \ref{sect:astr_validation}).

\subsection{Background subtraction}\label{subsect:background}

We ran the background subtraction step to remove possible residual background from the mosaic images.
To avoid contamination from bright extended sources in the input mosaic, we applied a segmentation mask generated by source detection run using \texttt{SourceExtractor++} \citep[\texttt{SE++,}][]{2020ASPC..527..461B, 2022arXiv221202428K}, which is a re-designed and extended version of the \texttt{SExtractor2} \citep[\texttt{SE2,}][]{1996A&AS..117..393B} software.
The masked mosaic was divided into a grid, for each cell of which we estimated the background as a combination of k-sigma clipping and median filtering. 
In each cell, the local background pixel distribution was clipped iteratively until convergence at $2\sigma$ around its mean. 
If after convergence the mean and the median background differed by more than $3\sigma$ (e.g., in crowded fields), the background on that cell was defined to be the mean of the clipped histogram. If the mean and the median differed by less than $3\sigma$, the background on that cell was estimated as the mode of the distribution,

\begin{equation}\label{eq_background}
    B = 2.5 \, M - 1.5 \, \mu,
\end{equation}
as was described in \citet{1996A&AS..117..393B}. $M$ and $\mu$ are the median and mean of the clipped histogram, respectively.

Eventual gaps (cells lacking non-flagged pixels to estimate the background) were filled with the distance-weighted average of the background values of the cells at the edges of the gap. Once the background was computed in each cell of the grid, we applied a median filter to suppress possible local overestimations due to residuals from bright stars.
The background model at the original resolution of the mosaic was determined by a spline-interpolation of the cell values, and was subtracted from the mosaic image.
Two main parameters control the background generation: the grid size, which we set to $256 \times 256$ pixels, and the size of the filter box, which we set to $3 \time 3$ cells. 

\section{Assembly of the source list}\label{sect:mb_detection}

\subsection{General considerations for source detection and deblending}\label{subsect:detection}

The strategy for detecting objects in the available imaging data is dictated by the \Euclid project requirements:
\begin{itemize}
\item Objects are only to be detected in the photometric bands coming directly from the \Euclid instruments, which are \IE, \YE, \JE, and \HE. VIS and NIR imaging data for a specific tile are observed over a short time (around $1\,\rm{h}$) and give consistent basis for object detection. EXT data can span a large range of epochs, even within the bands available for one tile. Due to the epoch differences this would result in astrometric offsets within the data and pose problems for the photometry.
\item We need to detect all objects in the \Euclid VIS band (\IE). The weak lensing measurements by OU-SHE require an object sample with a simple selection function. Deriving the weak lensing sample from objects detected outside of VIS, such as co-added or $\chi^2$ images in \IE+\YE/\JE/\HE would result in biases and systematic effects that cannot be recovered in the shear analysis.
\item We need to detect all objects in the \Euclid NIR data (\YE,  \JE, \HE). The catalogues from the MER PF are used as input to the slit-less spectroscopy analysis.
This requires a deep and complete detection process to find the direct image
counterparts of all objects that possibly have spectral traces or emission lines in
the slit-less spectroscopic NISP data and hence must include a detection in the NIR
bands.
\item The detection records whether a source was detected in VIS or in NIR.
\end{itemize}
Since we need to separate sources detected in VIS and in NIR we cannot work with a single, deep detection image such as VIS+NIR or a $\chi^2$ image as described in \cite{1999AJ....117...68S}. Instead, as is outlined in Fig.\  \ref{pipeline_workflow}, we took the following the approach:
\begin{itemize}
    \item We performed a source detection and deblending in VIS.
    \item We performed an independent source detection and deblending in NIR.
    \item We combined the VIS sources and the NIR sources into a common object list and segmentation map.
    \item The computation of object properties, such as photometry and morphology, was then performed on the combined object lists.
\end{itemize}

\subsection{Source detection}\label{subsect:}
\begin{figure*}[h!]
\centering
\includegraphics[width=1.0\textwidth]{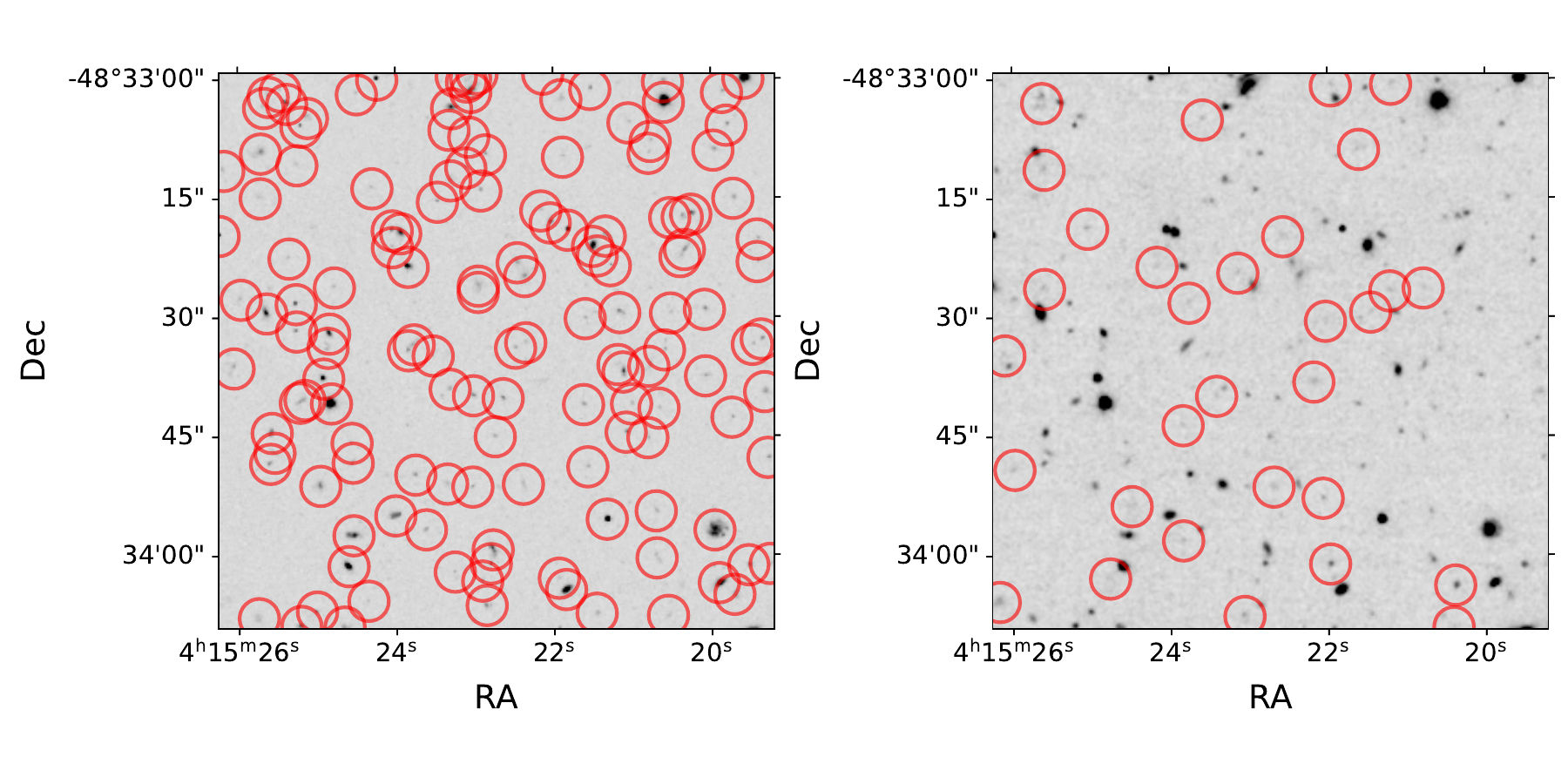}
\caption{Small cut-outs of the detection images $\IE$ (left) and NIR ($\YE+\JE+\HE$, right) in tile 102021017. The VIS- and NIR-detected sources are marked in the left and right panels, respectively.}
\label{fig:vis_nir_detection}
\end{figure*}
The source detection was done using SE++. The typical procedure for detecting objects in astronomical images is to:
\begin{itemize}
\item convolve the data and the RMS images with the segmentation filter in order to enhance the signal-to-noise ratio (S/N);
\item register all pixels with a $\rm{S/N}>{\rm \verb|detection-threshold|}$;
\item mark as an object the set of connected pixels with $N_{\rm pixel}>{\rm \verb|detection-minimum-area|}$.
\end{itemize}
With this method there are usually some false detections around large, extended galaxies. Using reasonable values for the parameters \verb|detection-threshold| and \verb|detection-minimum-area|, which are tailored to avoid false detections from noise peaks in the background, the detection cannot follow the extended objects far out into the sky background. The algorithm then picks up `objects' in the extended parts of large galaxies as individual sources.

To avoid these false positive detections, we have implemented an improved detection scheme in \texttt{SE++} that reduces the number of false detections around bright galaxies. In this detection
scheme there is in addition to the \verb|detection-threshold| and \verb|detection-minimum-area| discussed above a second threshold with:
\begin{itemize}
\item \verb|core-threshold-value|: the SNR threshold value;
\item \verb|core-minimum-area|: the number of pixels above the core-threshold-value an objects needs to have to validate the object;
\end{itemize}
With these two detection thresholds it is possible to have a rather low value for the \verb|detection-threshold| parameter. Then the detection areas of large
galaxies extend into the region of the fake objects. The second, higher threshold \verb|core-threshold-value| then eliminates the false
detections that result from the low value \verb|detection-threshold| at noise peaks in the sky background. The lower value of the detection-threshold in the `deep detection' leads to a larger
detection area (around) of the objects.

The VIS detection image is the background-subtracted mosaiced image described in Sects.\ \ref{subsect:mosaicing} and \ref{subsect:background}. For the detection in NIR we produced a deep NIR detection image from the \YE, \JE, and \HE background-subtracted mosaiced images using weighted summation. The different zero points of the input images were taken into account in the weighted summation to yield a common zero point for the resulting deep NIR detection image.
\begin{table}[t]
    \centering
    \caption{Important \texttt{SE++} parameters used for the VIS and NIR detections.}
    \begin{tabular}{lll}
       \hline
       \hline
       Parameter & VIS & NIR \\
       \hline
       \small\texttt{segmentation-filter} (Gaussian FWHM [pix])&$2.4$ & $5.0$\\
       \small\texttt{detection-threshold} & 0.9&  0.4\\
       \small\texttt{detection-minimum-area} & 10& 10\\
       \small\texttt{core-threshold-value} & 1.3&  0.55\\
       \small\texttt{core-minimum-area} & 10& 10\\
       \hline
    \end{tabular}
    \label{tab:detection_params}
\end{table}
Table~\ref{tab:detection_params} lists the most important parameters used in the VIS and NIR detection. The segmentation filters were chosen to have a width similar to the target population, which are faint galaxies. The other parameters were `calibrated' using simulated images
to produce a rate of false detections $<1\%$, which is a typical value for large-scale surveys.

\subsection{Deblending}\label{subsect:deblending}

Deblending is an essential procedure for separating overlapping sources in astronomical images. With its unprecedented sensitivity and resolution, \Euclid captures densely packed fields and intricate morphologies, making source disentanglement one of the key challenges in its data analysis.

Following the independent image segmentation in the VIS mosaic and the NIR stack, resulting from detection, deblending was executed in a parallel manner for both groups of source clusters. The aim is to put together a refined list of individual sources in the merging stage described in the next section. 

The deblending process was carried out using a dedicated software component based on \texttt{ASTErIsM} \citep{Tramacere2016}.
\texttt{ASTErIsM} is a Python package designed for automatic source detection and classification, leveraging two topometric clustering algorithms: \texttt{DBSCAN} \citep{ester1996proc} for identifying sources and \texttt{DENCLUE} \citep{hinneburg1998efficient, hinneburg2007denclue} for deblending overlapping sources and identifying galaxy morphological features such as spiral arms. Our internal PE, with a specific set of configuration parameters for the VIS mosaic and the NIR stack, inherits the operation of \texttt{ASTErIsM} as a deblender, which is based on the execution of a modified version of the \texttt{DENCLUE} density-based clustering algorithm, where the kernel-density estimation is substituted with a convolution of the image using a given kernel. 
This algorithm processes, as a function of the configuration parameters, a downsampled version of the segmentation clusters in terms of a distinct spatial scale. The adaptive approach allows for a robust separation of the overlapping sources within each cluster.

The two sets of parameters selected in the MER pipeline for deblending the clusters of sources detected in the VIS mosaic and the NIR stack were validated using the images and catalogues from the True Universe (TU) simulation \citep{EP-Serrano}. This was done by exploring a discrete sampling of the parameter space to single out the values for which the total number of over- and under-deblended sources is minimised in reference to a ground-truth segmentation map built from the TU catalogue. Two parameters were considered in this process due to their significant influence on the operation of the DENCLUE algorithm. One of them defines the typical scale of the sources blended within a cluster, while the other one establishes the converge criterion that identifies the local maximum corresponding to the centre of a deblended source.

\begin{figure}[h!]
\centering
\includegraphics[width=0.50\textwidth]{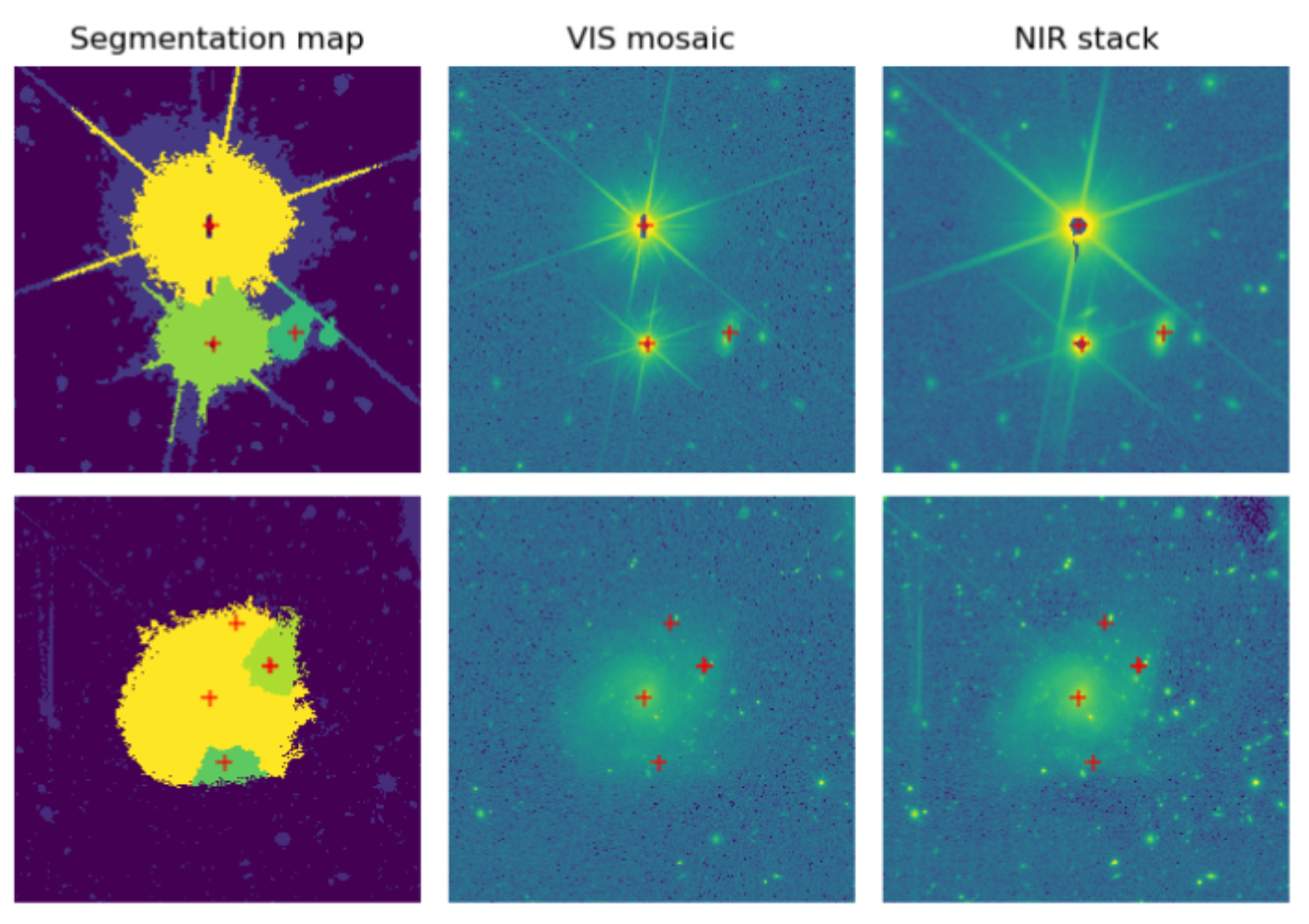}
\caption{Two examples of suboptimal deblending results produced by the OU-MER pipeline in problematic scenarios. In the top row, we can see how the segmentation map of a galaxy is highly perturbed by the close proximity of two bright stars. In the bottom row, we can observe how the deblending algorithm struggles when processing galaxies with intricate features such as spiral ones.}
\label{fig:deblending_complex_cases}
\end{figure}

 To illustrate some of the expected limitations of the pipeline's deblending step when handling complex cases, we present in Figure \ref{fig:deblending_complex_cases} two challenging examples, that go beyond the typical merging galaxies and/or dense environment scenario, where the pipeline provides suboptimal results due to the convoluted nature of the situation. In the top row we can observe how the segmentation map of a galaxy located in the vicinity of two bright stars is significantly impacted by a spike originating from one of them. This interference even causes the galaxy's segmentation map to incorrectly include another nearby source that was not properly deblended. The bottom row showcases an ambiguous case where the arms of a spiral galaxy have been partially over-deblended. This seems to be due to the galaxy's diffuse nature combined with the presence of internal features or background sources.

\subsection{Merging the VIS and NIR detections}\label{subsect:det_combination}

The results of the independent detection and deblending in VIS and NIR are one catalogue and one segmentation image in each of the VIS and NIR bands. These two data sets need to be combined to a single output catalogue and the corresponding segmentation image for the subsequent determination of object properties (photometry, morphology, etc.).

The VIS data have higher resolution ($0\overset{\prime\prime}{.}1$) than NIR ($0\overset{\prime\prime}{.}3$) and a better sampling. This leads to much better results for the detection and deblending. As a consequence, we combine the VIS and NIR datasets such that the merged catalogue and segmentation image contain all VIS objects plus the so-called NIR-only objects, which are the objects that are detected in NIR but not in VIS.

To identify the NIR-only objects we project the co-ordinates of all NIR-detected objects into the VIS segmentation image. NIR objects that are already detected in VIS are identified by evaluating the VIS segmentation image, which marks the pixels of all VIS detected sources. VIS segmentation values > 0 at the projected central NIR position are considered to be already detected in VIS, since they are already part of a VIS object. NIR objects with value 0 in the VIS segmentation image are not detected in VIS and thus form the subset of NIR-only sources. The final output catalogue is then the combined VIS and NIR-only catalogue.

To derive the combined VIS+NIR segmentation image, we copied the VIS segmentation image and added the detected pixels of the NIR-only sources from the NIR segmentation image to the now-combined segmentation image. In this operation we did not overwrite any VIS-detected pixels. Preserving all VIS-detected pixels changes and reduces the detection area of a few NIR-only objects, and we recomputed the properties (positions and shapes) of these. Also we discarded NIR-only objects that would end up having fewer than \verb|detection-minimum-area| pixels (see Tab. \ref{tab:detection_params}) after the combination of the segmentation maps.

Figure \ref{fig:vis_nir_detection} shows in the left and right panels a small cut-out from the VIS and NIR detection images, respectively. The red circles mark all sources detected on these cut-outs. The brighter objects are all detected in VIS and the NIR-detected sources constitute the faint and rather red sources. There are $141$ and $31$ sources detected on the VIS and NIR cut-out, respectively. As in this small cut-out about $20\%$ of all sources are detected in NIR.

It can sometimes happen in the combined VIS+NIR segmentation map that VIS and NIR-only sources are located side-by-side without any empty sky pixels in between. These artificial VIS-NIR blends are identified and marked by assigning a unique number to the corresponding objects in the final catalogue column \verb|PARENT_VISNIR|.

We are aware that current merging procedure may disfavour very interesting sources such as red objects that are very close to VIS objects. Some of those sources should be discovered in dedicated pipelines such as  the Strong Lensing Discovery Engine \citep[see][]{Q1-SP048}, but we are also planning to refine the merging procedure in the future.

\section{Mapping of the point spread functions and convolution kernels}\label{sect:PSF_calculation}

MER receives the PSFs measured by the VIS, NIR, and EXT PFs and propagates them to the MER mosaic grid to calculate a different PSF stamp for each individual source and band. VIS PSFs are measured for each VIS quadrant on a dedicated calibration processing run using the \texttt{PSFEx} software \citep{Bertin2011} and are assumed to remain constant for each VIS calibrated frame observation \citep{Q1-TP002}. NIR PSFs are calculated for each NISP detector and each calibrated frame \citep{Q1-TP003} using \texttt{PSFEx}.

MER propagates the VIS and NIR PSFs following a similar approach as described in Sect.~\ref{subsect:mosaicing}, with the following differences:
\begin{itemize}
\item the pixel values of the mosaicing step input images are set to zero, but the zero points, WCS information, flags, and RMS are left untouched.
\item PSF stamps are then added to the empty images at the deblended source positions, taking into account the VIS and NIR PSF spatial variations and the image zero points.
\item the \texttt{Swarp} resampling and co-adding software \citep{2002ASPC..281..228B} is run on the modified images and a mosaic with the averaged stamp PSFs is produced.
\item the averaged PSF stamps are extracted for each source and stored in a FITS file, which contains a grid with the PSF stamps and a table with their associated sky and pixel co-ordinates.
\end{itemize}

Source separations in the MER catalogue can be smaller than the input VIS and NIR PSF stamp sizes (less than \ang{;;3}). To avoid any overlap between the PSF stamps of nearby sources the whole process is repeated three times. Each run uses a different subset of sources whose separations are large enough to avoid PSF stamp overlaps. This generally produces a PSF stamp for more than 80\% of the sources. For the remaining sources, the closest PSF stamp is used, which by design is at a distance smaller than the PSF stamp size.

In its current state, the MER pipeline assumes that the PSF of a given band does not change with the source intrinsic colour. The propagation of wavelength-dependent PSFs could be introduced in the future if the VIS and NIR PFs are able to deliver their input PSFs with this information.

The EXT stage-2 PF calculates their stacked frames PSFs using a similar approach as described above. MER re-bins the EXT PSFs from the EXT stacked frame pixel size to the reference~$0\overset{\prime\prime}{.}1$ mosaic pixel size. The PSF stamp positions are based on the EXT stacked frame catalogue, which is generally not as deep as the MER catalogue, and as a result contains fewer sources. We selected the closest PSF stamp for each source in the MER catalogue.

\begin{table*}[h!]
    \centering
    \caption{Full width at half maximum (FWHM) extracted from the MER mosaics.}    
    \begin{tabular}{lccccccc}
        \hline
        \hline
        Survey & Band & \multicolumn{3}{c}{PSF FWHM} & \multicolumn{3}{c}{Gaia stars FWHM} \\
        & & Median & Min & Max & Median & Min & Max \\
        & & (\arcsec) &  (\arcsec) &  (\arcsec) &  (\arcsec) &  (\arcsec) &  (\arcsec) \\
        \hline
        EWS & \IE & 0.203 & 0.197 & 0.212 & 0.204 & 0.199 & 0.213 \\
        EWS & \YE & 0.475 & 0.456 & 0.484 & 0.492 & 0.474 & 0.502 \\
        EWS & \JE & 0.504 & 0.494 & 0.517 & 0.513 & 0.505 & 0.524 \\
        EWS & \HE & 0.542 & 0.536 & 0.547 & 0.552 & 0.539 & 0.570 \\
        UNIONS CFIS & $u$ & 1.00 & 0.84 & 1.16 & 0.99 & 0.84 & 1.17 \\
        UNIONS WHIGS & $g$ & 0.79 & 0.57 & 1.47 & 0.79 & 0.58 & 1.48 \\
        UNIONS CFIS & $r$ & 0.85 & 0.69 & 1.00 & 0.85 & 0.70 & 0.99 \\
        UNIONS Pan-STARRS & $i$ & 1.27 & 1.18 & 1.37 & 1.26 & 1.17 & 1.36 \\
        UNIONS WISHES & $z$ & 0.63 & 0.51 & 0.83 & 0.63 & 0.51 & 0.83 \\
        DES & $g$ & 1.28 & 1.02 & 1.40 & 1.28 & 1.01 & 1.43 \\
        DES & $r$ & 1.12 & 0.95 & 1.23 & 1.12 & 0.95 & 1.26 \\
        DES & $i$ & 1.07 & 1.00 & 1.47 & 1.07 & 0.99 & 1.47 \\
        DES & $z$ & 1.05 & 0.95 & 1.27 & 1.06 & 0.95 & 1.27 \\
        \hline
    \end{tabular}
    \tablefoot{PSF values are extracted from the MER mosaics. The different statistics are calculated combining the median FWHM values measured for each tile in the Q1 data release. Tiles with less than 20\% coverage in one of the \Euclid bands, as well as tiles associated with the Dark Cloud region, have not been considered.}
    \label{tab:PSF and stars FWHM}
\end{table*}

Table~\ref{tab:PSF and stars FWHM} compares the typical full width at half maximum (FWHM) values measured on the propagated PSF stamps and the detected \textit{Gaia} stars in the MER mosaics. The agreement between the median values is better than 1\%, with the exception of the NISP bands. The propagated PSFs for these bands are between 1.75\% and 3.5\% more compact than the \textit{Gaia} star profiles. Future versions of the NIR PF pipeline should improve the quality of the PSFs that MER receives as input. We expect that this will reduce the FWHM discrepancies that we currently find in the Q1 processing.

The FWHM values reported for the propagated VIS and NIR PSFs in Table~\ref{tab:PSF and stars FWHM} are 30\% to 35\% larger than the corresponding input PSFs measured on the VIS and NIR calibrated frames \citep{Q1-TP002,Q1-TP003}. This is a result of the stacking procedure described in Sect.~\ref{subsect:mosaicing}. In particular, the use of a BILINEAR interpolation kernel instead of a LANCZOS3 kernel introduces an extra 15\% broadening of the PSFs.

The multi-band photometry (described in Sect.~\ref{sect:photometry}) needs convolution kernels to homogenise the PSFs from the different bands. The algorithm we used to create the convolution kernels is based on Wiener filtering with a tunable regularisation parameter. For each detected object, we computed convolution kernels from higher- to lower-resolution bands using the object PSF stamps generated in the previous step. A complete description of the kernel creation algorithm can be found in \cite{Boucaud2016}.

\section{Multi-band photometry}\label{sect:photometry}

OU-MER must provide photometric measurements for all the detected sources. The pipeline has been designed to produce photometric estimates using various techniques, which can be combined in different ways to obtain the desired scientific information. Namely, the output catalogue contains: (i) aperture photometry, obtained using a customised version of the software \texttt{A-PHOT} \citep{Merlin2019}; (ii) template-fitting photometry, obtained using a customised version of the code \texttt{T-PHOT} \citep{Merlin2015, Merlin2016}; and (iii) model-fitting photometry, obtained using \texttt{SE++}. 

Concerning aperture photometry, the total flux of each object is estimated as the flux within a \citet{Kron1980} elliptical aperture on the detection image, which is the \IE mosaic for most of the sources, and the NIR stack for the additional NIR-detected sources (see Sect.~\ref{subsect:mosaicing}). This corresponds to the \texttt{FLUX\_AUTO} in \texttt{SE2}. Here, the Kron radius was computed as 
\begin{equation}
  R_{\rm Kron} =   \max \left(7 \, a, \frac{ \sum_{ij} r_{ij}\,f(i,j)}{\sum_{ij} f(i,j)}\right),
\end{equation}
where $a$ is the semi-major axis provided by the deblending algorithm during the deblending step, $r_{ij}$ is the distance of each pixel to the centroid of the object, $f(i,j)$ is its flux, and the summation is extended over a circular region typically of radius $6\, a$. The final Kron area for which the flux is computed is an elliptical area of semi-major axis $2.5\, R_{\rm Kron}$ and ellipticity, again provided by the deblending step. \texttt{A-PHOT} sub-divides the pixels crossed by the limit of the elliptical aperture into 10 sub-pixels, to provide a more precise estimate.

Aperture photometry was then measured for all bands on PSF-matched images, within a set of circular apertures with diameters fixed individually source by source as integer multiples of the FWHM of the worst-resolution band (typically an EXT band, with FWHM around $1''$). In this way, the same physical region of the sources was always sampled, providing a robust colour estimate. The smoothing was performed using convolution kernels obtained from the PSFs estimated in the manner described in Sect. \ref{sect:PSF_calculation}. To determine the total flux in each band, one must combine the value obtained on the detection band with these aperture fluxes, as follows \citep[e.g.,][]{Merlin2022}:
\begin{equation}
 f_{\rm tot,band}=f_{\rm ap,band} \, \frac{f_{\rm tot,det}}{f_{\rm ap,det}}.   
\end{equation}

The uncertainty budget can be computed with an analogous formula, where it is safe to assume that the ratio $f_{\rm tot,det}/f_{\rm ap,det}$ is a constant and there is no need to propagate their errors. The uncertainty, $e_{\rm ap,band}$, was computed by \texttt{A-PHOT} as the square root of the quadratic sum of the pixels in the PSF-matched RMS map, within the considered aperture.
All the measurements were done without performing a further local background subtraction, since two background subtraction steps had already been performed on the images.

Template-fitting photometry is a well-established technique used to exploit spatial and morphological information from high-resolution images to estimate fluxes on lower-resolution images of the same FoV, using cut-outs of sources from the former (smoothed with a convolution kernel) as priors for the latter, to mitigate the effect of blending of sources. While it was initially designed to fit infrared data (from e.g., \textit{Spitzer}) with HST priors, it has proven to be useful in the \Euclid context as well. The crispness and depth of the \IE data can provide exquisite priors for EXT data, and can be used for NIR data as well, although in that case the difference in resolution is not dramatic. Fluxes are obtained by solving a $\chi^2$ minimisation problem, 
\begin{equation}
\chi^2 = \sum_{m,n} \left [ \frac{I(m,n) - \sum_i^N F_i(m,n)P_i(m,n)}{\sigma(m,n)}\right ]^2,
\end{equation}

\noindent where $m$ and $n$ are pixel indexes, $N$ is the number of sources being fitted, $P$ are the values of the pixels of the normalised high-resolution cut-outs convolved to the measurement image resolution, $I$ are the values of the pixels in the measurement image, and $\sigma$ are the values of the pixels of its RMS map \citep[see][for details]{Merlin2015}. The error budget of the fitting process is obtained from the covariance matrix of the problem, as the square root of the diagonal element relative to that source (i.e. its variance). 
In principle, one could fit the full image at once, but in practice this would be too computationally demanding given the large spatial dimensions of the OU-MER tiles. The pipeline therefore exploits a feature of \texttt{T-PHOT} that allows one to perform individual fits looping on the sources, building a cell around each one including its contaminants, and removing if after it has been fitted. This algorithm has proven to yield a significant saving of computational time, while keeping optimal accuracy. 

\begin{figure}
    \centering
    \includegraphics[width=0.45\textwidth]{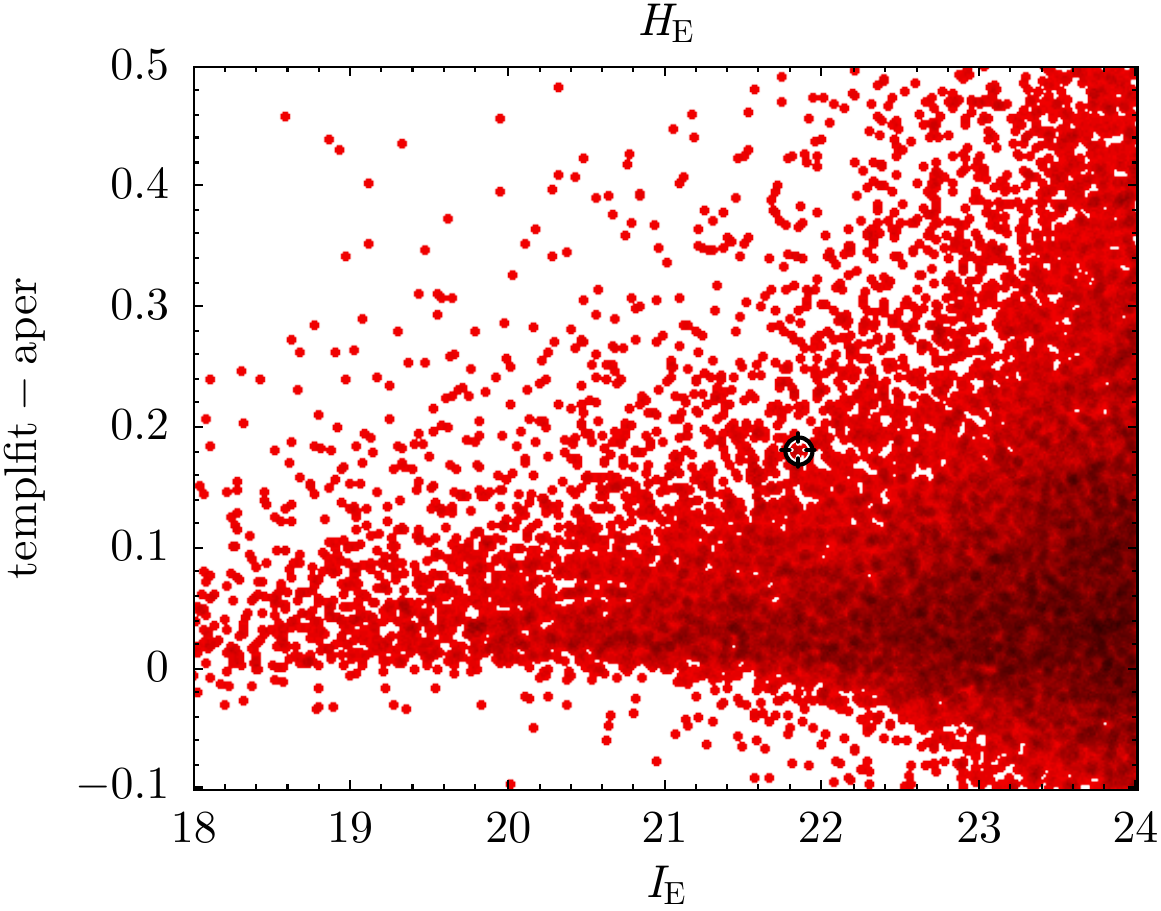}
    \includegraphics[width=0.45\textwidth]{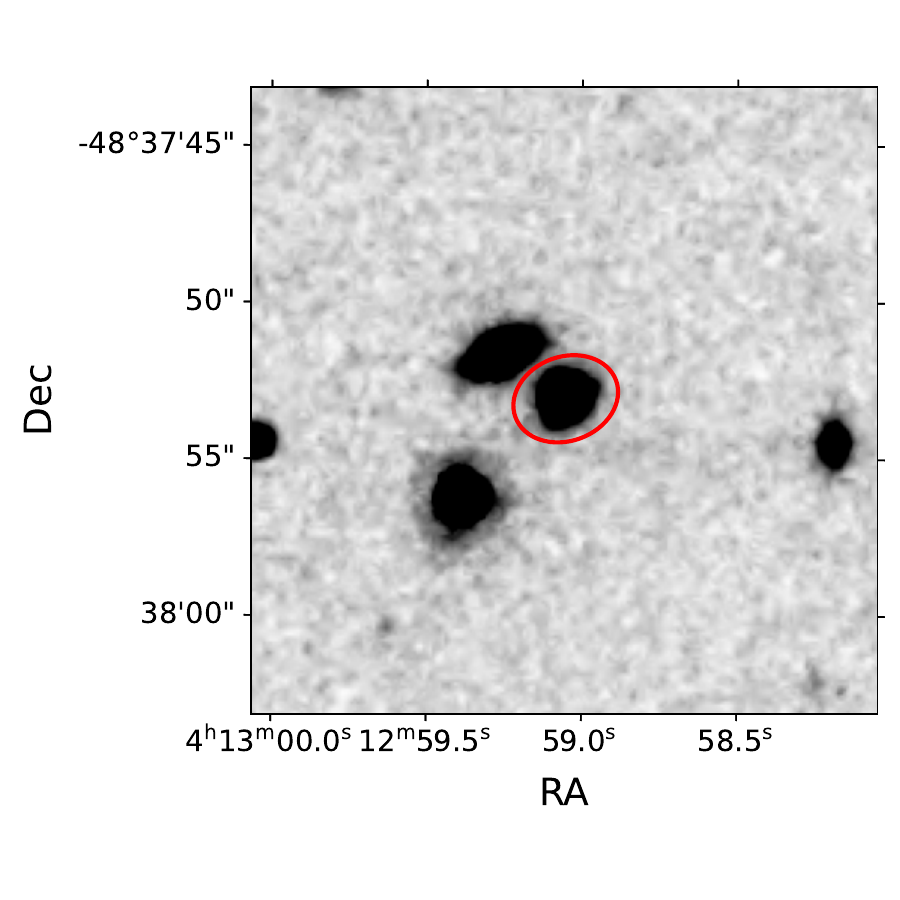}    
    \caption{Upper panel: Difference in \HE total magnitudes as obtained with template-fitting and aperture photometry as a function of \IE detection magnitude, for extended objects (\texttt{POINT\_LIKE\_PROB}$<0.7$, see Sect. \ref{subsect:sg_separation}) in a random OU-MER tile. Template-fitting magnitudes are typically fainter, because the contaminating light from blended sources is removed in the fitting process. Lower panel: As an example, the source selected with the black circle in the upper panel in the corresponding \HE OU-MER mosaic. The AB magnitude of the pixels within the white aperture (which is certainly still contaminated by the nearby objects), as given by the \texttt{SAO ds9} software, is 19.96; the OU-MER catalogue values are 19.81$\pm$0.02 (total from colour in 2FWHM aperture) and 19.99$\pm0.01$ (template-fitting).}
    \label{tphot_example}
\end{figure}

\begin{figure}
    \centering
    \includegraphics[width=0.5\textwidth]{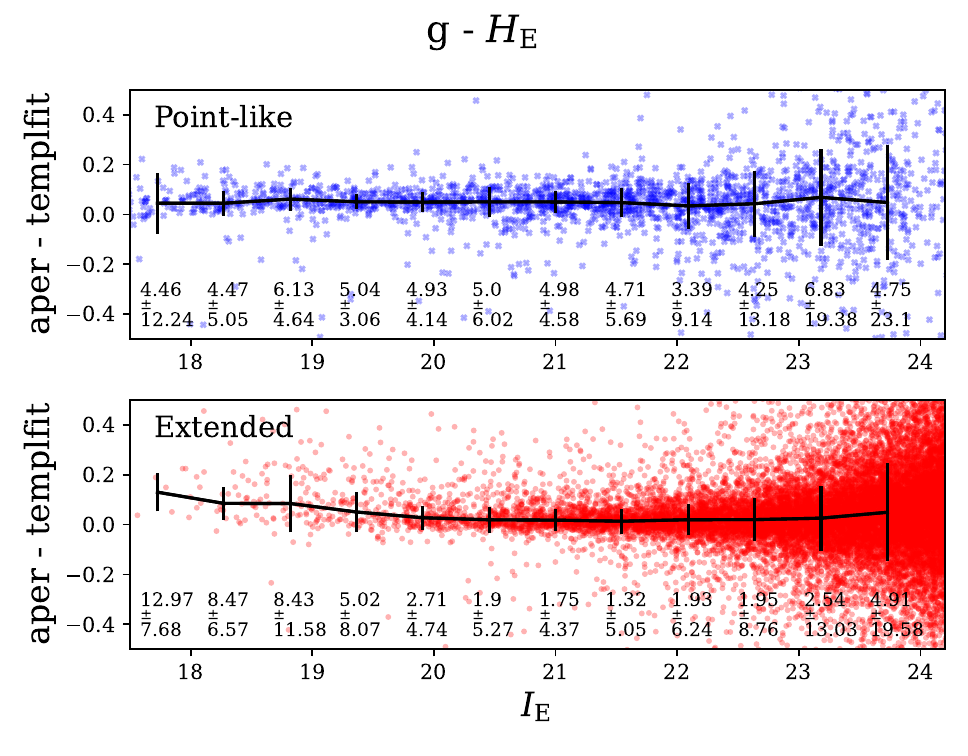}
    \caption{Difference in $g$-\HE colours obtained from aperture and template fitting photometry for a random OU-MER tile, as a function of \IE magnitude. Upper panel: Point-like sources, defined as objects with \texttt{POINT\_LIKE\_PROB>0.7}. Lower panel: Extended sources. For each magnitude bin, the percent 3$\sigma$-clipped median and standard deviation are reported in the plot. Weighted average values for the whole sample are 4.92$\pm$10.39 (point-like sources) and 3.29$\pm$13.25 (extended sources).}
    \label{color}
\end{figure}

The pipeline provides the direct output of the template-fitting runs, which can be taken as a proxy for the total flux of the objects. Additionally, fluxes measured in the detection band smoothed to the resolution of each other band are also provided; they can be used to obtain an estimate of colours complementary to that obtained with aperture photometry, and used to scale the detection Kron flux to get another total flux value in each band. Typically, one expects $f_{\rm templ,band} \leq f_{\rm tot,band}$, where the latter is the total flux from aperture photometry as described above; this is because the contamination from nearby sources, when present, should be drastically decreased. Figure \ref{tphot_example} shows the situation for a random OU-MER tile, and the case of one discrepant galaxy. However, the colours obtained with the two techniques should obviously be reasonably consistent. Figure \ref{color} shows the compared $g$ - \HE colours ($g$ is from DECAM) obtained from aperture and template fitting photometry in the same tile, for point-like (blue crosses, \texttt{POINT\_LIKE\_PROB}$>0.7$) and extended sources (red dots, \texttt{POINT\_LIKE\_PROB}$\leq0.7$; see Sect. \ref{subsect:sg_separation}), ensuring this is indeed the case. It should be stressed that for Q1 we found that for point-like sources (bright stars in particular), contrary to what was just discussed for extended sources, the total fluxes computed using aperture photometry were found to be more accurate than template-fitting fluxes in NIR bands, most likely due to their higher sensitivity to PSF related issues (see Sect. \ref{subsect:photometry}). Currently, OU-PHZ uses aperture colours to compute photometric redshifts \citep[see][]{Q1-TP005}.

We point out that template-fitting, where real cut-outs of high-resolution images are used to infer the expected shape of low-resolution images, is different from model-fitting, where galaxy profiles are fitted using analytical models. Indeed, the pipeline also provides photometric values  derived by fitting a parametrised model to the multi-band imaging data. For Q1 we have chosen to fit S\'ersic models \citep{1963BAAA....6...41S} to all detected objects. This is done with \texttt{SE++} in an iterative procedure that involves VIS and NIR data to fix the S\'ersic parameters and then EXT data for photometry only. Further details are given in Sect.\ \ref{subsect:sersic}.

It is worth recalling that, as it is custom practice, the provided error budgets are a measure of the uncertainty due to the observational noise level, as encoded in the RMS map. In other words, it does not include an estimate of any systematic deviation to which a measurement might be prone (e.g., because of contamination from nearby sources), and might therefore be an underestimation of the actual uncertainty on the real flux of the sources. The flux within the segmented area of each source is also provided (corresponding to \verb|FLUX_ISO| in \texttt{SE2}). Finally, a PSF-fitting estimate is provided for the VIS band only, again using \texttt{T-PHOT}.

All the photometric measurements are stored in the MER photometric catalogues (Sect. \ref{sect:catalog}), a column-by-column description of which is referenced in Appendix \ref{appendix:a}. It is not always trivial, though, to handle the amount of information stored in such a huge container. In order to help the user to become familiar with our products and to handle the photometric information measured by OU-MER, we provide a Photometry Cookbook\footnote{\url{http://st-dm.pages.euclid-sgs.uk/-/data-product-doc/-/jobs/162527/artifacts/build/merdpd/merphotometrycookbook.html}} , explaining the basic operations that need to be performed in order to correctly use MER data.

\section{Morphological characterisation}\label{sect:morphology}

For each galaxy used in the weak lensing analysis up to the limiting magnitude the following information is provided: (i) the classification star-galaxy; (ii) at least one size measurement. An object radius is provided via the detection and photometry pipelines. Three kinds of morphological estimators are included in the MER catalogue:
\begin{itemize}
    \item Non-parametric parameters. Those are Concentration, Asymmetry, and Smoothness \citep[CAS,][]{Conselice2014,Tohill2021}, the Gini index \citep{Lotz2004}, and the second order moment at 20\% flux, i.e the M20 index;
    \item Parametric parameters computed through model fitting \citep{Bretonniere-EP13,Bretonniere-EP26};
    \item Deep learning based morphologies calibrated on visual classifications.
\end{itemize}

As per the photometric catalogue, we provide a Morphology Cookbook\footnote{\url{http://st-dm.pages.euclid-sgs.uk/-/data-product-doc/-/jobs/162527/artifacts/build/merdpd/mermorphologycookbook.html}} aimed at better understanding how to correctly use the information stored in the MER morphology catalogue.

\subsection{Non-parametric indices for object morphology}\label{subsect:morpho_cas}
The original implementation of CAS, Gini, and M20 parameters as \Euclid PEs was coded from \texttt{ASTErIsM}, a code for deblending objects and morphological measurements \citep{Tramacere2016}. The CAS parameters are non-parametric parameters carrying morphological characteristic in a three-dimensional space that segregate main morphological features. They are computed on \Euclid $\IE$ images covering the segmentation map of each object.

\subsubsection{Concentration index, $C$}
The concentration of light index, $C$ \citep{Conselice2014}, is used as a method of quantifying how much light is in the centre of a galaxy as opposed to its outer parts. It correlates strongly with S\'ersic $n$ values,
\begin{equation}
    C := 5 \log_{10}\frac{r_\text{80}}{r_\text{20}},
    \label{equ_C}
\end{equation}
where $r_\text{80}, r_\text{20}$ are the circular apertures enclosing $80 \%$ and $ 20 \%$ of the total galaxy flux.
The associated error is defined as
\begin{equation}
C_\mathrm{err} :=  5 \logten \frac{r_{80}}{r_{20}} - 5 \logten
\frac{r_{80}-0.5}{r_{20}+0.5}.
\end{equation}

\begin{figure}
    \centering
    \includegraphics[width=0.25\textwidth]{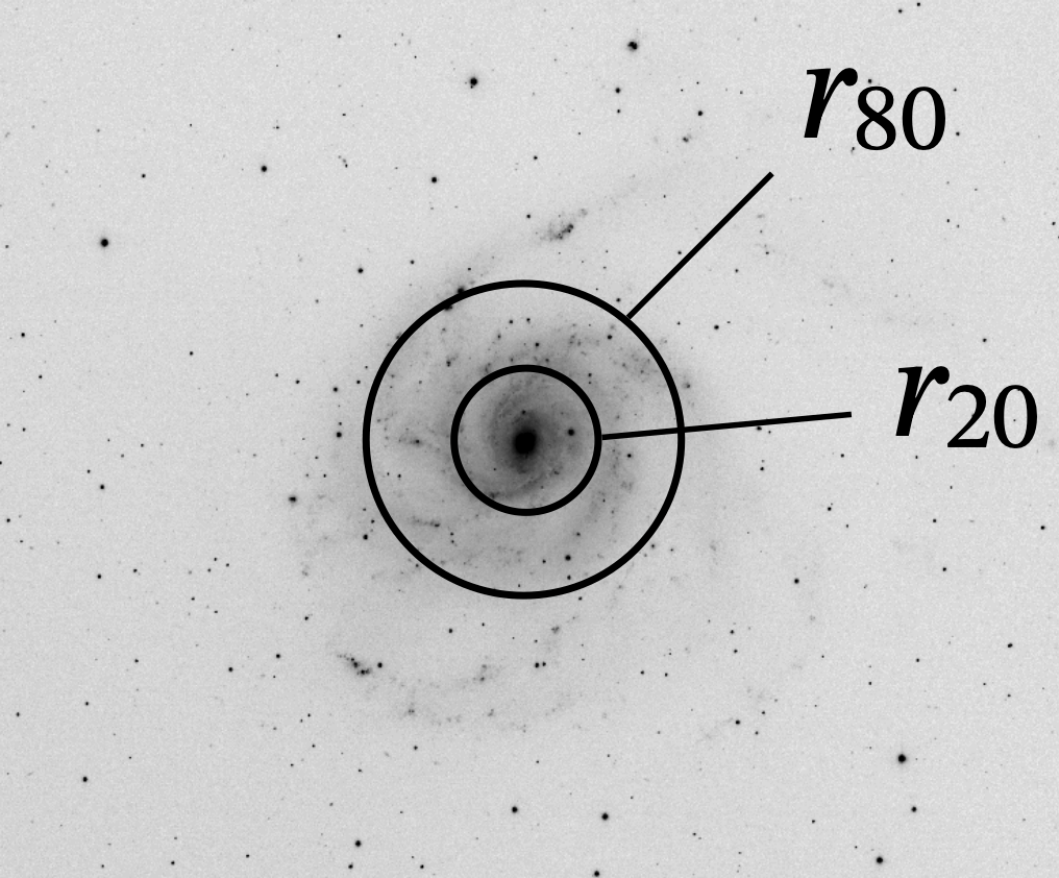}
    \caption{Concentration calculation schematics, here $r_{80}$ and $r_{20}$ are circular radii encompassing 80\% and 20\% of the galaxy flux (cf text and Eq.~\ref{equ_C}).}
    \label{concentration}
\end{figure}

\subsubsection{Asymmetry, $A$}
\begin{figure}
    \centering
    \includegraphics[width=0.47\textwidth]{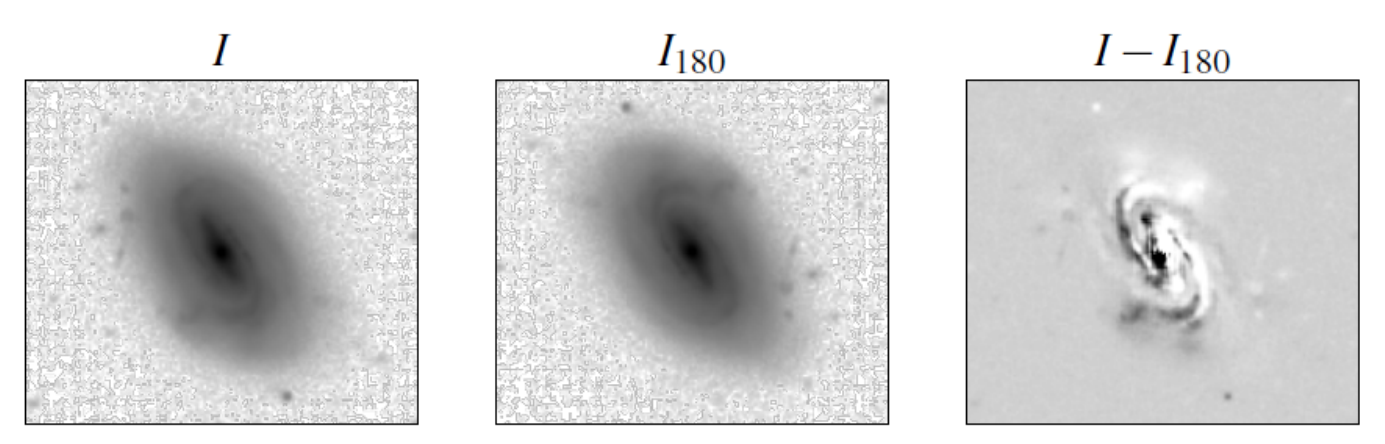}
    \caption{Asymmetry calculation schematics. The original image ($I$) is rotated by \ang{180.;;} ($R$), the asymmetry index, $A$, was computed from a combination of the two (cf text and Eq.~\ref{equ_A}).}
    \label{asymmetry}
\end{figure}

The asymmetry index, $A$ \citep{Conselice2014}, measures how asymmetric a galaxy is after rotating along the line of sight centre axis of the galaxy by \ang{180.;;} (Fig.~\ref{asymmetry}). The analytic definition calculating $A$ is given by
\begin{equation}
    A := \min\left(\frac{\sum |I_0 - I_{180}|}{\sum|I_0|}\right) -  \min\left( \frac{\sum |B_0-B_{180}|}{\sum|I_0|}\right),
    \label{equ_A}
\end{equation}
where $I_0$ is the original galaxy image, and $I_{180}$ is the image after rotating it from its centre by \ang{180.;;}. The measurement of the asymmetry parameter includes a careful treatment of the background noise in the same way that the galaxy itself is analysed by using a blank background area ($B_0$), and finding the location for the centre of rotation. The area $B_0$ is a blank part of the sky near the galaxy. The centre of rotation is not defined a priori, but tweaked iteratively around the starting value to yield the minimum value of $A$. The error on $A$ is

\begin{equation}
A_\mathrm{err} := \frac{\sigma[B(x,y)] }{ f+ \sqrt{|f|}}.
\end{equation}

\subsubsection{Clumpiness (smoothness) index, $S$}

The clumpiness (or smoothness) index, S \citep{Conselice2014}, is used to describe the fraction
of light in a galaxy that is contained in clumps.  The smoothness can by
computed as

\begin{equation}
S:=  \frac{f_s-f_b}{f},
\end{equation}
with
\begin{eqnarray}
f_s&:=& \sum I(x,y) - I(x,y)_\mathrm{blurr},\\
f_b&:=& \sum B(x,y) - B(x,y)_\mathrm{blurr},\\
f &:=& \sum I(x,y),
\end{eqnarray}
where $I(x,y)$ are the image pixels chopped within $r_{\rm inner} = 0.1
\eta$ of the object flux peak, $\eta$ is the petrosian radius
(Petrosian, 1976), $ I(x,y)_\mathrm{blurr}$ is the chopped image
convolved with a gaussian kernel of $\sigma =  0.1 \eta$, $B(x,y)$, and
$B(x,y)_\mathrm{blurr}$ are the noise maps treated identically.
Finally, the associated error is

\begin{equation}
S_\mathrm{err} := \frac{10 \sqrt{| n f_s |}}{n(f+\sqrt{|f|})},
\end{equation}
where $n$ is the number of unmasked pixels.

\subsubsection{Gini index}
The Gini index \citep{Lotz2004} originates from the world of economics, where it has been defined originally to evaluate inequalities of incomes in a population. If everyone earns the same income, Gini\,=\,0, if one person concentrate all incomes Gini\,=\,1. Applied on an astronomical image, the calculation is done on pixels covered by the segmentation map of objects. A galaxy in this case is considered a system with $n$ pixels each with a flux $f_i$, where $i$ ranges from 1 to $n$. The Gini index is then measured by

\begin{equation}
    \text{Gini} := \frac{1}{|\bar{f}|n(n-1)}\sum_i^n (2i-n-1)|f_i| \, ,
\end{equation}
where $\bar{f}$ is the average pixel flux value.

\subsubsection{Moment-20 index, M20}
The mathematical definition of the M20 index is
\begin{equation}
    \text{M20}:= \log_{10}\left(\frac{\sum_i M_i}{M_\text{tot}}\right) \quad\text{while}\quad \sum_i^n i f_i < 0.2 f_\text{tot}.
\end{equation}
The value of $M_\text{tot}$ is
\begin{equation}
  M_\text{tot} = \sum_i^n M_i = \sum_i^n f_i [(x_i-x_c)^2 + (y_i-y_c)^2], 
\end{equation}
where ($x_c$,$y_c$) is the flux-weighted mean position of the galaxy. In the case of M20, this centre is defined as the location where the value of $M_\text{tot}$ is minimal.

\subsection{Parametric morphology}\label{subsect:sersic}
To derive some parametric morphology, we have decided to fit two-dimensional S\'ersic profiles \citep{1963BAAA....6...41S} in all photometric bands to all objects. As a function of the angular radial $r$, the S\'ersic function defines the variation in the light intensity as
\begin{equation}
    I(r) = I_\mathrm{e} \exp\left\{ -b_n\left[\left( \frac{r}{R_\mathrm{e}}\right) ^{1/n} -1\right]\right\},
    \label{eq-sersic}
\end{equation}
with $R_\mathrm{e}$ the major-axis of the elliptical profile that encloses half of the total light, $I_\mathrm{e} = I(R_\mathrm{e})$ the light intensity at $R_\mathrm{e}$, $n$ the S\'ersic index characterising the steepness of the profile, and $b_n$ a normalisation parameter that depends solely on $n$. Fitting 2D elliptical light profiles to the \Euclid imaging data adds the position angle and the axis ratio.

The S\'ersic fitting was performed using \texttt{SE++} \citep{2022arXiv221202428K}, 
which, from version 0.21\footnote{\url{https://github.com/astrorama/SourceXtractorPlusPlus}} on, includes model fitting as an isolated process without the preceding object detection process. All necessary parameters such as objects' centres, the fitting area, and initial values for the model parameters are provided in an input table to the model fitting process.

The model fitting in the OU-MER PF has a dual purpose and provides, in addition to the S\'ersic parameters, also the S\'ersic flux in all photometric bands as an independent photometry in addition the ones introduced in Sect.\ \ref{sect:photometry}. We did this in an iterative process:
\begin{itemize}
\item The first model fitting operation on VIS and NIR data computed all structural S\'ersic parameters and the VIS and NIR fluxes:
\begin{itemize}
\item We fitted to the VIS (\IE) data the corresponding S\'ersic radius ($R_{\rm VIS}$), ellipticity ($e_{\rm VIS}$), and index ($n_{\rm VIS}$);
\item We also fitted to the NIR (\YE/\JE/\HE) data the independent S\'ersic radius ($R_{\rm NIR}$), ellipticity ($e_{\rm NIR}$), and index ($n_{\rm NIR}$) for the NIR bands.
\item The VIS and NIR  S\'ersic models share the same angle, $\alpha_{\rm VIS,NIR}$.
\item In addition the integrated S\'ersic flux in the VIS and NIR bands ($f_{\rm {S\text{\'e}rsic},\IE}$ $f_{\rm {S\text{\'e}rsic},\YE}$, $f_{\rm {S\text{\'e}rsic},\JE}$ and $f_{\rm {S\text{\'e}rsic},\HE}$) were computed as well.
\end{itemize}
\item The second model fitting operation runs on the EXT data $g, r, i, z$. It uses the S\'ersic parameters, $R_{\rm VIS}$, $e_{\rm VIS}$, $n_{\rm VIS}$, and $\alpha_{\rm VIS,NIR}$, as constant values to compute the integrated S\'ersic flux in those EXT bands.
\item In both fitting operations we used the object centres provided as an input as the fixed centres for the S\'ersic models.
\end{itemize}
For all fitted quantities we provide the error obtained from the square root of the diagonal element of the covariance matrix. The approach of using an iterative approach for VIS plus NIR for the morphology and then EXT for photometry is rather unusual, and configuration for deriving two independent set of S\'ersic parameters with a common orientation angle is complex. As an example of this set-up we share a multi-band data set\footnote{\url{https://cloud.physik.lmu.de/index.php/s/3K4KemBsw5y9yqd}}. As imaging data we provide cut-outs from a Q1 tile. There are also the various configuration files for both model fitting operations described above and the \texttt{SE++} commands to run the \texttt{SE++} model fitting on the data. A scientific evaluation of the parametric morphology provided in the OU-MER catalogues is given in \cite{Q1-SP040}.

\subsection{Machine-learning-based morphology - Zoobot}\label{subsect:morpho_zoobot}

The MER catalogue contains visual morphologies following the \texttt{GalaxyZoo} (GZ) classification tree. This includes broad morphological classes, such as `featured' or `smooth', as well as more detailed features such as the count and tightness of spiral arms or the strength of stellar bars. For the Q1 data release, the MER catalogue includes visual morphologies for galaxies brighter than \IE$<20.5$ or with a segmentation area larger than $1\,200$ pixels, which represent around $1\%$ of galaxies in the photometric catalogue. This conservative selection is performed to guarantee that the structure is unambiguously detected but could be extended to fainter and smaller sources in future releases.

Galaxies are classified with the deep foundation model \texttt{Zoobot} \citep{Walmsley2023zoobot}.
For this work, we use the pre-trained ConvNext Nano model ($\rm 22.5$ million parameters) available on HuggingFace\footnote{\url{https://huggingface.co/mwalmsley/zoobot-encoder-convnext_nano}}. This is pre-trained on the GZ Evo dataset~\citep{2022mla..confE..29W,walmsleyScalingLawsGalaxy2024}, which includes $\rm 820\,000$ images and over $\rm 100$ million volunteer votes drawn from every major Galaxy Zoo campaign: GZ2~\citep{2013MNRAS.435.2835W}, GZ UKIDSS ~\citep{mastersGalaxyZooMorphologies2024}, GZ Hubble~\citep{2017MNRAS.464.4176W}, GZ CANDELS~\citep{2017MNRAS.464.4420S}, GZ DECaLS/DESI~\citep{2022MNRAS.509.3966W,2023MNRAS.526.4768W}, and GZ Cosmic Dawn (in prep).

To optimise the classifications for \Euclid, we fine-tune \texttt{Zoobot} with an additional 1.56 million labels obtained on \Euclid stamps through a dedicated GZ campaign carried out during August 2024. Specific details about the  methodology and reliability of the classifications are presented in~\cite{Q1-SP047}.  

\subsection{Star-galaxy separation}\label{subsect:sg_separation}

A simple star-galaxy (S/G) classifier has been historically implemented in the MER pipeline in order to identify the point-like detected objects on which the PSF characterisation could be performed by SHE. In addition to the \verb|POINT_LIKE_FLAG|, a point-like probability (\verb|POINT_LIKE_PROB|) was also requested in the output catalogue. Note that this classifier is heavily biased towards a high purity, and thus has a low completeness. The method is inspired by the \verb|SPREAD_MODEL| method provided by \texttt{SE2}, which is used in \citet{Desai2012} and \citet{SpreadModel2018}. 

Our method uses \verb|MU_MAX| $-$ \verb|MAG_AUTO| as a proxy for \verb|SPREAD_MODEL|; \verb|MU_MAX| being the peak surface brightness above the background. Thus, our estimator \verb|MU_MAX| $-$ \verb|MAG_AUTO| is related to the concentration of light at the peak versus the total magnitude.
This parameter has been used in \citet{Jauzac2012}, \citet{sharon2022}, and \citet{Estrada2023} as input for the S/G classification.

\begin{figure}
    \centering
    \includegraphics[width=0.5\textwidth]{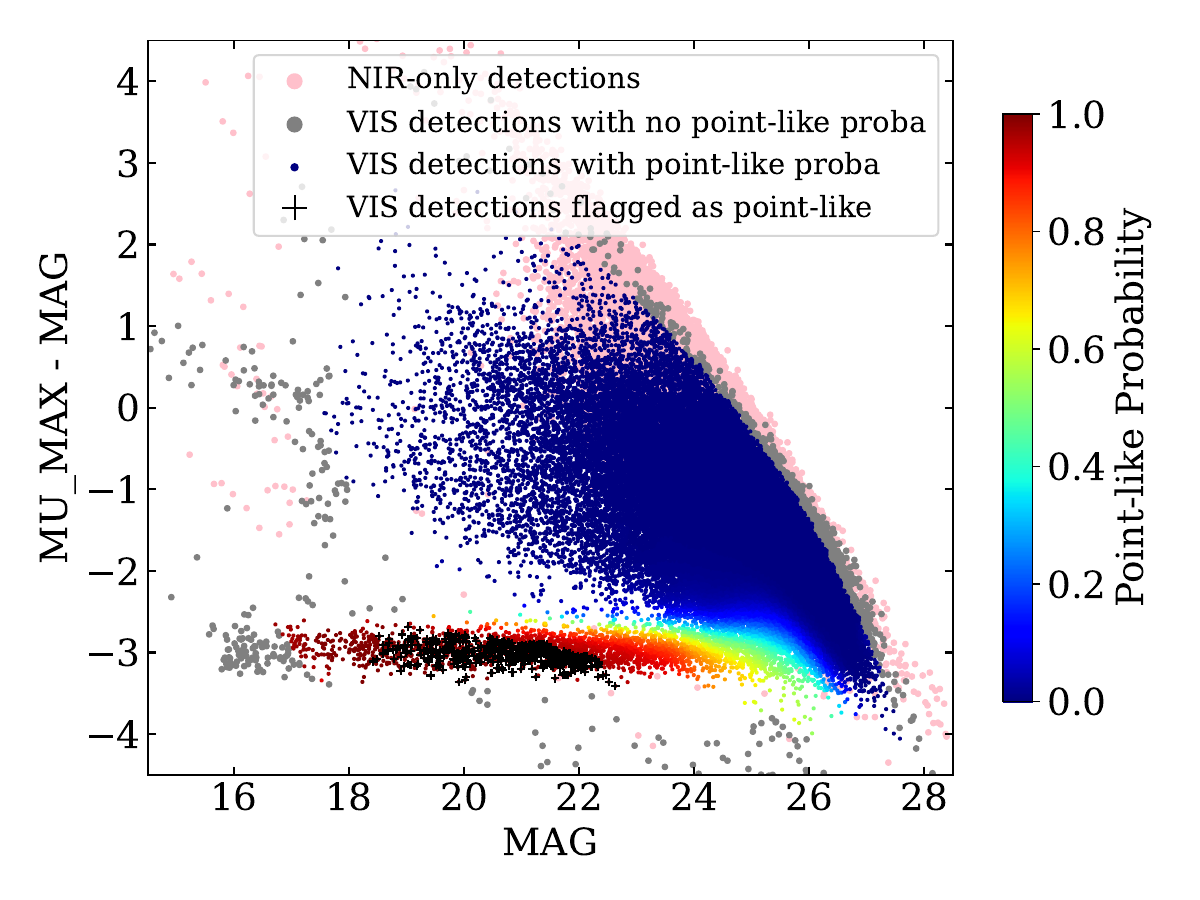}
    \caption{VIS and NIR detections in the {\tt MU\_MAX} $-$ {\tt MAG\_AUTO} plane for real data from tile 102021495. Stars are prominently present in the bottom horizontal branch. Black crosses are sources identified as stars and the colours code the probability of being a point source.}
    \label{fig:tile_sg_sep_pdf}
\end{figure}

Before flight, the point-like probability has been estimated in the plane of Fig.~\ref{fig:tile_sg_sep_pdf}  ($x_m=\texttt{MAG},y_m=\texttt{MU\_MAX}-\tt{MAG}$) using official simulated data by merely counting the number of galaxies and stars in each bin of this 2D space. Afterwards, the point-like probability of any real observation is computed by interpolating in this 2D array at the ($x_m,y_m$) position of the observation.

As shown in Fig.~\ref{fig:tile_sg_sep_pdf}, the S/G separation is only performed in VIS-detected objects. The current rule-of-thumb to select stars in the MER catalogue is: (a) select very bright (even saturated) sources at VIS magnitudes $\IE < 17$; and (b) select \verb|POINT_LIKE_PROB|~$>0.96$. Note that the \verb|POINT_LIKE_FLAG| is defined as \verb|POINT_LIKE_PROB|~$>0.96$ and bits 1 to 4 of \verb|DET_QUALITY_FLAG|~$=0$ so this flag can also be used. 

As a more refined object classification, differentiating between star, galaxies and QSO, is performed by PHZ \citep{Q1-TP005} using the full photometric coverage and the spectral energy distribution information of each source, we did not use  other S/G separation and classification methods  \citep{odewahn2004,soumagnac2015,slater2020} at the MER level.

\section{The MER catalogue}\label{sect:catalog}

The last step of the MER pipeline merges all the measurements performed in the previous steps and creating the output catalogues. The format of the MER PF output catalogues comply with the \Euclid SGS standard. 
The main output of the cataloguing PE is the following:
\begin{itemize}
    \item \verb|EUC_MER_FINAL-CAT|: a FITS table, storing the official \Euclid source ID, i.e., the \verb|OBJECT_ID|, the source co-ordinates, the quality flags, and all the information related to the photometry measurements. Some DEEP regions could contain imaging data from additional EXT filters. In those cases the additional EXT photometry measurements are stored in a separated FITS table, namely the \verb|EUC_MER_FINAL-DEEP-CAT|.
    \item \verb|EUC_MER_FINAL-MORPH-CAT|: a FITS table storing all the morphological measurements performed in the MER pipeline. Each source in this table shares the \verb|OBJECT_ID| information with all the other catalogues produced within this step.
    \item \verb|EUC_MER_FINAL-CUTOUTS-CAT|: a FITS table containing information on the corners of the source cut-outs. As for the \verb|EUC_MER_FINAL-MORPHO-CAT|, all the sources share the \verb|OBJECT_ID| information with the rest of the output catalogues.
    \item \verb|EUC_MER_FINAL-SEGMAP|: a FITS binary image representing the map showing the connected pixels of the objects detected on the corresponding detection mosaics (VIS and NIR).
\end{itemize}
Appendix~\ref{appendix:a} gives a the detailed description of the content of the MER PF output catalogues. 

External data coverage is heterogeneous across the different sky regions observed by \Euclid. In particular, the Q1 release focuses on the processing of the EWS, including data from the DES (South) and UNIONS (North, Gwyn et al. in prep.) surveys, namely WHIGS, CFIS, Pan-STARRS, and WHISHES. Table~\ref{tab:band_list_q1} shows the list of bands available in the Q1 release.

\begin{table*}[h!]
    \centering
    \caption{List of photometric bands available in the Q1 release.}
    \begin{tabular}{lll}
       \hline
       \hline
       Survey & Band & Filter label \\
       \hline
       EWS                & \IE                        & \verb|VIS| \\
       EWS                & \JE                        & \verb|J| \\
       EWS                & \HE                        & \verb|H| \\
       EWS                & \YE                        & \verb|Y| \\
       EWS                & Stack of \JE, \HE and \YE  & \verb|NIR_STACK| \\
       UNIONS CFIS        & $u$                          & \verb|U_EXT_MEGACAM| \\
       UNIONS CFIS        & $r$                          & \verb|R_EXT_MEGACAM| \\
       UNIONS Pan-STARRS  & $i$                          & \verb|I_EXT_PANSTARRS| \\
       UNIONS WHIGS       & $g$                          & \verb|G_EXT_HSC| \\
       UNIONS WISHES      & $z$                          & \verb|Z_EXT_HSC| \\
       DES                & $g$                          & \verb|G_EXT_DECAM| \\
       DES                & $r$                          & \verb|R_EXT_DECAM| \\
       DES                & $i$                          & \verb|I_EXT_DECAM| \\
       DES                & $z$                          & \verb|Z_EXT_DECAM| \\
       \hline
    \end{tabular}
    \tablefoot{The `Filter label' columns, shows how the filter is labelled in the MER catalogue data model.}
    \label{tab:band_list_q1}
\end{table*}

\subsection{E(B-V) estimates}\label{subsect:galext}

The distribution of galaxies on large scales probed by the \Euclid satellite is affected by Galactic absorption and therefore the study of the galaxy clustering has to be corrected by the different survey depths, which could induce artefacts in the analysis. Moreover, the errors in this correction should be taken into account in the covariance matrix.

OU-MER provides the extinction $E(B-V)$ as estimated by the R1.20 \textit{Planck} data release and estimates an error in the $E(B-V)$. This is done via the \verb|LE3_GALEXT_ED| PF, a Python program that imports the public \texttt{healpy}\footnote{\url{https://healpix.jpl.nasa.gov/}} in order to read the extinction map described in \cite{Planck2013}. The map provides values for $E(B-V)$ and $\tau$, the optical depth of dust as measured at 353\,GHz, together with its associated error, $\tau_{\text{err}}$. The error associated with $E(B-V)$ can be calculated via the ratio between~$\tau_{\text{err}}$~and~$\tau$,

\begin{equation}
E(B-V)_{\text{err}} = E(B-V) \, \frac{\tau_{\text{err}}}{\tau}.
\label{equ_EBV}
\end{equation}

The algorithm takes a list of sky positions in a specified co-ordinate system and computes the dust extinction $E(B-V)$ based on the reference extinction map, assigning an $E(B-V)$ measurement and its associated error to each source in the catalogue.

\subsection{Identification of spurious sources}\label{subsect:spurious_proba}

We have developed a machine-learning tool to identify and remove spurious (i.e not real) sources in the photometric catalogues. This belongs to the general class of classification problems, since we need to assign a binary flag ($1$ for spurious objects, $0$ otherwise) to each source.  
For this task, we adopt the Random Forest Classifier (RFC) algorithm \citep{Breiman2001}. We trained the algorithm on a reference simulated catalogue of $\rm 430\,000$ sources, to which we already assigned a spurious flag, matching the catalogue to the \Euclid pre-launch True Universe (TU) simulation \citep{SIM2024}. Sources that are matched to at least one TU source (either a star or a galaxy) within a radius of $0\overset{\prime\prime}{.}2$ are flagged as real (spurious flag set to 0), while sources that are not matched to any TU object within $0\overset{\prime\prime}{.}8$ are considered to be spurious (spurious flag set to 1). We exclude sources with matching radii between $0\overset{\prime\prime}{.}2$ and $0\overset{\prime\prime}{.}8$, to minimise the uncertainty of the label assignments. We kept $2\%$ and $20\%$ of the original sample, randomly selected, as validation and testing set, respectively, to check the performance of the classification algorithm. 
The RFC uses a set of $N$ input features to determine whether a source is spurious or not, and the training allows us to find an optimal function from the input $N$-dimensional parameter space to the final binary flag. We selected, as input features, template fitting (TPhot) fluxes and S/N of photometric bands \IE, \YE, \JE, and \HE from \Euclid, and the $g$, $r$, $i$, and $z$ subset from ground-based instruments. In addition, we adopted a set of discrete features: the quality flags estimated for each of the above-mentioned bands (\verb|FLAG_<band>|), the detection quality flag (\verb|DET_QUALITY_FLAG|) and a flag telling if the source comes from a blended parent (\verb|DEBLENDED_FLAG|). In total these are $26$ input features.
RFC allows us to extract two types of information:
\begin{itemize}
    \item the importance ranking of each input feature, which determines the most relevant properties in classifying the sources as spurious or real, resulting in the ground bands $g$, $r$, $i$, and $z$ obtaining the highest ranking;
    \item the actual spurious probability (\verb|SPURIOUS_PROB|) for each object in the photometric catalogue. 
\end{itemize}
Given \verb|SPURIOUS_PROB|, we identify a source to be spurious by adopting a probablity threshold of $50 \%$. With such a threshold, we identify up to $90 \%$ of the true spurious sources in our test sample, while only $0.13 \%$ of real galaxies or stars are wrongly labelled as spurious. A more conservative threshold of $75 \%$ allows us to increase the purity (only $0.04 \%$ of real sources are lost), at the cost of a lower completeness of the spurious sample (only $80 \%$ of them are correctly identified and removed). We found $50 \%$ to be a good compromise between purity and completeness, and this represents our default choice.

\subsection{Bright star masking}
\begin{figure*}[h!]
\centering
\includegraphics[width=0.95\textwidth]{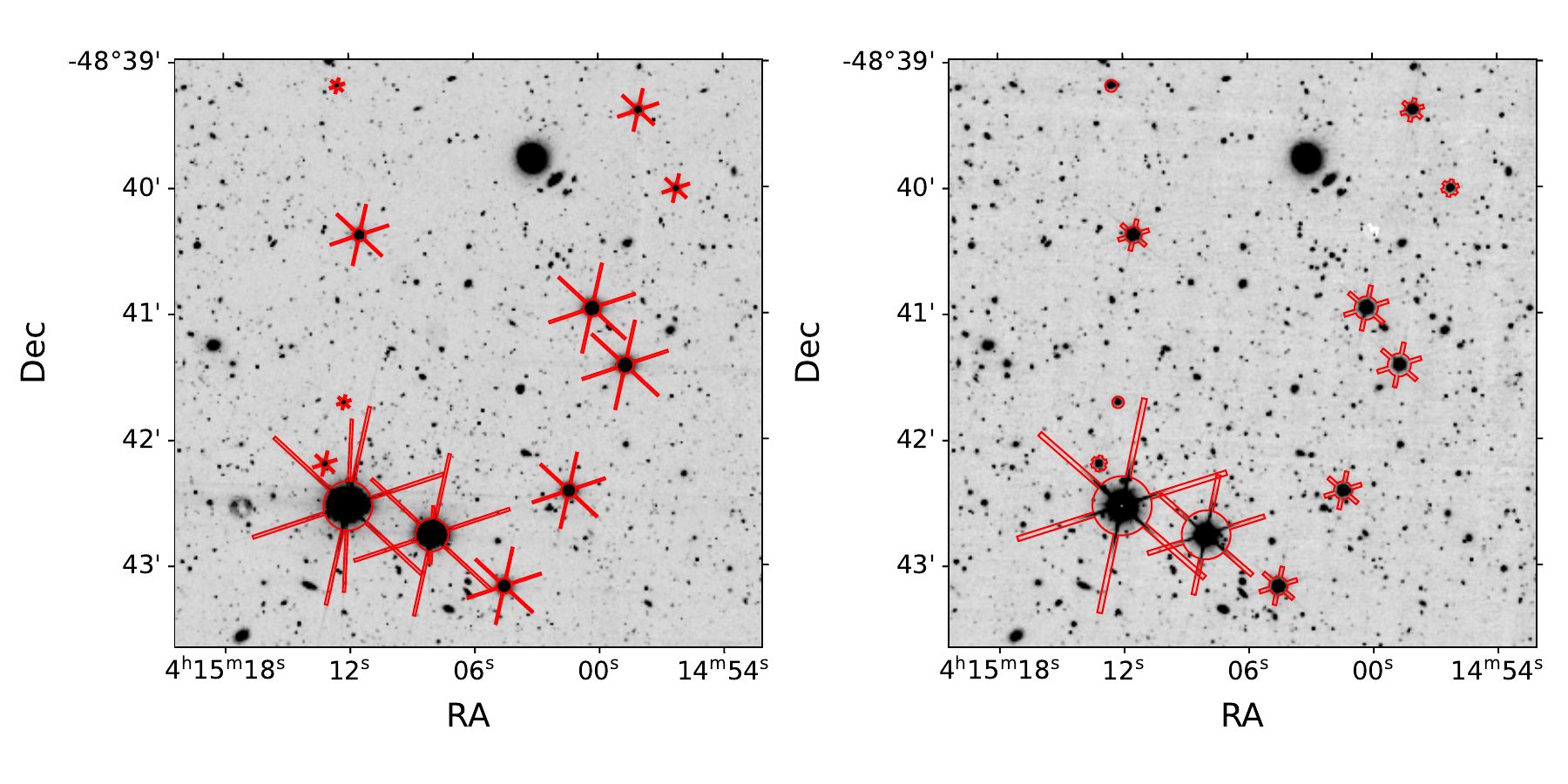}
\caption{Bright star masks (regions marked in red) in two small cut-outs of the $\text{\IE}$  (left) and $\text{\HE}$ (right) band image of tile 102021017. The central ring and the length of the diffraction and blooming features scale with brightness of the star.}
\label{fig:bs_masking}
\end{figure*}

As a survey satellite, \Euclid observes bright stars up to 8th magnitude brightness. The image of bright stars is characterised by a large and often saturated centre, together with diffraction spikes and possibly a blooming signature from the VIS detectors. These features are difficult to handle in the detection, deblending, and photometry. To mitigate this we map the contours of bright stars on the co-added VIS and NIR images as polygons.

Starting from positions provided by \textit{Gaia} we map the imprint of stars on the individual VIS and NIR calibrated images using simple geometric forms:
\begin{itemize}
    \item a central circle for the stellar disc;
    \item rectangles for the three diffraction spikes;
    \item a rectangle for the VIS blooming pattern.
\end{itemize}
The size of these simple forms scales with the brightness of the stars. For every \textit{Gaia} star we build the union of the simple forms as a polygon and then project the polygon from each calibrated image to the co-added mosaic. The final polygon for each \textit{Gaia} star is then set as the union of the polygons projected from the individual VIS and NIR calibrated images. As output we provide for all  co-added measurement images a list of the combined polygons for the \textit{Gaia} stars on that image  as a \texttt{JSON}\footnote{https://www.json.org/json-en.html} file.

Figure \ref{fig:bs_masking} shows with the red lines the bright star masks in two small cut-outs of the $\IE$ (left) and $\HE$ (right) band image of tile 102021017. The different colours denote the brightness bracket of star, which scales the size of the geometric forms constituting the masks. All objects within the bright star masks in $\IE$ or any of the bright star masks in $\YE$, $\JE$ or $\HE$ are marked in the column \verb|DET_QUALITY_FLAG| in two separate bits.

\subsection{Matching of \textit{Gaia} sources}\label{subsect:gaia_matching}

We automatically matched the \Euclid objects with the \textit{Gaia} DR3 sources \citep{GAIADR3}. To optimise the matching result, we propagated the positions of the \textit{Gaia} objects from the DR3 epoch J2016.0 to the reference time of the MER catalogue, which was averaged from the VIS and NIR detection image reference times. The maximum allowed distance of $0\overset{\prime\prime}{.}3$, and the ID of the matched \textit{Gaia} source, is stored in the catalogue column \verb|GAIA_ID|. The column \verb|GAIA_MATCH_QUALITY| stores the matching quality as the squared distance between the \textit{Gaia} and \Euclid positions. Depending on the Galactic latitude, the number of \textit{Gaia} matches is between $800$ and $3000$ objects in the final catalogues.

\section{Validation pipeline}\label{sect:validation}
The ultimate purpose of the validation is to assure that output data of the MER pipeline are within the specification and requirements and contribute to reach the science goals of the \Euclid project. Data validation helps also to identify processing issues or in general data quality problems. In the OU-MER validation we measure key properties of the MER outputs, such as object properties and image characteristics. If possible we compare and extrapolate the object properties with ground-truth information provided by the \textit{Gaia} satellite.

The OU-MER validation was run automatically as a separate pipeline directly after the processing pipeline.  The Q1 dataset is made up of three distinct fields, namely EDF-N, EDF-S, and EDF-F \citep{Q1-TP001}, covered by several tiles each. We aggregated the information of all the tiles covering a given field, in order to analyse the statistical trend of the parameters computed in the validation pipeline. This aggregated information is an essential tool to monitor and improve the MER processing. In the following sections, we present the most important validation results either for individual selected tiles or the aggregate numbers for a dataset such as EDF-S.

\subsection{Astrometry}\label{sect:astr_validation}
\begin{figure}[h!]
\centering
\includegraphics[width=0.5\textwidth]{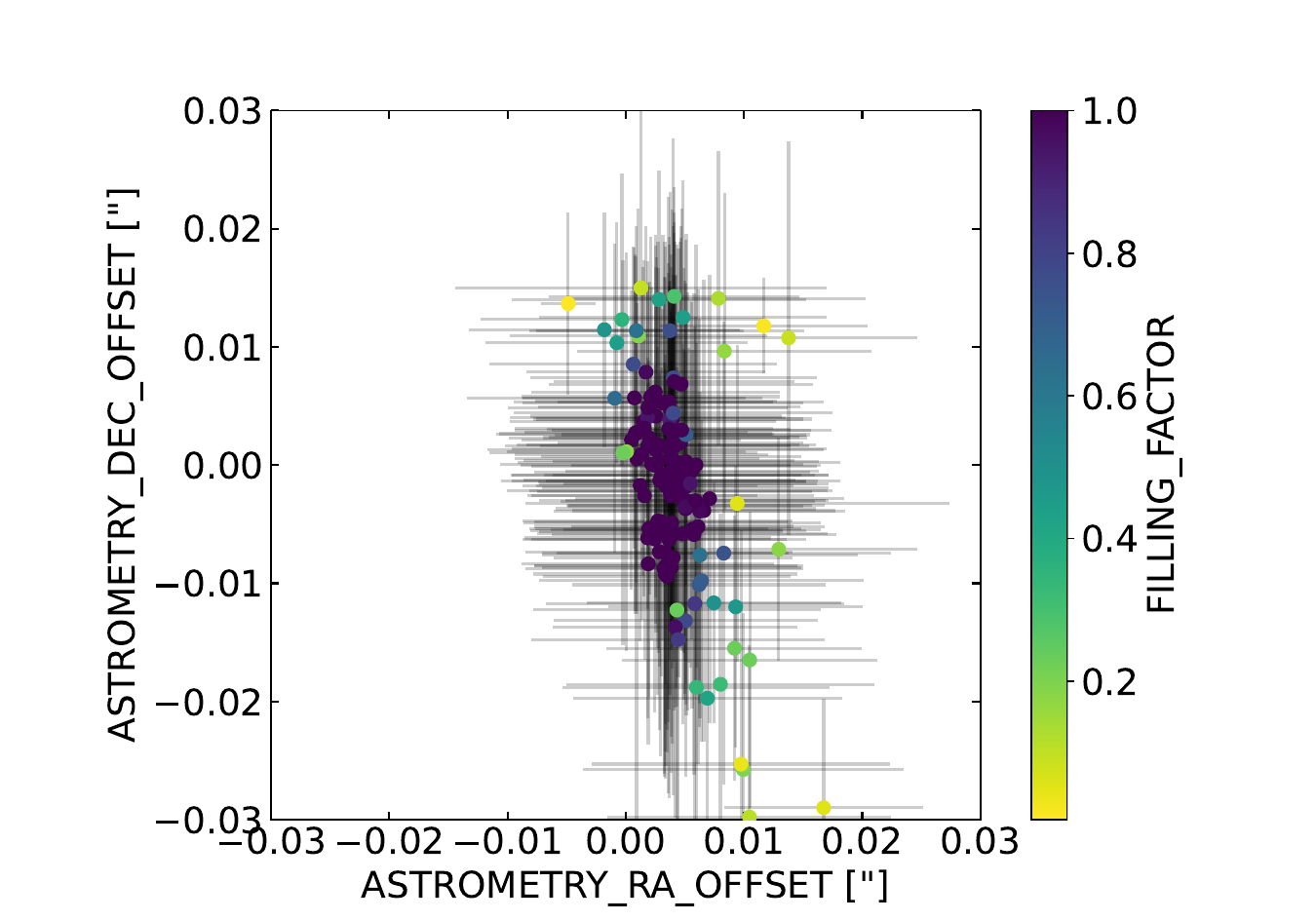}
\caption{Distribution of the astrometric offsets (median and NMAD) for all tiles in the EDF-S~field.}
\label{fig:q1_astro}
\end{figure}

As a validation of the positional accuracy of the objects in our final catalogue we measured, for each tile, the mean positional offset of the \textit{Gaia} objects in the catalogue with respect to their \textit{Gaia} reference position. Fig.~\ref{fig:q1_astro} shows the robust statistics (median plus NMAD) for these offsets as determined for all tiles of the EDF-S field. 
Tiles with a \verb|FILLING_FACTOR| (Sect. \ref{sect:detec_validation}) larger than $0.5$ have an offset smaller than $0."01$ in both right ascension and declination, which is the requirement for the absolute accuracy. The ensemble of these tiles in EDF-S has the mean value of $3.4$ and $-0.3\,\mathrm{mas}$ with a rms of $2.2$ and $5.6\,\mathrm{mas}$ in right ascension and declination, respectively. These offsets originate from systematic residuals in the VIS astrometry \citep{Q1-TP002}, which also explains the larger offsets for the tiles at the border of the field with a smaller \verb|FILLING_FACTOR|, which do not evenly sample these residuals.

\subsection{Image Depth}
\begin{figure}[h!]
\centering
\includegraphics[width=0.5\textwidth]{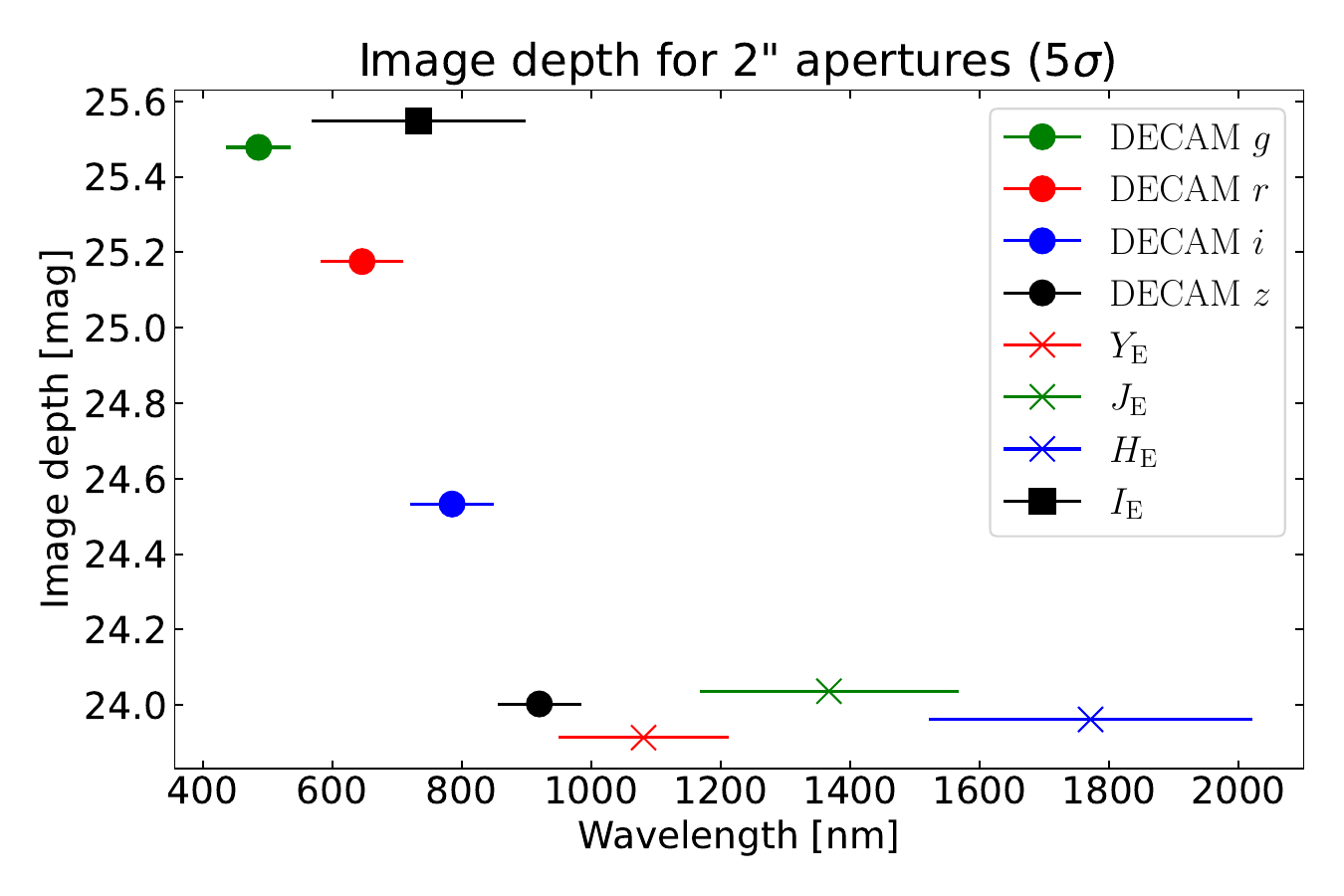}
\caption{$5\,\sigma$ depth depth measured in all imaging data of tile 102021017 using flux measurement in $2''$ apertures in the sky background.}
\label{fig:q1_depth}
\end{figure}

We determined the depth of the imaging data by measuring the flux in several thousand randomly selected circular apertures. We restricted the measurements to background areas by skipping over apertures that have non-zero values in the segmentation image or contain empty image areas. From the scatter of the flux values in these apertures, we computed the image depth in the corresponding filter.

As an example Fig. \ref{fig:q1_depth} shows the $5\,\sigma$ depth in the multi-band data of tile 102021017 measured in apertures of $2^{\prime\prime}$ diameter. We have chosen a $2^{\prime\prime}$ diameter since corresponds to the typical $2\,{\rm FHWM}$ diameter in the worst resolution band (see Sect. \ref{sect:photometry}). The depth estimates in other Q1 tiles are very similar to the ones shown in Fig.\ \ref{fig:q1_depth}, especially for the data observed with the \Euclid satellite $\IE$, $\YE$, $\JE$, and $\HE$.

Note that these depth estimates are done on aligned images that are resampled from their original pixel scale, which certainly has an effect on these measurements. Also the variable exposure time within fully covered images  and especially in the border tiles of Q1 may not fully be represented in a single depth value per band. Consequently we do not see these estimates as very accurate and absolute values. We use them as relative estimates for a set of tiles such as that released in Q1 to control the data quality.

\subsection{Detection}\label{sect:detec_validation}
\begin{figure*}[t!]
\centering
\includegraphics[width=1.0\textwidth]{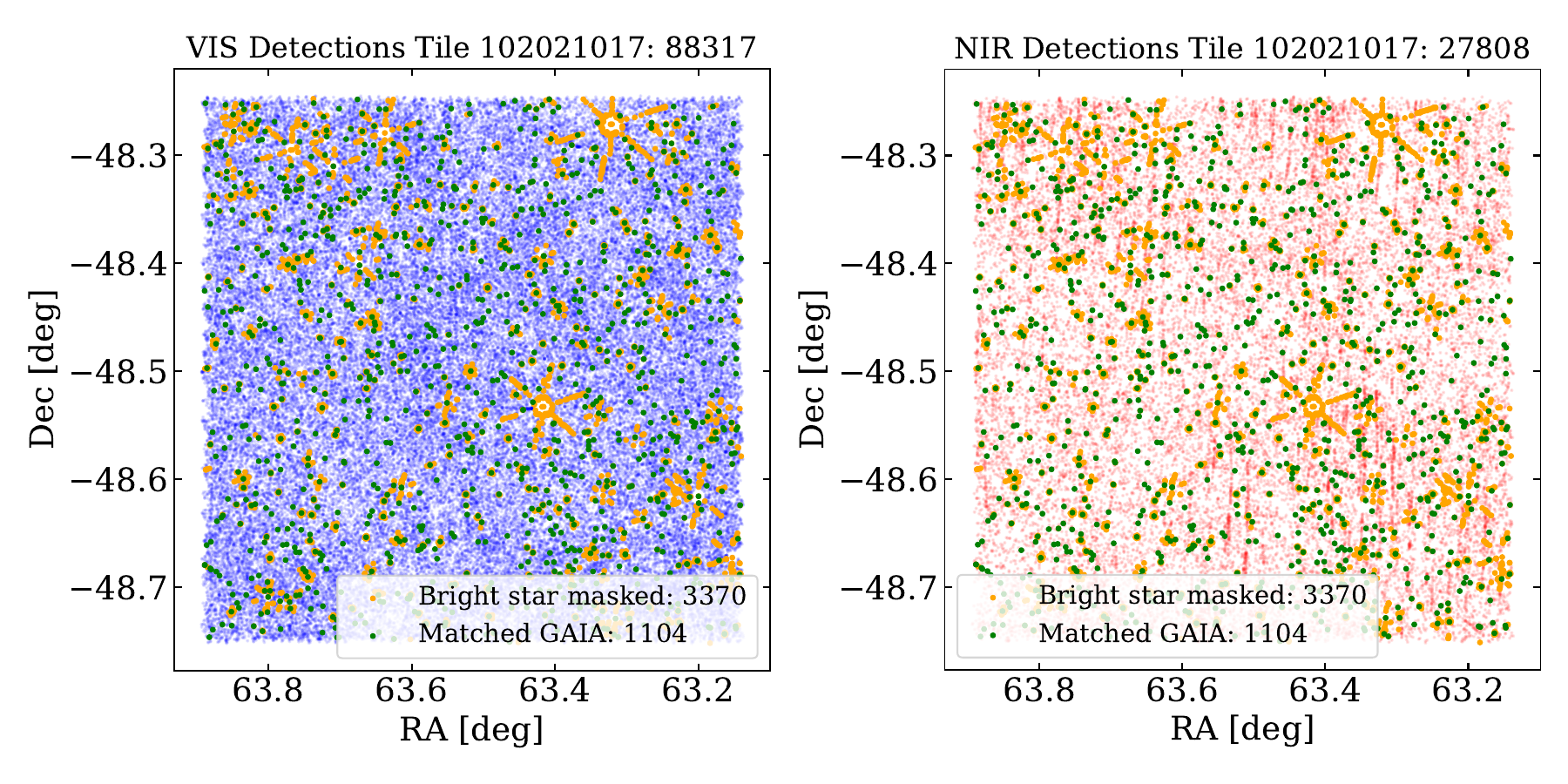}
\caption{Positions of all detected objects in tile 102021017. The colours mark the various object categories as indicated on the right.}
\label{fig:detection_validation}
\end{figure*}

In this part we collect for each tile all characteristic numbers associated with the object detection, cross-matching, and object marking. We compute the object density in the \texttt{HealPix} pixels of the tile core area and determine the \verb|FILLING_FACTOR| of the tile, which is the fraction of \texttt{HealPix} pixels containing at least one detected object. For interactive checks we illustrate these numbers in images such as Fig.\ \ref{fig:detection_validation}, which shows the positions of the VIS and NIR detect numbers for tile 102021017 in EDF-S on the left and right panel, respectively. The bright star masked objects as well as the objects identified from \textit{Gaia} are marked in both panels. The vertical ‘lines’ of NIR-detected objects in the right panel are caused by an insufficient masking of pixels affected by persistence \citep[see][]{Q1-TP003}. For an entire Q1 dataset the detection validation numbers help to identify border tiles with a small \verb|FILLING_FACTOR| or data quality issues such as shallow or incomplete coverage in a detection band when comparing the number of VIS- and NIR-detected objects.

\subsection{PSF}

The PSF propagation method described in Sect.~\ref{sect:PSF_calculation} is validated using \textit{Gaia} stars with
\verb|CLASSPROB_DSC_COMBMOD_STAR| > 0.99. In addition, a magnitude cut of
\verb|MAG_G_GAIA| > 18 is applied to avoid possible saturation effects (see \hyperlink{https://gea.esac.esa.int/archive/documentation/GDR3/Gaia_archive/chap_datamodel/sec_dm_main_source_catalogue/ssec_dm_gaia_source.html}{\textit{Gaia} DR3 Data Model} for details on the \textit{Gaia} source catalogue definitions). For Q1, 
depending on the field, this results on a typical sample of 300 to 600 \textit{Gaia} stars
per MER tile.

We extracted stamps in the MER mosaics around the \textit{Gaia} star positions corrected
for their proper motion. We used small stamps, $2\overset{\prime\prime}{.}1 \times 2\overset{\prime\prime}{.}1$ for
the \Euclid bands and $4\overset{\prime\prime}{.}5 \times 4\overset{\prime\prime}{.}5$ for the EXT mosaics, to minimise contamination from nearby sources. The star stamps were then fitted to a two-dimension
Gaussian function and the fit FWHM values were compared
with the propagated PSF stamps at the star positions. PSF stamps were fitted with exactly
the same method. Ideally one would expect a 1 to 1 correlation between the two
fitted values. Any systematic shift or large scatter in the correlation would imply
that either the propagation method is not correct or that the input PSFs are not
applicable to the input image data.

\begin{figure}[h!]
\centering
\includegraphics[width=0.47\textwidth]{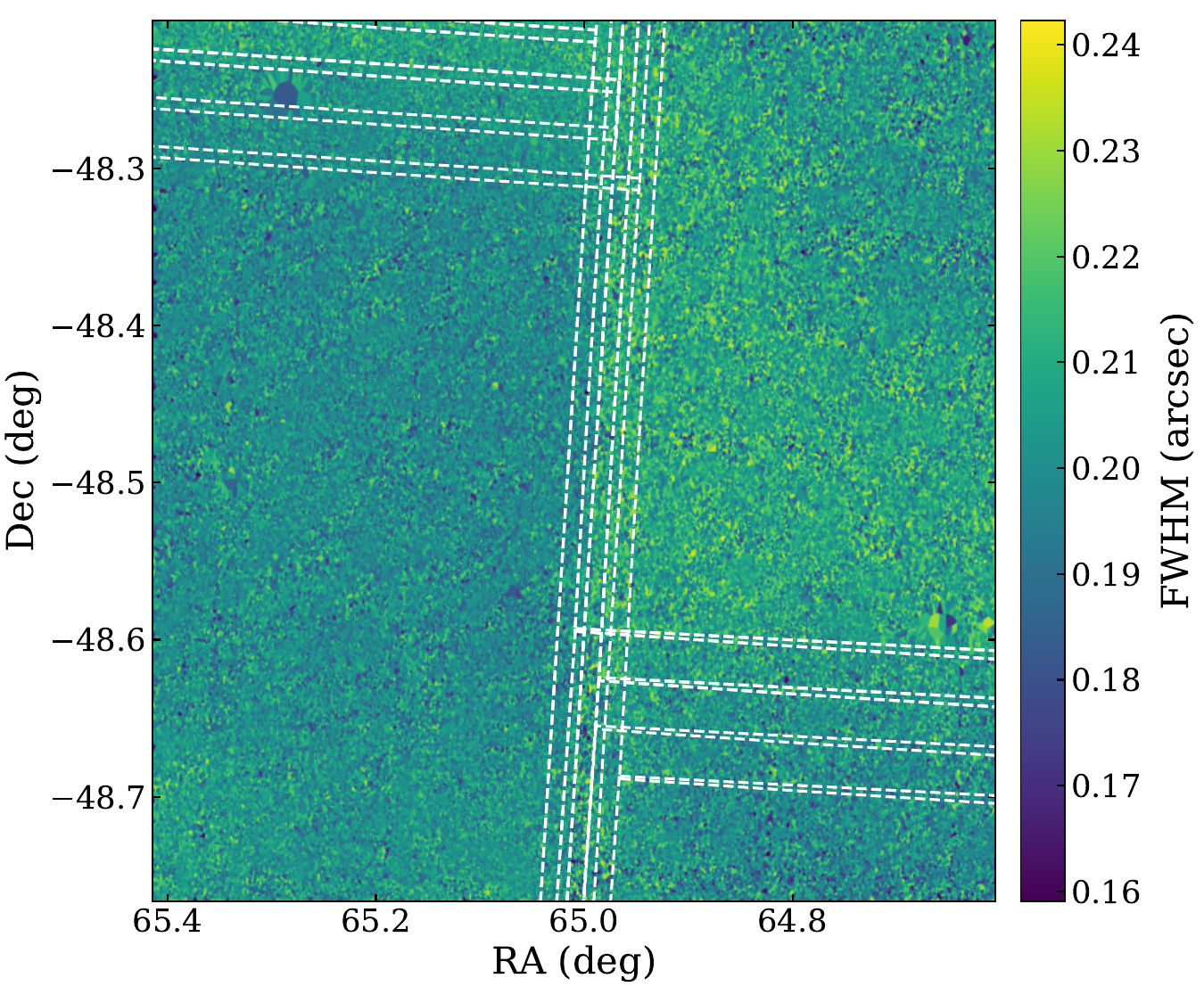}
\caption{Spatial variation in the MER catalogue-PSF for the VIS band on an example MER tile in the EDF-S field (tile index 102021019). FWHM values are measured fitting a two-dimensional Gaussian model to each propagated PSF stamp. Dashed white lines indicate the outer spatial footprint of the VIS calibrated frames that intersect the tile. Large-scale variations can be traced back to PSF changes along the VIS focal plane \citep[see][for more details]{Q1-TP002}.}
\label{fig:VIS PSF FWHM map}
\end{figure}

\begin{figure}[h!]
\centering
\includegraphics[width=0.45\textwidth]{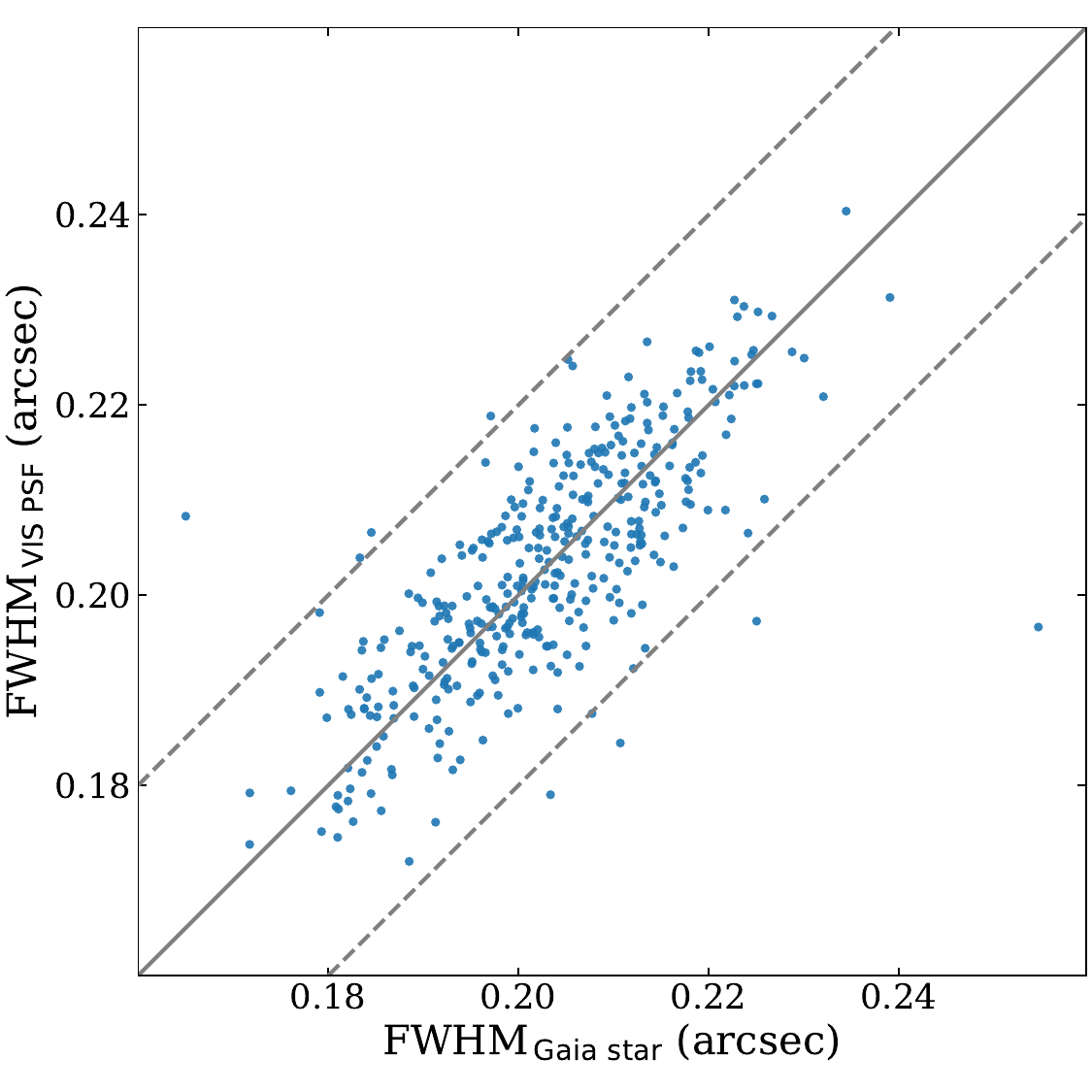}

\caption{Comparison between the \textit{Gaia} star profiles extracted from the VIS mosaic in tile 102021019 and the propagated VIS PSFs stamps at the \textit{Gaia} star positions. FWHM values are measured fitting a two-dimensional Gaussian model to the PSF stamps and the \textit{Gaia} star profiles. The identity line (continuous) and $\pm 2\%$ FWHM difference lines (dashed) show that both measurements agree within 2\% in most of the cases.}
\label{fig:vis psf vs gaia}
\end{figure}

Figure~\ref{fig:VIS PSF FWHM map} shows the propagated VIS PSF FWHM values for an example tile in the EDF-S
field. The FWHM could change by more than 20\%, depending on the source spatial
position. Large-scale variations can be attributed to PSF changes in the VIS
instrument field of view \citep{Q1-TP002}. Figure~\ref{fig:vis psf vs gaia} shows the comparison between the
propagated PSF stamps and the \textit{Gaia} stars extracted from the VIS mosaic. The
agreement is very good with minimal excursions from the $\pm2\%$ offset lines.
As discussed in Sect. \ref{sect:PSF_calculation}, the results for the NIR bands are not as good, but they
should improve in DR1 with newer NIR input PSFs. The agreement for the EXT
bands is generally also very good. In these cases the PSF spatial variations
can be attributed to varying seeing conditions in the images that were used
to produce the mosaics.

\subsection{Photometry}\label{subsect:photometry}

The photometric accuracy of the MER pipeline in each of the three Q1 datasets can be assessed by means of the magnitude offsets provided by the MER \textit{Gaia} validation. The offsets are computed for each band (namely \IE, \YE, \JE, and \HE along with the EXT-UNIONS filters for the case of EDF-N, and \IE, \YE, \JE, and \HE plus the EXT-DECAM filters for the case of EDF-N and EDF-F) and for two different photometric approaches (\verb|APHOT| and \verb|TPHOT|), taking into account the non-saturated sources with a magnitude $\IE < 19.5$ that have been detected by the MER pipeline and that have been successfully cross-matched against the \textit{Gaia} DR3 catalogue.

\begin{figure}[h!]
\centering
\includegraphics[width=0.50\textwidth]{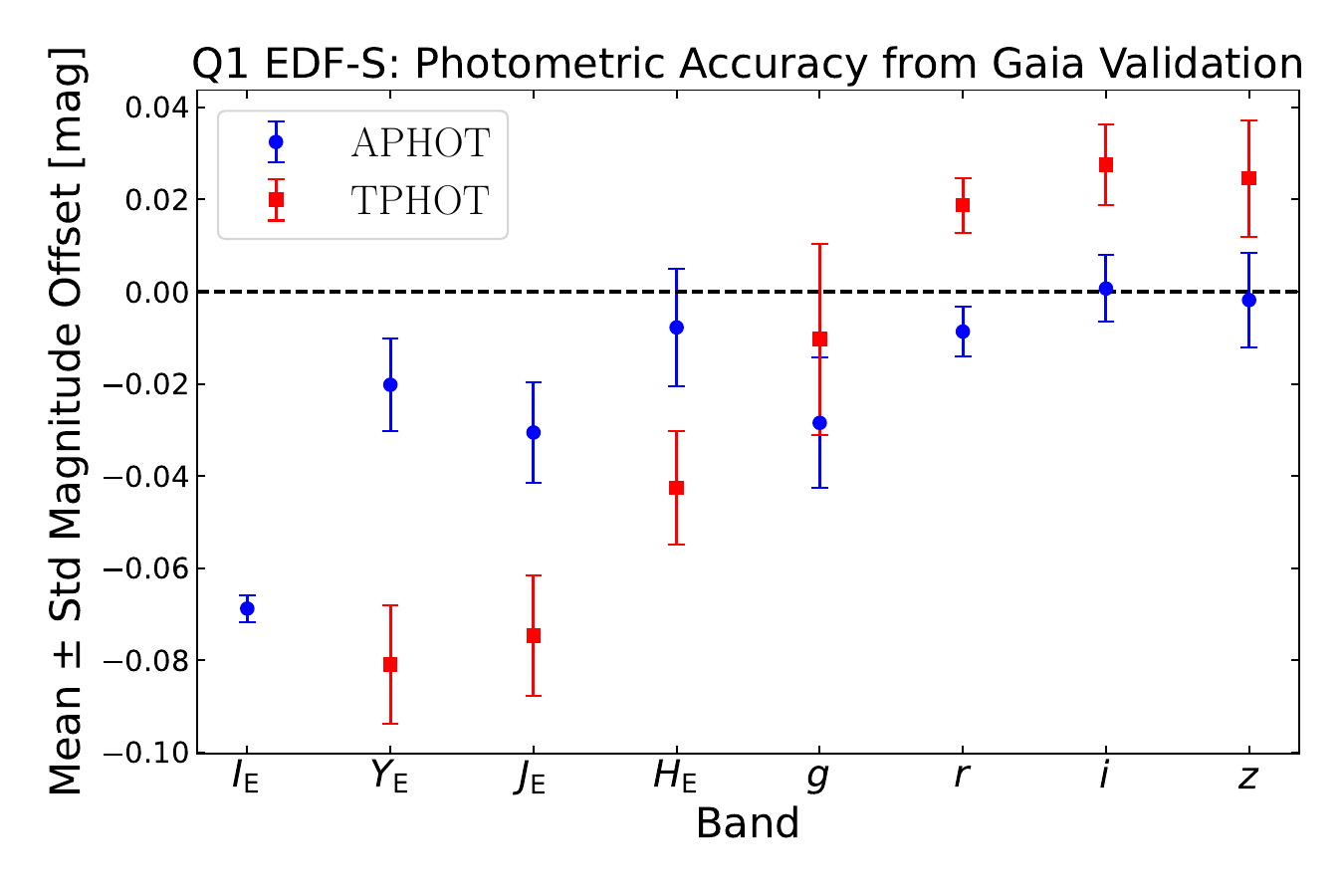}
\caption{Mean magnitude offsets for aperture (\texttt{A-PHOT}) and template fitting (\texttt{T-PHOT}) photometry in EDF-S derived from the results computed by the validation pipeline against cross-matched \textit{Gaia} sources for the \Euclid (\IE, \YE, \JE and \HE) and EXT-DECAM bands ($g$, $r$, $i$ and $z$). The error bars presented in this plot represent the standard deviations of the specific samples utilised to calculate each mean magnitude offset. This provides valuable insight into the inherent scatter within each one of these individual samples.}
\label{fig:photometry_validation_edfs}
\end{figure}

A photometric offset can be calculated for each source using a set of transformation functions generated to provide an expected analytic value for the magnitude difference between a given \Euclid band and the total $g$ magnitude in the \textit{Gaia} system.
Finally, the median magnitude offset of the all selected sources is computed, in order to have an aggregate quantity to describe the overall photometric accuracy per tile. For each of the three Q1 fields, this mean magnitude offset was computed after applying a $3 \, \sigma$-clipping over the per-tile values, a procedure that also allowed for the identification of outlier tiles where the specific magnitude offset value was not aligned with the global trend of the field (this typically being a direct consequence of major distorting elements that affected the normal execution of the MER pipeline). We find overall reasonable agreement, with offsets at the percent level varying from band to band. For instance, as it can be observed in Fig. \ref{fig:photometry_validation_edfs} for the case of EDF-S, we find mean magnitude offsets of $-0.0688\pm0.0029$ mag in \IE, $-0.0202\pm0.0100$ mag in \YE, $-0.0306\pm0.0109$ mag in \JE and $-0.0078\pm0.0128$ mag in \HE, respectively in what concerns aperture photometry (\verb|APHOT|), while, for the case of template fitting photometry (\verb|TPHOT|), we report mean magnitude offsets of $-0.0809\pm0.0128$ mag in \YE, $-0.0746\pm0.0131$ mag in \JE and $-0.0426\pm0.0124$ mag in \HE, respectively. For the case of the EXT-DECAM bands included in EDF-S, we find mean magnitude offsets of $-0.0285\pm0.0142$ mag in $g$, $-0.0086\pm0.0054$ mag in $r$, $0.0007\pm0.0072$ mag in $i$ and $-0.0018\pm0.0102$ mag in $z$, respectively in what concerns \verb|APHOT|, while for the case of \verb|TPHOT|, we report mean magnitude offsets of $-0.0103\pm0.0208$ mag in $g$, $0.0188\pm0.0059$ mag in $r$, $0.0275\pm0.0087$ mag in $i$ and $0.0246\pm0.0126$ mag in $z$, respectively. 
However, we point out that these offsets values have been continuously changing in the various phases of the observational campaign. Their causes  are under investigation, with difference in photometric techniques, PSF, and calibration related issues being the most probable culprits.
More detailed information, including data for other fields and bands, can be found in the Q1 release notes associated with this paper (see Sect \ref{subsect:release}).

It should be kept in mind that all of the validation involves only bright stars, which are not the main targets of the survey and therefore of the pipeline; checks against archival catalogues of galaxies have been performed, yielding reasonable agreement, but the uncertainties are too large to allow for a thorough evaluation of the accuracy. We have verified with random direct checks on the final mosaics that while for bright stars the aperture photometry estimate yields the best estimates, template fitting performs better for extended objects, not only helping to beat blending issues, but providing more accurate measurements in all cases for galaxies.

\subsection{Point-like probability}\label{subsect:pointlike}

\begin{figure}[h!]
\centering
\includegraphics[width=0.55\textwidth]{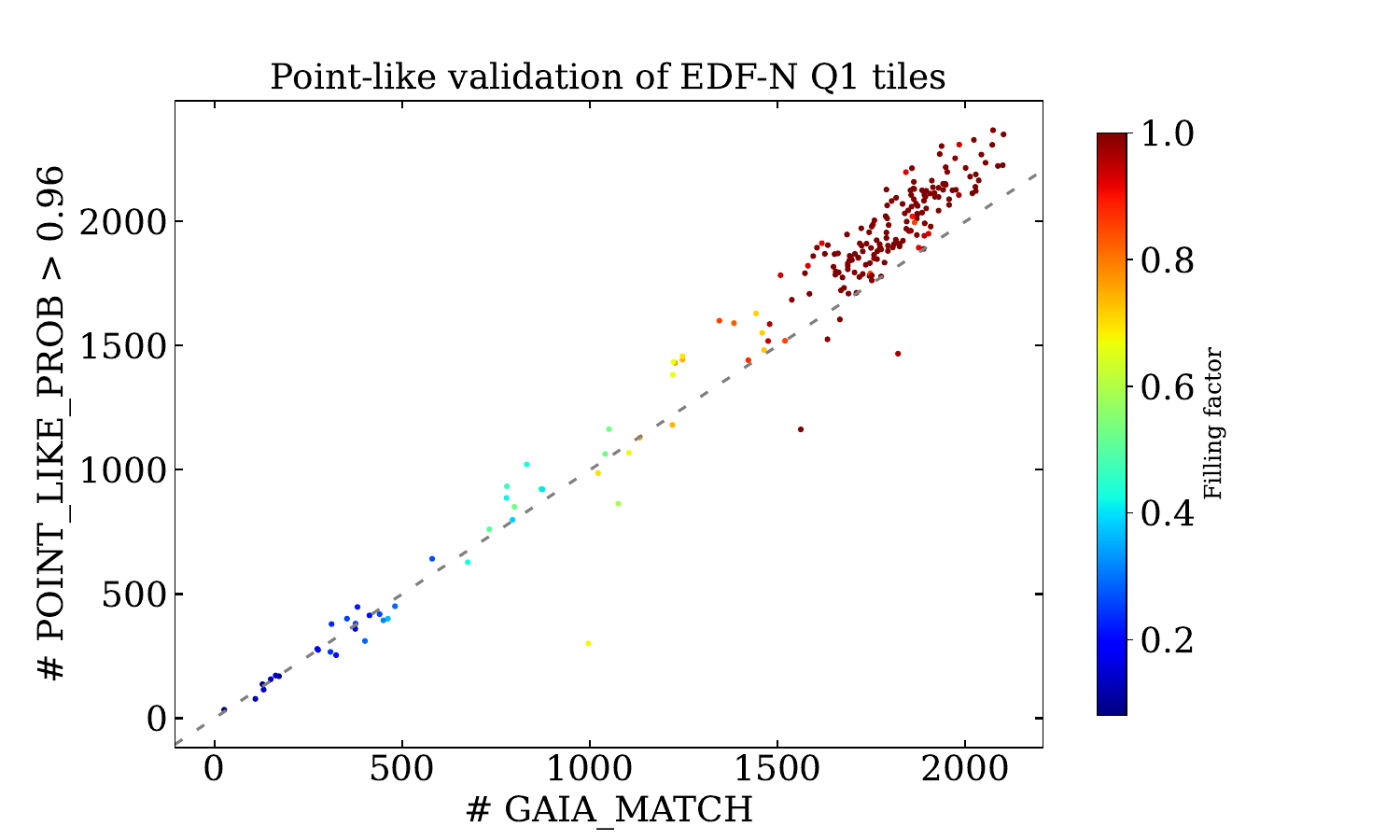}

\caption{Comparison for each EDF-N tile between the number of \textit{Gaia} objects found and the number of objects with a high probability to be point-like ($\texttt{POINT\_LIKE\_PROB}  > 0.96 $).}
\label{fig:point_like_prob validation}
\end{figure}

In order to identify possible errors in the identification of point-like objects, we compared for each tile the number of \textit{Gaia} matches with the number of objects with \verb|POINT_LIKE_PROB|~$>0.96$. As can be seen in Fig. \ref{fig:point_like_prob validation}, these two parameters are roughly equal. In EDF-N (the galactic latitude is $30\overset{\circ}{.}0$), they can be both small if the tile is not fully observed, or both around 2000 if the filling factor is close to 1. Any strong deviation from the dashed grey line ($y=x$) indicates a tile that needs further analysis (it is often a good indicator that some solar X ray burst has hit the VIS detectors).

\begin{figure}[h!]
\centering
\includegraphics[width=0.5\textwidth]{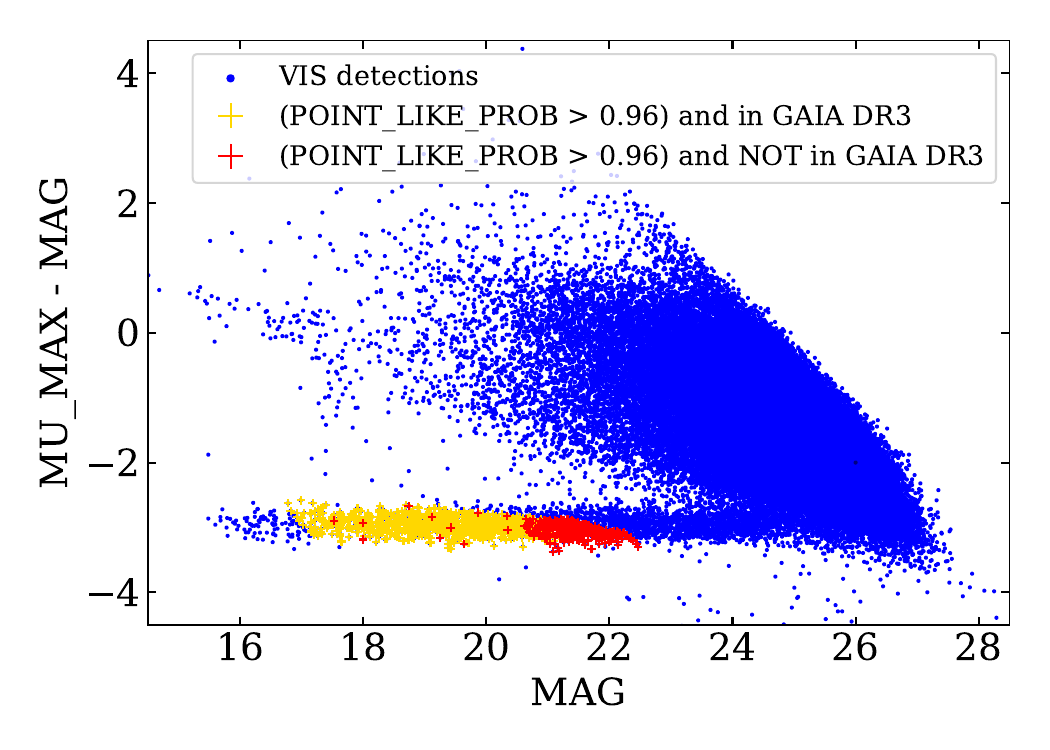}

\caption{Tile 102021017: VIS detections with a high probability to be point-like ($\texttt{POINT\_LIKE\_PROB}  > 0.96 $) identified in \textit{Gaia}. Most of them (in yellow) are found up to $\rm MAG = 21$. Red points (MER point-like detections not found in \textit{Gaia}) are mostly at fainter magnitudes, beyond the \textit{Gaia} detection limit.}
\label{fig:tile_valid_sg_sep_pdf}
\end{figure}

\label{subsect:gaia_matching}
At the tile level we validated \verb|POINT_LIKE_PROB| against \textit{Gaia} matches identified in the MER final catalogue (See Sect.\ \ref{subsect:gaia_matching}) as shown in Fig.\ \ref{fig:tile_valid_sg_sep_pdf} for tile 102021017. There are 1104 objects with \textit{Gaia} matches and 1070 with a high point-like probability (\verb|POINT_LIKE_PROB|~$>0.96$ ). From these 1070 there are 744 with and 326 without \textit{Gaia} matches. The unmatched objects are concentrated beyond the limiting magnitude of \textit{Gaia}. 

\subsection{Q1 release notes and known issues}\label{subsect:release}

Q1 data come with a list of caveats and known issues identified also thanks to the effort of EC scientists outside the perimeter of OU-MER and, in general, the SGS. An exhaustive description of how to deal with OU-MER products in the Q1 dataset can be found in the Q1 Explanatory Supplement\footnote{\url{https://euclid.esac.esa.int/dr/q1/expsup}}. 

\section{Conclusions and outlook}\label{sect:future}

In this paper we have presented a general overview of the OU-MER pipeline, with a particular focus on the work done for the data released as part of  Q1. The Q1 exercise has allowed us to identify some areas in which to focus our work towards future, larger data releases such as DR1.

The main goal of OU-MER is to provide photometric and morphological measurements to both the customer OUs in the \Euclid SGS (OU-PHZ, OU-SIR, OU-SHE) and the general scientific community. The main effort towards the upcoming Data Release 1 (DR1) is to improve the consistency of our photometry. The validation pipeline has shown that the current results for photometry are reasonably good, although with some room for improvement.
As part of the continuing pipeline development, we are currently conducting various tests to reach a full understanding of some issues that are outlined in the Q1 release notes (details in the Q1 Explanatory Supplement linked in Sect. \ref{subsect:release}).
The most important assessment on the quality of the OU-MER pipeline output products is going to come from the downstream units such as OU-PHZ, -SHE, and -LE3 when using larger quantities of \Euclid data in a cosmology analysis such as the one planned for DR1 and later releases.

\begin{acknowledgements}
\AckEC\\
\AckQone\\
Based on data from UNIONS, a scientific collaboration using
three Hawaii-based telescopes: CFHT, Pan-STARRS, and Subaru
\url{www.skysurvey.cc}\,.'' Correspondingly, please add ``Based on
data from the Dark Energy Camera (DECam) on the Blanco 4-m Telescope
at CTIO in Chile \url{https://www.darkenergysurvey.org}\,.''  if you
use MER data from the southern hemisphere.  If you use Gaia data, please
acknowledge: ``This work uses results from the ESA mission {\it Gaia},
whose data are being processed by the Gaia Data Processing and
Analysis Consortium \url{https://www.cosmos.esa.int/gaia}\
\end{acknowledgements}

\bibliographystyle{aa}
\bibliography{mer_v2}

\begin{thebibliography}{64}
\expandafter\ifx\csname natexlab\endcsname\relax\def\natexlab#1{#1}\fi

\bibitem[{{Abbott} {et~al.}(2021){Abbott}, {Adam{\'o}w}, {Aguena}, {Allam}, {Amon}, {Annis}, {Avila}, {Bacon}, {Banerji}, {Bechtol}, {Becker}, {Bernstein}, {Bertin}, {Bhargava}, {Bridle}, {Brooks}, {Burke}, {Carnero Rosell}, {Carrasco Kind}, {Carretero}, {Castander}, {Cawthon}, {Chang}, {Choi}, {Conselice}, {Costanzi}, {Crocce}, {da Costa}, {Davis}, {De Vicente}, {DeRose}, {Desai}, {Diehl}, {Dietrich}, {Drlica-Wagner}, {Eckert}, {Elvin-Poole}, {Everett}, {Evrard}, {Ferrero}, {Fert{\'e}}, {Flaugher}, {Fosalba}, {Friedel}, {Frieman}, {Garc{\'\i}a-Bellido}, {Gaztanaga}, {Gelman}, {Gerdes}, {Giannantonio}, {Gill}, {Gruen}, {Gruendl}, {Gschwend}, {Gutierrez}, {Hartley}, {Hinton}, {Hollowood}, {Honscheid}, {Huterer}, {James}, {Jeltema}, {Johnson}, {Kent}, {Kron}, {Kuehn}, {Kuropatkin}, {Lahav}, {Li}, {Lidman}, {Lin}, {MacCrann}, {Maia}, {Manning}, {Maloney}, {March}, {Marshall}, {Martini}, {Melchior}, {Menanteau}, {Miquel}, {Morgan}, {Myles}, {Neilsen}, {Ogando}, {Palmese}, {Paz-Chinch{\'o}n}, {Petravick},
  {Pieres}, {Plazas}, {Pond}, {Rodriguez-Monroy}, {Romer}, {Roodman}, {Rykoff}, {Sako}, {Sanchez}, {Santiago}, {Scarpine}, {Serrano}, {Sevilla-Noarbe}, {Smith}, {Smith}, {Soares-Santos}, {Suchyta}, {Swanson}, {Tarle}, {Thomas}, {To}, {Tremblay}, {Troxel}, {Tucker}, {Turner}, {Varga}, {Walker}, {Wechsler}, {Weller}, {Wester}, {Wilkinson}, {Yanny}, {Zhang}, {Nikutta}, {Fitzpatrick}, {Jacques}, {Scott}, {Olsen}, {Huang}, {Herrera}, {Juneau}, {Nidever}, {Weaver}, {Adean}, {Correia}, {de Freitas}, {Freitas}, {Singulani}, {Vila-Verde}, \& {Linea Science Server}}]{Abbott2021}
{Abbott}, T.~M.~C., {Adam{\'o}w}, M., {Aguena}, M., {et~al.} 2021, \apjs, 255, 20

\bibitem[{Aihara {et~al.}(2017)Aihara, Arimoto, Armstrong, Arnouts, Bahcall, Bickerton, Bosch, Bundy, Capak, Chan, Chiba, Coupon, Egami, Enoki, Finet, Fujimori, Fujimoto, Furusawa, Furusawa, Goto, Goulding, Greco, Greene, Gunn, Hamana, Harikane, Hashimoto, Hattori, Hayashi, Hayashi, Hełminiak, Higuchi, Hikage, Ho, Hsieh, Huang, Huang, Ikeda, Imanishi, Inoue, Iwasawa, Iwata, Jaelani, Jian, Kamata, Karoji, Kashikawa, Katayama, Kawanomoto, Kayo, Koda, Koike, Kojima, Komiyama, Konno, Koshida, Koyama, Kusakabe, Leauthaud, Lee, Lin, Lin, Lupton, Mandelbaum, Matsuoka, Medezinski, Mineo, Miyama, Miyatake, Miyazaki, Momose, More, More, Moritani, Moriya, Morokuma, Mukae, Murata, Murayama, Nagao, Nakata, Niida, Niikura, Nishizawa, Obuchi, Oguri, Oishi, Okabe, Okamoto, Okura, Ono, Onodera, Onoue, Osato, Ouchi, Price, Pyo, Sako, Sawicki, Shibuya, Shimasaku, Shimono, Shirasaki, Silverman, Simet, Speagle, Spergel, Strauss, Sugahara, Sugiyama, Suto, Suyu, Suzuki, Tait, Takada, Takata, Tamura, Tanaka, Tanaka, Tanaka, Tanaka,
  Terai, Terashima, Toba, Tominaga, Toshikawa, Turner, Uchida, Uchiyama, Umetsu, Uraguchi, Urata, Usuda, Utsumi, Wang, Wang, Wong, Yabe, Yamada, Yamanoi, Yasuda, Yeh, Yonehara, \& Yuma}]{Aihara2017}
Aihara, H., Arimoto, N., Armstrong, R., {et~al.} 2017, PASJ, 70

\bibitem[{{Bertin}(2011)}]{Bertin2011}
{Bertin}, E. 2011, in ASP Conf. Ser., Vol. 442, Astronomical Data Analysis Software and Systems XX, ed. I.~N. {Evans}, A.~{Accomazzi}, D.~J. {Mink}, \& A.~H. {Rots}, 435

\bibitem[{{Bertin} \& {Arnouts}(1996)}]{1996A&AS..117..393B}
{Bertin}, E. \& {Arnouts}, S. 1996, \aaps, 117, 393

\bibitem[{{Bertin} {et~al.}(2002){Bertin}, {Mellier}, {Radovich}, {Missonnier}, {Didelon}, \& {Morin}}]{2002ASPC..281..228B}
{Bertin}, E., {Mellier}, Y., {Radovich}, M., {et~al.} 2002, in ASP Conf. Ser., Vol. 281, Astronomical Data Analysis Software and Systems XI, ed. D.~A. {Bohlender}, D.~{Durand}, \& T.~H. {Handley}, 228

\bibitem[{{Bertin} {et~al.}(2020){Bertin}, {Schefer}, {Apostolakos}, {{\'A}lvarez-Ayll{\'o}n}, {Dubath}, \& {K{\"u}mmel}}]{2020ASPC..527..461B}
{Bertin}, E., {Schefer}, M., {Apostolakos}, N., {et~al.} 2020, in ASP Conf. Ser., Vol. 527, Astronomical Data Analysis Software and Systems XXIX, ed. R.~{Pizzo}, E.~R. {Deul}, J.~D. {Mol}, J.~{de Plaa}, \& H.~{Verkouter}, 461

\bibitem[{{Boucaud} {et~al.}(2016){Boucaud}, {Bocchio}, {Abergel}, {Orieux}, {Dole}, \& {Hadj-Youcef}}]{Boucaud2016}
{Boucaud}, A., {Bocchio}, M., {Abergel}, A., {et~al.} 2016, \aap, 596, A63

\bibitem[{{Breiman}(2001)}]{Breiman2001}
{Breiman}, L. 2001, Machine Learning, 45, 5

\bibitem[{{Conselice}(2014)}]{Conselice2014}
{Conselice}, C.~J. 2014, \araa, 52, 291

\bibitem[{Desai {et~al.}(2012)Desai, Armstrong, Mohr, Semler, Liu, Bertin, Allam, Barkhouse, Bazin, Buckley-Geer, Cooper, Hansen, High, Lin, Lin, Ngeow, Rest, Song, Tucker, \& Zenteno}]{Desai2012}
Desai, S., Armstrong, R., Mohr, J.~J., {et~al.} 2012, ApJ, 757, 83

\bibitem[{Ester {et~al.}(1996)Ester, Kriegel, Sander, Xu, Simoudis, Han, \& Fayyad}]{ester1996proc}
Ester, M., Kriegel, H., Sander, J., {et~al.} 1996, Proc. 2nd Int. Conf. on Knowledge Discovery and Data Mining (KDD’96)

\bibitem[{{Estrada} {et~al.}(2023){Estrada}, {Mercurio}, {Vulcani}, {Rodighiero}, {Nonino}, {Annunziatella}, {Rosati}, {Grillo}, {Caminha}, {Angora}, {Biviano}, {Brescia}, {De Lucia}, {Demarco}, {Girardi}, {Gobat}, \& {Lemaux}}]{Estrada2023}
{Estrada}, N., {Mercurio}, A., {Vulcani}, B., {et~al.} 2023, \aap, 671, A146

\bibitem[{{Euclid Collaboration: Aussel} {et~al.}(2025){Euclid Collaboration: Aussel}, {Tereno}, {Schirmer}, {et~al.}}]{Q1-TP001}
{Euclid Collaboration: Aussel}, H., {Tereno}, I., {Schirmer}, M., {et~al.} 2025, A\&A, submitted (Euclid Q1 SI), arXiv:2503.15302

\bibitem[{{Euclid Collaboration: Bretonni{\`e}re} {et~al.}(2022){Euclid Collaboration: Bretonni{\`e}re}, {Huertas-Company}, {Boucaud}, {et~al.}}]{Bretonniere-EP13}
{Euclid Collaboration: Bretonni{\`e}re}, H., {Huertas-Company}, M., {Boucaud}, A., {et~al.} 2022, \aap, 657, A90

\bibitem[{{Euclid Collaboration: Bretonni{\`e}re} {et~al.}(2023){Euclid Collaboration: Bretonni{\`e}re}, {Kuchner}, {Huertas-Company}, {et~al.}}]{Bretonniere-EP26}
{Euclid Collaboration: Bretonni{\`e}re}, H., {Kuchner}, U., {Huertas-Company}, M., {et~al.} 2023, \aap, 671, A102

\bibitem[{{Euclid Collaboration: Copin} {et~al.}(2025){Euclid Collaboration: Copin}, {Fumana}, {Mancini}, {et~al.}}]{Q1-TP006}
{Euclid Collaboration: Copin}, Y., {Fumana}, M., {Mancini}, C., {et~al.} 2025, A\&A, submitted (Euclid Q1 SI), arXiv:2503.15307

\bibitem[{{Euclid Collaboration: Cropper} {et~al.}(2025){Euclid Collaboration: Cropper}, {Al-Bahlawan}, {Amiaux}, {et~al.}}]{EuclidSkyVIS}
{Euclid Collaboration: Cropper}, M., {Al-Bahlawan}, A., {Amiaux}, J., {et~al.} 2025, A\&A, 697, A2

\bibitem[{{Euclid Collaboration: Jahnke} {et~al.}(2025){Euclid Collaboration: Jahnke}, {Gillard}, {Schirmer}, {et~al.}}]{EuclidSkyNISP}
{Euclid Collaboration: Jahnke}, K., {Gillard}, W., {Schirmer}, M., {et~al.} 2025, A\&A, 697, A3

\bibitem[{{Euclid Collaboration: McCracken} {et~al.}(2025){Euclid Collaboration: McCracken}, {Benson}, {Dolding}, {et~al.}}]{Q1-TP002}
{Euclid Collaboration: McCracken}, H.~J., {Benson}, K., {Dolding}, C., {et~al.} 2025, A\&A, submitted (Euclid Q1 SI), arXiv:2503.15303

\bibitem[{{Euclid Collaboration: Mellier} {et~al.}(2025){Euclid Collaboration: Mellier}, {Abdurro'uf}, {Acevedo~Barroso}, {et~al.}}]{EuclidSkyOverview}
{Euclid Collaboration: Mellier}, Y., {Abdurro'uf}, {Acevedo~Barroso}, J., {et~al.} 2025, A\&A, 697, A1

\bibitem[{{Euclid Collaboration: Polenta} {et~al.}(2025){Euclid Collaboration: Polenta}, {Frailis}, {Alavi}, {et~al.}}]{Q1-TP003}
{Euclid Collaboration: Polenta}, G., {Frailis}, M., {Alavi}, A., {et~al.} 2025, A\&A, submitted (Euclid Q1 SI), arXiv:2503.15304

\bibitem[{{Euclid Collaboration: Quilley} {et~al.}(2025){Euclid Collaboration: Quilley}, {Damjanov}, {de Lapparent}, {et~al.}}]{Q1-SP040}
{Euclid Collaboration: Quilley}, L., {Damjanov}, I., {de Lapparent}, V., {et~al.} 2025, A\&A, submitted (Euclid Q1 SI), arXiv:2503.15309

\bibitem[{{Euclid Collaboration: Scaramella} {et~al.}(2022){Euclid Collaboration: Scaramella}, {Amiaux}, {Mellier}, {et~al.}}]{Scaramella-EP1}
{Euclid Collaboration: Scaramella}, R., {Amiaux}, J., {Mellier}, Y., {et~al.} 2022, \aap, 662, A112

\bibitem[{{Euclid Collaboration: Serrano} {et~al.}(2024){Euclid Collaboration: Serrano}, {Hudelot}, {Seidel}, {et~al.}}]{EP-Serrano}
{Euclid Collaboration: Serrano}, S., {Hudelot}, P., {Seidel}, G., {et~al.} 2024, \aap, 690, A103

\bibitem[{{Euclid Collaboration: Tucci} {et~al.}(2025){Euclid Collaboration: Tucci}, {Paltani}, {Hartley}, {et~al.}}]{Q1-TP005}
{Euclid Collaboration: Tucci}, M., {Paltani}, S., {Hartley}, W.~G., {et~al.} 2025, A\&A, accepted (Euclid Q1 SI), arXiv:2503.15306

\bibitem[{{Euclid Collaboration: Walmsley} {et~al.}(2025{\natexlab{a}}){Euclid Collaboration: Walmsley}, {Holloway}, {Lines}, {et~al.}}]{Q1-SP048}
{Euclid Collaboration: Walmsley}, M., {Holloway}, P., {Lines}, N.~E.~P., {et~al.} 2025{\natexlab{a}}, A\&A, submitted (Euclid Q1 SI), arXiv:2503.15324

\bibitem[{{Euclid Collaboration: Walmsley} {et~al.}(2025{\natexlab{b}}){Euclid Collaboration: Walmsley}, {Huertas-Company}, {Quilley}, {et~al.}}]{Q1-SP047}
{Euclid Collaboration: Walmsley}, M., {Huertas-Company}, M., {Quilley}, L., {et~al.} 2025{\natexlab{b}}, A\&A, submitted (Euclid Q1 SI), arXiv:2503.15310

\bibitem[{{Euclid Quick Release Q1}(2025)}]{Q1cite}
{Euclid Quick Release Q1}. 2025, \url{https://doi.org/10.57780/esa-2853f3b}

\bibitem[{{Gaia Collaboration} {et~al.}(2023){Gaia Collaboration}, {Vallenari}, {Brown}, {Prusti}, {de Bruijne}, {Arenou}, {Babusiaux}, {Biermann}, {Creevey}, {Ducourant}, {Evans}, {Eyer}, {Guerra}, {Hutton}, {Jordi}, {Klioner}, {Lammers}, {Lindegren}, {Luri}, {Mignard}, {Panem}, {Pourbaix}, {Randich}, {Sartoretti}, {Soubiran}, {Tanga}, {Walton}, {Bailer-Jones}, {Bastian}, {Drimmel}, {Jansen}, {Katz}, {Lattanzi}, {van Leeuwen}, {Bakker}, {Cacciari}, {Casta{\~n}eda}, {De Angeli}, {Fabricius}, {Fouesneau}, {Fr{\'e}mat}, {Galluccio}, {Guerrier}, {Heiter}, {Masana}, {Messineo}, {Mowlavi}, {Nicolas}, {Nienartowicz}, {Pailler}, {Panuzzo}, {Riclet}, {Roux}, {Seabroke}, {Sordo}, {Th{\'e}venin}, {Gracia-Abril}, {Portell}, {Teyssier}, {Altmann}, {Andrae}, {Audard}, {Bellas-Velidis}, {Benson}, {Berthier}, {Blomme}, {Burgess}, {Busonero}, {Busso}, {C{\'a}novas}, {Carry}, {Cellino}, {Cheek}, {Clementini}, {Damerdji}, {Davidson}, {de Teodoro}, {Nu{\~n}ez Campos}, {Delchambre}, {Dell'Oro}, {Esquej},
  {Fern{\'a}ndez-Hern{\'a}ndez}, {Fraile}, {Garabato}, {Garc{\'\i}a-Lario}, {Gosset}, {Haigron}, {Halbwachs}, {Hambly}, {Harrison}, {Hern{\'a}ndez}, {Hestroffer}, {Hodgkin}, {Holl}, {Jan{\ss}en}, {Jevardat de Fombelle}, {Jordan}, {Krone-Martins}, {Lanzafame}, {L{\"o}ffler}, {Marchal}, {Marrese}, {Moitinho}, {Muinonen}, {Osborne}, {Pancino}, {Pauwels}, {Recio-Blanco}, {Reyl{\'e}}, {Riello}, {Rimoldini}, {Roegiers}, {Rybizki}, {Sarro}, {Siopis}, {Smith}, {Sozzetti}, {Utrilla}, {van Leeuwen}, {Abbas}, {{\'A}brah{\'a}m}, {Abreu Aramburu}, {Aerts}, {Aguado}, {Ajaj}, {Aldea-Montero}, {Altavilla}, {{\'A}lvarez}, {Alves}, {Anders}, {Anderson}, {Anglada Varela}, {Antoja}, {Baines}, {Baker}, {Balaguer-N{\'u}{\~n}ez}, {Balbinot}, {Balog}, {Barache}, {Barbato}, {Barros}, {Barstow}, {Bartolom{\'e}}, {Bassilana}, {Bauchet}, {Becciani}, {Bellazzini}, {Berihuete}, {Bernet}, {Bertone}, {Bianchi}, {Binnenfeld}, {Blanco-Cuaresma}, {Blazere}, {Boch}, {Bombrun}, {Bossini}, {Bouquillon}, {Bragaglia}, {Bramante}, {Breedt},
  {Bressan}, {Brouillet}, {Brugaletta}, {Bucciarelli}, {Burlacu}, {Butkevich}, {Buzzi}, {Caffau}, {Cancelliere}, {Cantat-Gaudin}, {Carballo}, {Carlucci}, {Carnerero}, {Carrasco}, {Casamiquela}, {Castellani}, {Castro-Ginard}, {Chaoul}, {Charlot}, {Chemin}, {Chiaramida}, {Chiavassa}, {Chornay}, {Comoretto}, {Contursi}, {Cooper}, {Cornez}, {Cowell}, {Crifo}, {Cropper}, {Crosta}, {Crowley}, {Dafonte}, {Dapergolas}, {David}, {David}, {de Laverny}, {De Luise}, \& {De March}}]{GAIADR3}
{Gaia Collaboration}, {Vallenari}, A., {Brown}, A.~G.~A., {et~al.} 2023, \aap, 674, A1

\bibitem[{{G{\'o}rski} {et~al.}(2005){G{\'o}rski}, {Hivon}, {Banday}, {Wandelt}, {Hansen}, {Reinecke}, \& {Bartelmann}}]{2005ApJ...622..759G}
{G{\'o}rski}, K.~M., {Hivon}, E., {Banday}, A.~J., {et~al.} 2005, \apj, 622, 759

\bibitem[{Hinneburg \& Gabriel(2007)}]{hinneburg2007denclue}
Hinneburg, A. \& Gabriel, H.-H. 2007, in International symposium on intelligent data analysis, Springer, 70--80

\bibitem[{Hinneburg {et~al.}(1998)Hinneburg, Keim, {et~al.}}]{hinneburg1998efficient}
Hinneburg, A., Keim, D.~A., {et~al.} 1998, An efficient approach to clustering in large multimedia databases with noise, Vol.~98 (Bibliothek der Universit{\"a}t Konstanz Konstanz, Germany)

\bibitem[{{Ibata} {et~al.}(2017){Ibata}, {McConnachie}, {Cuillandre}, {Fantin}, {Haywood}, {Martin}, {Bergeron}, {Beckmann}, {Bernard}, {Bonifacio}, {Caffau}, {Carlberg}, {C{\^o}t{\'e}}, {Cabanac}, {Chapman}, {Duc}, {Durret}, {Famaey}, {Fabbro}, {Gwyn}, {Hammer}, {Hill}, {Hudson}, {Lan{\c{c}}on}, {Lewis}, {Malhan}, {di Matteo}, {McCracken}, {Mei}, {Mellier}, {Navarro}, {Pires}, {Pritchet}, {Reyl{\'e}}, {Richer}, {Robin}, {S{\'a}nchez-Janssen}, {Sawicki}, {Scott}, {Scottez}, {Spekkens}, {Starkenburg}, {Thomas}, \& {Venn}}]{Ibata2017}
{Ibata}, R.~A., {McConnachie}, A., {Cuillandre}, J.-C., {et~al.} 2017, \apj, 848, 128

\bibitem[{Ivezić {et~al.}(2019)Ivezić, Kahn, Tyson, Abel, Acosta, Allsman, Alonso, AlSayyad, Anderson, Andrew, P.~Angel, Angeli, Ansari, Antilogus, Araujo, Armstrong, Arndt, Astier, Aubourg, Auza, Axelrod, Bard, Barr, Barrau, Bartlett, Bauer, Bauman, Baumont, Bechtol, Bechtol, Becker, Becla, Beldica, Bellavia, Bianco, Biswas, Blanc, Blazek, Blandford, Bloom, Bogart, Bond, Booth, Borgland, Borne, Bosch, Boutigny, Brackett, Bradshaw, Brandt, Brown, Bullock, Burchat, Burke, Cagnoli, Calabrese, Callahan, Callen, Carlin, Carlson, Chandrasekharan, Charles-Emerson, Chesley, Cheu, Chiang, Chiang, Chirino, Chow, Ciardi, Claver, Cohen-Tanugi, Cockrum, Coles, Connolly, Cook, Cooray, Covey, Cribbs, Cui, Cutri, Daly, Daniel, Daruich, Daubard, Daues, Dawson, Delgado, Dellapenna, Peyster, Val-Borro, Digel, Doherty, Dubois, Dubois-Felsmann, Durech, Economou, Eifler, Eracleous, Emmons, Neto, Ferguson, Figueroa, Fisher-Levine, Focke, Foss, Frank, Freemon, Gangler, Gawiser, Geary, Gee, Geha, Gessner, Gibson, Gilmore,
  Glanzman, Glick, Goldina, Goldstein, Goodenow, Graham, Gressler, Gris, Guy, Guyonnet, Haller, Harris, Hascall, Haupt, Hernandez, Herrmann, Hileman, Hoblitt, Hodgson, Hogan, Howard, Huang, Huffer, Ingraham, Innes, Jacoby, Jain, Jammes, Jee, Jenness, Jernigan, Jevremović, Johns, Johnson, Johnson, Jones, Juramy-Gilles, Jurić, Kalirai, Kallivayalil, Kalmbach, Kantor, Karst, Kasliwal, Kelly, Kessler, Kinnison, Kirkby, Knox, Kotov, Krabbendam, Krughoff, Kubánek, Kuczewski, Kulkarni, Ku, Kurita, Lage, Lambert, Lange, Langton, Guillou, Levine, Liang, Lim, Lintott, Long, Lopez, Lotz, Lupton, Lust, MacArthur, Mahabal, Mandelbaum, Markiewicz, Marsh, Marshall, Marshall, May, McKercher, McQueen, Meyers, Migliore, Miller, Mills, Miraval, Moeyens, Moolekamp, Monet, Moniez, Monkewitz, Montgomery, Morrison, Mueller, Muller, Arancibia, Neill, Newbry, Nief, Nomerotski, Nordby, O’Connor, Oliver, Olivier, Olsen, O’Mullane, Ortiz, Osier, Owen, Pain, Palecek, Parejko, Parsons, Pease, Peterson, Peterson, Petravick, Petrick,
  Petry, Pierfederici, Pietrowicz, Pike, Pinto, Plante, Plate, Plutchak, Price, Prouza, Radeka, Rajagopal, Rasmussen, Regnault, Reil, Reiss, Reuter, Ridgway, Riot, Ritz, Robinson, Roby, Roodman, Rosing, Roucelle, Rumore, Russo, Saha, Sassolas, Schalk, Schellart, Schindler, Schmidt, Schneider, Schneider, Schoening, Schumacher, Schwamb, Sebag, Selvy, Sembroski, Seppala, Serio, Serrano, Shaw, Shipsey, Sick, Silvestri, Slater, Smith, Smith, Sobhani, Soldahl, Storrie-Lombardi, Stover, Strauss, Street, Stubbs, Sullivan, Sweeney, Swinbank, Szalay, Takacs, Tether, Thaler, Thayer, Thomas, Thornton, Thukral, Tice, Trilling, Turri, Berg, Berk, Vetter, Virieux, Vucina, Wahl, Walkowicz, Walsh, Walter, Wang, Wang, Warner, Wiecha, Willman, Winters, Wittman, Wolff, Wood-Vasey, Wu, Xin, Yoachim, \& Zhan}]{Ivezic2019}
Ivezić, {\v{Z}}., Kahn, S.~M., Tyson, J.~A., {et~al.} 2019, \apj, 873, 111

\bibitem[{Jauzac {et~al.}(2012)Jauzac, Jullo, Kneib, Ebeling, Leauthaud, Ma, Limousin, Massey, \& Richard}]{Jauzac2012}
Jauzac, M., Jullo, E., Kneib, J.-P., {et~al.} 2012, MNRAS, 426, 3369

\bibitem[{{Kron}(1980)}]{Kron1980}
{Kron}, R.~G. 1980, ApJ, 43, 305

\bibitem[{{K{\"u}mmel} {et~al.}(2022{\natexlab{a}}){K{\"u}mmel}, {{\'A}lvarez-Ayll{\'o}n}, {Bertin}, {Dubath}, {Gavazzi}, {Hartley}, \& {Schefer}}]{2022arXiv221202428K}
{K{\"u}mmel}, M., {{\'A}lvarez-Ayll{\'o}n}, A., {Bertin}, E., {et~al.} 2022{\natexlab{a}}, arXiv e-prints, arXiv:2212.02428

\bibitem[{{K{\"u}mmel} {et~al.}(2022{\natexlab{b}}){K{\"u}mmel}, {Vassallo}, {Dabin}, \& {Gracia Carpio}}]{2022ASPC..532..329K}
{K{\"u}mmel}, M., {Vassallo}, T., {Dabin}, C., \& {Gracia Carpio}, J. 2022{\natexlab{b}}, in ASP Conf. Ser., Vol. 532, Astronomical Data Analysis Software and Systems XXX, ed. J.~E. {Ruiz}, F.~{Pierfedereci}, \& P.~{Teuben}, 329

\bibitem[{{Lotz} {et~al.}(2004){Lotz}, {Primack}, \& {Madau}}]{Lotz2004}
{Lotz}, J.~M., {Primack}, J., \& {Madau}, P. 2004, \aj, 128, 163

\bibitem[{{Magnier} {et~al.}(2020){Magnier}, {Schlafly}, {Finkbeiner}, {Tonry}, {Goldman}, {R{\"o}ser}, {Schilbach}, {Casertano}, {Chambers}, {Flewelling}, {Huber}, {Price}, {Sweeney}, {Waters}, {Denneau}, {Draper}, {Hodapp}, {Jedicke}, {Kaiser}, {Kudritzki}, {Metcalfe}, {Stubbs}, \& {Wainscoat}}]{Magnier2020}
{Magnier}, E.~A., {Schlafly}, E.~F., {Finkbeiner}, D.~P., {et~al.} 2020, \apjs, 251, 6

\bibitem[{Masters {et~al.}(2024)Masters, Galloway, Fortson, Lintott, Read, Scarlata, Simmons, Walmsley, \& Willett}]{mastersGalaxyZooMorphologies2024}
Masters, K.~L., Galloway, M., Fortson, L., {et~al.} 2024, Research Notes of the AAS, 8, 198, arXiv:2408.10160

\bibitem[{{Merlin} {et~al.}(2022){Merlin}, {Bonchi}, {Paris}, {Belfiori}, {Fontana}, {Castellano}, {Nonino}, {Polenta}, {Santini}, {Yang}, {Glazebrook}, {Treu}, {Roberts-Borsani}, {Trenti}, {Birrer}, {Brammer}, {Grillo}, {Calabr{\`o}}, {Marchesini}, {Mason}, {Mercurio}, {Morishita}, {Strait}, {Boyett}, {Leethochawalit}, {Nanayakkara}, {Vulcani}, {Bradac}, \& {Wang}}]{Merlin2022}
{Merlin}, E., {Bonchi}, A., {Paris}, D., {et~al.} 2022, \apjl, 938, L14

\bibitem[{{Merlin} {et~al.}(2016){Merlin}, {Bourne}, {Castellano}, {Ferguson}, {Wang}, {Derriere}, {Dunlop}, {Elbaz}, \& {Fontana}}]{Merlin2016}
{Merlin}, E., {Bourne}, N., {Castellano}, M., {et~al.} 2016, \aap, 595, A97

\bibitem[{{Merlin} {et~al.}(2015){Merlin}, {Fontana}, {Ferguson}, {Dunlop}, {Elbaz}, {Bourne}, {Bruce}, {Buitrago}, {Castellano}, {Schreiber}, {Grazian}, {McLure}, {Okumura}, {Shu}, {Wang}, {Amor{\'\i}n}, {Boutsia}, {Cappelluti}, {Comastri}, {Derriere}, {Faber}, \& {Santini}}]{Merlin2015}
{Merlin}, E., {Fontana}, A., {Ferguson}, H.~C., {et~al.} 2015, \aap, 582, A15

\bibitem[{{Merlin} {et~al.}(2019){Merlin}, {Pilo}, {Fontana}, {Castellano}, {Paris}, {Roscani}, {Santini}, \& {Torelli}}]{Merlin2019}
{Merlin}, E., {Pilo}, S., {Fontana}, A., {et~al.} 2019, \aap, 622, A169

\bibitem[{{Odewahn} {et~al.}(2004){Odewahn}, {de Carvalho}, {Gal}, {Djorgovski}, {Brunner}, {Mahabal}, {Lopes}, {Moreira}, \& {Stalder}}]{odewahn2004}
{Odewahn}, S.~C., {de Carvalho}, R.~R., {Gal}, R.~R., {et~al.} 2004, \aj, 128, 3092

\bibitem[{{Planck Collaboration}(2014)}]{Planck2013}
{Planck Collaboration}. 2014, \aap, 571, 37

\bibitem[{Serrano {et~al.}(2024)Serrano, Hudelot, Seidel, Pollack, Jullo, Torradeflot, Benielli, Fahed, Auphan, Carretero, Aussel, Casenove, Castander, Davies, Fourmanoit, Huot, Kara, Keihänen, Kermiche, Okumura, Zoubian, Ealet, Boucaud, Bretonnière, Casas, Clément, Duncan, George, Kiiveri, Kurki-Suonio, Kümmel, Laugier, Mainetti, Mohr, Montoro, Neissner, Rosset, Schirmer, Tallada-Crespí, Tonello, Venhola, Verderi, Zacchei, Aghanim, Altieri, Amara, Andreon, Auricchio, Azzollini, Baccigalupi, Baldi, Bardelli, Basset, Battaglia, Bernardeau, Bodendorf, Bonino, Branchini, Brescia, Brinchmann, Camera, Candini, Capobianco, Carbone, Casas, Castellano, Castignani, Cavuoti, Cimatti, Cledassou, Colodro-Conde, Congedo, Conselice, Conversi, Copin, Corcione, Courbin, Courtois, Crocce, Cropper, Da~Silva, Degaudenzi, De~Lucia, Di~Giorgio, Dinis, Dubath, Dupac, Dusini, Farina, Farrens, Ferriol, Frailis, Franceschi, Franzetti, Galeotta, Garilli, Gillard, Gillis, Giocoli, Granett, Grazian, Grupp, Guzzo, Haugan, Hoar,
  Hoekstra, Holmes, Hook, Hormuth, Hornstrup, Jahnke, Joachimi, Kiessling, Kitching, Kohley, Kunz, Le~Boulc’h, Liebing, Ligori, Lilje, Lindholm, Lloro, Maino, Maiorano, Mansutti, Marcin, Marggraf, Markovic, Martinelli, Martinet, Marulli, Massey, Maurogordato, Medinaceli, Mei, Melchior, Mellier, Meneghetti, Merlin, Meylan, Moresco, Morris, Moscardini, Munari, Nakajima, Niemi, Nutma, Padilla, Paltani, Pasian, Pedersen, Percival, Pettorino, Pires, Polenta, Poncet, Popa, Pozzetti, Raison, Rebolo, Renzi, Rhodes, Riccio, Romelli, Roncarelli, Rossetti, Rusholme, Saglia, Sakr, Sánchez, Sapone, Sartoris, Sauvage, Schneider, Schrabback, Scodeggio, Secroun, Sirignano, Sirri, Skottfelt, Stanco, Starck, Steinwagner, Taylor, Teplitz, Tereno, Toledo-Moreo, Tutusaus, Valentijn, Valenziano, Vassallo, Veropalumbo, Wang, Weller, Zamorani, Zucca, Biviano, Bozzo, Di~Ferdinando, Farinelli, Graciá-Carpio, Mauri, Scottez, Tenti, Akrami, Allevato, Ballardini, Blanchard, Borgani, Borlaff, Bruton, Burigana, Cappi, Carvalho, Castro,
  Cañas-Herrera, Chambers, Cooray, Coupon, Davini, de~la Torre, Desai, Desprez, Díaz-Sánchez, Di~Domizio, Dole, Vigo, Escoffier, Ferrero, Finelli, Gabarra, Ganga, Garcia-Bellido, Gaztanaga, Giacomini, Gozaliasl, Gregorio, Hildebrandt, Huertas-Company, Ilbert, Muñoz, Kajava, Kansal, Kirkpatrick, Legrand, Loureiro, Macias-Perez, Magliocchetti, Maoli, Martins, Matthew, Maurin, Metcalf, Migliaccio, Monaco, Morgante, Nadathur, Nucita, Pöntinen, Popa, Porciani, Potter, Reimberg, Schneider, Sereno, Shulevski, Simon, Mancini, Stadel, Tewes, Teyssier, Toft, Tucci, Valiviita, Viel, \& Zinchenko}]{SIM2024}
Serrano, S., Hudelot, P., Seidel, G., {et~al.} 2024, \aap, 690, A103

\bibitem[{{S{\'e}rsic}(1963)}]{1963BAAA....6...41S}
{S{\'e}rsic}, J.~L. 1963, Boletin de la Asociacion Argentina de Astronomia La Plata Argentina, 6, 41

\bibitem[{Sevilla-Noarbe {et~al.}(2018)Sevilla-Noarbe, Hoyle, Marchã, Soumagnac, Bechtol, Drlica-Wagner, Abdalla, Aleksić, Avestruz, Balbinot, Banerji, Bertin, Bonnett, Brunner, Carrasco-Kind, Choi, Giannantonio, Kim, Lahav, Moraes, Nord, Ross, Rykoff, Santiago, Sheldon, Wei, Wester, Yanny, Abbott, Allam, Brooks, Carnero-Rosell, Carretero, Cunha, da Costa, Davis, de Vicente, Desai, Doel, Fernandez, Flaugher, Frieman, Garcia-Bellido, Gaztanaga, Gruen, Gruendl, Gschwend, Gutierrez, Hollowood, Honscheid, James, Jeltema, Kirk, Krause, Kuehn, Li, Lima, Maia, March, McMahon, Menanteau, Miquel, Ogando, Plazas, Sanchez, Scarpine, Schindler, Schubnell, Smith, Smith, Soares-Santos, Sobreira, Suchyta, Swanson, Tarle, Thomas, Tucker, Walker, \& (The DES Collaboration)}]{SpreadModel2018}
Sevilla-Noarbe, I., Hoyle, B., Marchã, M.~J., {et~al.} 2018, MNRAS, 481, 5451

\bibitem[{Sharon {et~al.}(2022)Sharon, Cerny, Rigby, Florian, Bayliss, Dahle, Gladders, \& Mahler}]{sharon2022}
Sharon, K., Cerny, C., Rigby, J.~R., {et~al.} 2022, HST-Based Lens Model of the First Extragalactic JWST Science Target, SDSS J1226+2152, in Preparation for TEMPLATES

\bibitem[{{Simmons} {et~al.}(2017){Simmons}, {Lintott}, {Willett}, {Masters}, {Kartaltepe}, {H{\"a}u{\ss}ler}, {Kaviraj}, {Krawczyk}, {Kruk}, {McIntosh}, {Smethurst}, {Nichol}, {Scarlata}, {Schawinski}, {Conselice}, {Almaini}, {Ferguson}, {Fortson}, {Hartley}, {Kocevski}, {Koekemoer}, {Mortlock}, {Newman}, {Bamford}, {Grogin}, {Lucas}, {Hathi}, {McGrath}, {Peth}, {Pforr}, {Rizer}, {Wuyts}, {Barro}, {Bell}, {Castellano}, {Dahlen}, {Dekel}, {Ownsworth}, {Faber}, {Finkelstein}, {Fontana}, {Galametz}, {Gr{\"u}tzbauch}, {Koo}, {Lotz}, {Mobasher}, {Mozena}, {Salvato}, \& {Wiklind}}]{2017MNRAS.464.4420S}
{Simmons}, B.~D., {Lintott}, C., {Willett}, K.~W., {et~al.} 2017, \mnras, 464, 4420

\bibitem[{{Slater} {et~al.}(2020){Slater}, {Ivezi{\'c}}, \& {Lupton}}]{slater2020}
{Slater}, C.~T., {Ivezi{\'c}}, {\v{Z}}., \& {Lupton}, R.~H. 2020, \aj, 159, 65

\bibitem[{{Soumagnac} {et~al.}(2015){Soumagnac}, {Abdalla}, {Lahav}, {Kirk}, {Sevilla}, {Bertin}, {Rowe}, {Annis}, {Busha}, {Da Costa}, {Frieman}, {Gaztanaga}, {Jarvis}, {Lin}, {Percival}, {Santiago}, {Sabiu}, {Wechsler}, {Wolz}, \& {Yanny}}]{soumagnac2015}
{Soumagnac}, M.~T., {Abdalla}, F.~B., {Lahav}, O., {et~al.} 2015, \mnras, 450, 666

\bibitem[{{Szalay} {et~al.}(1999){Szalay}, {Connolly}, \& {Szokoly}}]{1999AJ....117...68S}
{Szalay}, A.~S., {Connolly}, A.~J., \& {Szokoly}, G.~P. 1999, \aj, 117, 68

\bibitem[{{Tohill} {et~al.}(2021){Tohill}, {Ferreira}, {Conselice}, {Bamford}, \& {Ferrari}}]{Tohill2021}
{Tohill}, C., {Ferreira}, L., {Conselice}, C.~J., {Bamford}, S.~P., \& {Ferrari}, F. 2021, \apj, 916, 4

\bibitem[{Tramacere {et~al.}(2016)Tramacere, Paraficz, Dubath, Kneib, \& Courbin}]{Tramacere2016}
Tramacere, A., Paraficz, D., Dubath, P., Kneib, J.-P., \& Courbin, F. 2016, MNRAS, 463, 2939

\bibitem[{Walmsley {et~al.}(2023)Walmsley, Allen, Aussel, Bowles, Gregorowicz, Slijepcevic, Lintott, Scaife, Jabłońska, Karchev, Lanzieri, Mohan, O’Ryan, Saiguhan, Suárez, Guerra-Varas, \& Velu}]{Walmsley2023zoobot}
Walmsley, M., Allen, C., Aussel, B., {et~al.} 2023, Journal of Open Source Software, 8, 5312, publisher: The Open Journal

\bibitem[{Walmsley {et~al.}(2024)Walmsley, Bowles, Scaife, Makechemu, Gordon, Ferguson, Mann, Pearson, Popp, Bovy, Speagle, Dickinson, Fortson, Géron, Kruk, Lintott, Mantha, Mohan, O'Ryan, \& Slijepevic}]{walmsleyScalingLawsGalaxy2024}
Walmsley, M., Bowles, M., Scaife, A. M.~M., {et~al.} 2024, Scaling Laws for Galaxy Images

\bibitem[{{Walmsley} {et~al.}(2023){Walmsley}, {G{\'e}ron}, {Kruk}, {Scaife}, {Lintott}, {Masters}, {Dawson}, {Dickinson}, {Fortson}, {Garland}, {Mantha}, {O'Ryan}, {Popp}, {Simmons}, {Baeten}, \& {Macmillan}}]{2023MNRAS.526.4768W}
{Walmsley}, M., {G{\'e}ron}, T., {Kruk}, S., {et~al.} 2023, \mnras, 526, 4768

\bibitem[{{Walmsley} {et~al.}(2022{\natexlab{a}}){Walmsley}, {Lintott}, {G{\'e}ron}, {Kruk}, {Krawczyk}, {Willett}, {Bamford}, {Kelvin}, {Fortson}, {Gal}, {Keel}, {Masters}, {Mehta}, {Simmons}, {Smethurst}, {Smith}, {Baeten}, \& {Macmillan}}]{2022MNRAS.509.3966W}
{Walmsley}, M., {Lintott}, C., {G{\'e}ron}, T., {et~al.} 2022{\natexlab{a}}, \mnras, 509, 3966

\bibitem[{{Walmsley} {et~al.}(2022{\natexlab{b}}){Walmsley}, {Slijepcevic}, {Bowles}, \& {Scaife}}]{2022mla..confE..29W}
{Walmsley}, M., {Slijepcevic}, I., {Bowles}, M.~R., \& {Scaife}, A. 2022{\natexlab{b}}, in Machine Learning for Astrophysics, 29

\bibitem[{{Willett} {et~al.}(2017){Willett}, {Galloway}, {Bamford}, {Lintott}, {Masters}, {Scarlata}, {Simmons}, {Beck}, {Cardamone}, {Cheung}, {Edmondson}, {Fortson}, {Griffith}, {H{\"a}u{\ss}ler}, {Han}, {Hart}, {Melvin}, {Parrish}, {Schawinski}, {Smethurst}, \& {Smith}}]{2017MNRAS.464.4176W}
{Willett}, K.~W., {Galloway}, M.~A., {Bamford}, S.~P., {et~al.} 2017, \mnras, 464, 4176

\bibitem[{{Willett} {et~al.}(2013){Willett}, {Lintott}, {Bamford}, {Masters}, {Simmons}, {Casteels}, {Edmondson}, {Fortson}, {Kaviraj}, {Keel}, {Melvin}, {Nichol}, {Raddick}, {Schawinski}, {Simpson}, {Skibba}, {Smith}, \& {Thomas}}]{2013MNRAS.435.2835W}
{Willett}, K.~W., {Lintott}, C.~J., {Bamford}, S.~P., {et~al.} 2013, \mnras, 435, 2835

\end{thebibliography}

\begin{appendix}
\section{The OU MER Data Model}
\label{appendix:a}

All \Euclid data products must comply with the SGS Data Model (DM) and are documented in the MER Data Product Description Document\footnote{\url{https://euclid.esac.esa.int/dr/q1/dpdd}} (DPDD). The DPDD also states, for each product, if the product itself is ingested into the \Euclid Science Archive System \citep[SAS,][]{Q1-TP001}.
In this section we provide a brief summary of the OU-MER pipeline deliverables.

\subsection{Output images and segmentation maps}
\label{appendix:images}

Together with the photometric and morphological catalogues, the OU-MER pipeline outputs the set of images on which the photometric and morphological analysis is performed. Those images areas follows.

\begin{itemize}
    \item The background-subtracted mosaics, one for each input band. The associated data product in the DPDD is  \textsf{DpdMerBksMosaic}.
    \item The mosaics used to perform the object detection, one for VIS (\verb|VIS|) and one for the stack of the three NIR bands (\verb|NIR_STACK|). The associated data product in the DPDD is \textsf{DpdMerDetectionMosaic}.
    \item A map showing the connected pixels of the objects detected on the corresponding detection mosaics (\verb|VIS|+\verb|NIR_STACK|). The associated data product in the DPDD is \textsf{DpdMerSegmentationMap}.
\end{itemize}

\begin{figure}
    \centering
    \includegraphics[width=0.45\textwidth]{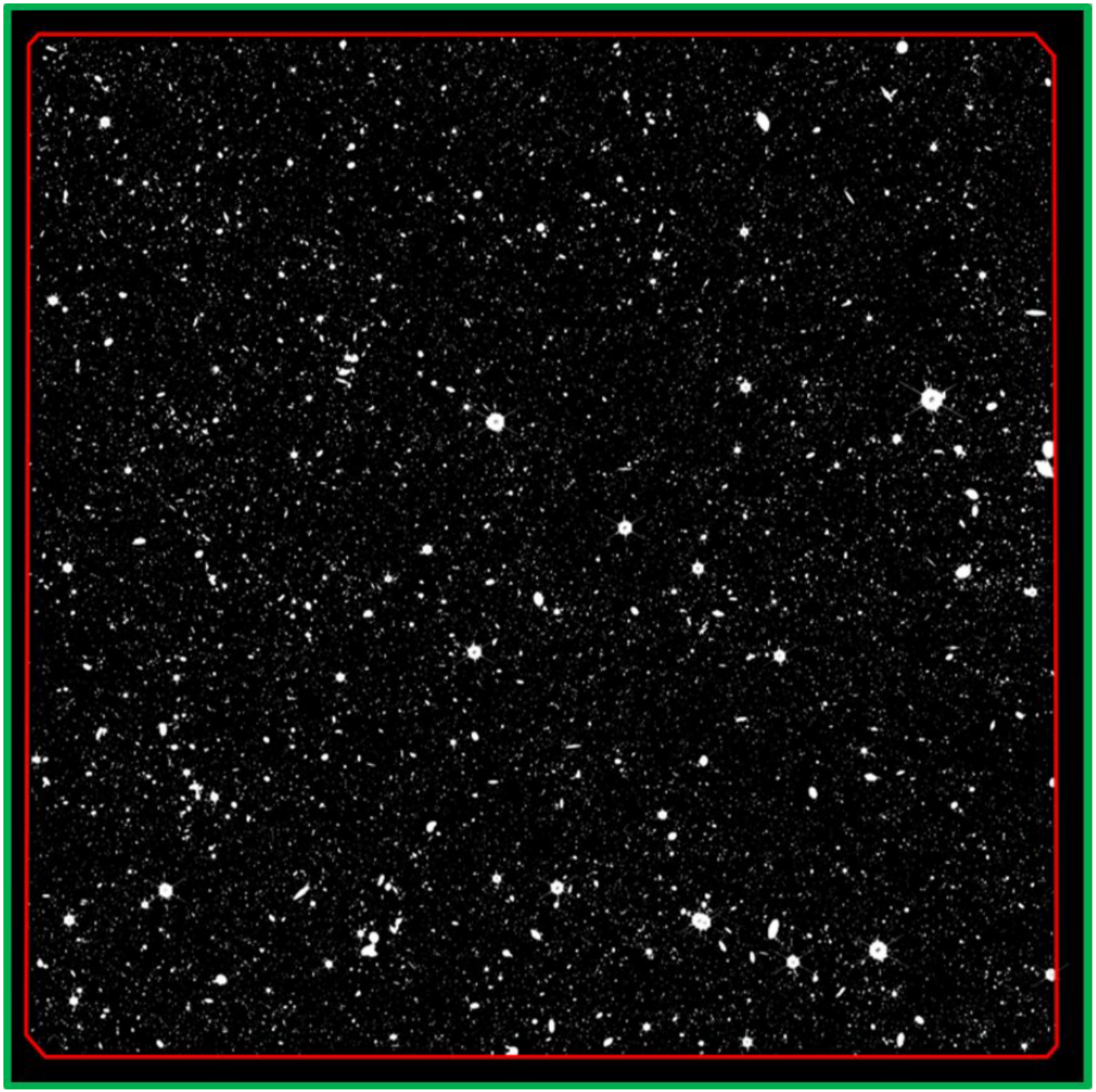}
    \caption{Outer and inner tile regions relative to an OU-MER mosaic. The green frame shows the MER tile outer region, while the red frame is the inner MER tile region. The segmentation map and the catalogue refer to the inner tile region; all the sources between the outer and inner regions are not catalogued. A value of $-1$ is assigned to pixels related to sources outside the inner tile region. A standard MER tile (outer region) covers an area of $32' \times 32'$. The border between the outer and inner region has a depth of $2'$.}
    \label{mosaic_vs_segmap}
\end{figure}

Note that the segmentation map, together with the catalogue, refers to the inner region of a MER tile (see Sect. \ref{subsect:tile}), while the mosaics cover the whole tile region. Figure \ref{mosaic_vs_segmap} shows the inner and outer tile region of a MER tile with respect to a reference mosaic.

\subsection{Naming convention and data format}\label{appendix:naming}

The names of the data files associated with the MER data products were formed using the following format:

\begin{quote}
\verb|EUC_MER_[TYPE]_TILE[INDEX]-[RND]_[TIME].[EXT]|,    
\end{quote}

where \verb|TYPE| is one of the file types described below, \verb|INDEX| is the associated MER tile index (e.g., \verb|102022523|), \verb|RND| is a random string that it is used to avoid possible file name duplications (e.g., \verb|15E4F0|), \verb|TIME| is the file creation timestamp (e.g., \verb|20241121T114727.022303Z|), and \verb|EXT| is the file extension (\verb|fits.gz| or \verb|json|).

The following are the file types used in the MER PF (see Table~\ref{tab:band_list_q1} for a complete list of band label values):

\begin{itemize}
    \item \verb|BGSUB-MOSAIC-<band>|: Mosaic image data.
    \item \verb|MOSAIC-<band>-RMS|: Mosaic rms data.
    \item \verb|MOSAIC-<band>-FLAG|: Mosaic flag data.
    \item \verb|BGMOD-<band>|: Mosaic subtracted background model.
    \item \verb|FILTER-TRANSMISSION-<band>|: Mosaic average filter transmission wavelength.
    \item \verb|DETECTOR-LAYERING-<band>|:  Mosaic input layers information.
    \item \verb|CATALOG-PSF-<band>|: Mosaic catalogue PSF.
    \item \verb|GRID-PSF-<band>|: Mosaic grid PSF.
    \item \verb|STAR-MASKS-<band>|: Mosaic bright star spatial masks.
    \item \verb|FINAL-SEGMAP|: Detection segmentation map.
    \item \verb|FINAL-CAT|:  Catalogue with the main photometry information.
    \item \verb|FINAL-MORPH-CAT|:  Catalogue with the morphology information.
    \item \verb|FINAL-CUTOUTS-CAT|: Catalogue with the source cut-out co-ordinates.
    \item \verb|FINAL-DEEP-PHOTO|: Catalogue with extra photometry information available only on the Euclid DEEP fields.
\end{itemize}

The DPDD provides more details about the content of the different MER data files.

\subsection{Column-by-column catalogue description}

The catalogues produced by OU MER pipeline list more that $3\,200$ columns. The structure of  the photometry catalogues (\verb|EUC_MER_FINAL-CAT| and \verb|EUC_MER_FINAL-DEEP-CAT|) is summarised in sections \href{http://st-dm.pages.euclid-sgs.uk/data-product-doc/dmq1/merdpd/dpcards/mer_finalcatalog.html#main-catalog-fits-file}{Main catalogue FITS file} and \href{http://st-dm.pages.euclid-sgs.uk/data-product-doc/dmq1/merdpd/dpcards/mer_finalcatalog.html#deep-field-photometry-catalog-fits-file}{DEEP field photometry catalogue FITS file} of the MER DPDD. Section \href{http://st-dm.pages.euclid-sgs.uk/data-product-doc/dmq1/merdpd/dpcards/mer_finalcatalog.html#morphology-catalog-fits-file}{Morphology catalogue FITS file} describes the morphology catalogue (\verb|EUC_MER_FINAL-MORPHO-CAT|), while section \href{http://st-dm.pages.euclid-sgs.uk/data-product-doc/dmq1/merdpd/dpcards/mer_finalcatalog.html#cutouts-catalog-fits-file}{Cutouts catalogue FITS file} lists the columns available in the cut-out catalogue (\verb|EUC_MER_FINAL-CUTOUTS-CAT|).
\end{appendix}

\end{document}